%% file: spinReviewTables.tex
\documentclass[aps, prx, showpacs, twocolumn, superscriptaddress, notitlepage, longbibliography, floatfix, nofootinbib]{revtex4-2}
\usepackage{graphicx}
\usepackage{amsmath}
\usepackage{hyperref}
\usepackage{multirow}
\usepackage[T1]{fontenc}

\usepackage{stix}

\usepackage[section]{placeins}

\usepackage[table, dvipsnames]{xcolor}
\definecolor{mygray}{gray}{0.7}

\usepackage{dcolumn}

\usepackage{siunitx}

\usepackage{booktabs}

\usepackage{enumitem}

\makeatletter
\newenvironment{tablehere}[1][]%
  {\par\addvspace{\baselineskip}\noindent
   \begin{minipage}{\linewidth}\def\@captype{table}
   \centering\small}%
  {\end{minipage}\par\addvspace{\baselineskip}}
\makeatother

\ifdefined\DeclareUnicodeCharacter
  \DeclareUnicodeCharacter{2219}{$\cdot$}
    \DeclareUnicodeCharacter{2212}{$-$}
   \DeclareUnicodeCharacter{22C5}{$\cdot$}
    \DeclareUnicodeCharacter{03BC}{$\mu$} 
\fi

\makeatletter
\newcommand*\labelcounter[2]{\begingroup
  \protected@edef\@currentlabel{\csname p@#1\endcsname\csname the#1\endcsname}%
  \label{#2}\endgroup}
\newcommand*\refsetcounter[2]{\setcounter{#1}{#2}%
  \protected@edef\@currentlabel{\csname p@#1\endcsname\csname the#1\endcsname}%
  }
\makeatother

\makeatletter
\newcommand{\customlabel}[2]{%
   \protected@write \@auxout {}{\string \newlabel {#1}{{#2}{\thepage}{#2}{#1}{}} }%
   \hypertarget{#1}{#2} %
}
\makeatother
\providecommand{\phantomsection}{}%
\AtBeginDocument{\let\textlabel\label}%
\makeatletter
\newcommand{\mylabel}[2]{ 
\raisebox{.7\normalbaselineskip}{\phantomsection}%
  \def\@currentlabel{#1}\textlabel{#2}}
\makeatother

\newcommand{\estimated}{\textit{(estimated) }}
\newcommand{\derived}{\textit{(derived) }}

\newcommand{\TRabi}{T_2^\mathrm{Rabi}}
\newcommand{\TEcho}{T_2^\mathrm{Echo}}
\newcommand{\TDynD}{T_2^\mathrm{DynD}}

\newif\ifIncremental
\Incrementalfalse

\newif\ifFirstTableOnly

\newif\ifSupplementalOnly

\ifIncremental %

\FirstTableOnlytrue
\graphicspath{{./figuresAndTables/Added/}{./figuresAndTables/}} %
\makeatletter
\def\input@path{{./figuresAndTables/Added/}}
\makeatother

\else %
\FirstTableOnlyfalse
\graphicspath{{./figuresAndTables/}}
\makeatletter
\def\input@path{{./figuresAndTables/}}
\makeatother

\fi

\newcommand{\myKey}[1]{$\whitearrowupfrombar$\textit{#1}}
\renewcommand{\myKey}[1]{\textit{#1}}
\renewcommand{\myKey}[1]{`{#1}'}

\newcommand{\key}[1]{$\whitearrowupfrombar$\textit{#1}}
\newcommand{\val}[1]{\rotatebox[origin=c]{180}{$\whitearrowupfrombar$}\textit{#1}}
\renewcommand{\key}[1]{\myKey{#1}}
\renewcommand{\val}[1]{\myKey{#1}}

\newcommand{\recheck}[1]{{#1}}

\newcommand{\separate}[1]
{
 #1
}

\newcounter{rowcount}

\SupplementalOnlyfalse

\begin{document}
\author{Peter Stano}
\affiliation{RIKEN Center for Emergent Matter Science (CEMS), Wako, Saitama 351-0198, Japan}
\affiliation{Institute of Physics, Slovak Academy of Sciences, Dubravska cesta 9, 845 11 Bratislava, Slovakia}
\author{Daniel Loss}
\affiliation{RIKEN Center for Emergent Matter Science (CEMS), Wako, Saitama 351-0198, Japan}
\affiliation{RIKEN Center for Quantum Computing (RQC), Wako, Saitama 351-0198, Japan}
\affiliation{Department of Physics, University of Basel, Klingelbergstrasse 82, CH-4056 Basel, Switzerland}
\date{\today}
\ifSupplementalOnly
\else
\title{
Review of performance metrics of spin qubits in gated semiconducting nanostructures
}
\begin{abstract}
This Technical Review collects values of selected performance characteristics of semiconductor spin qubits defined in electrically controlled nanostructures. The characteristics are envisioned to serve as a community source for the values of figures of merit with agreed-on definitions allowing the comparison of different spin qubit platforms. We include characteristics on the qubit coherence, speed, fidelity, and the qubit size of multi-qubit devices. The focus is on collecting and curating the values of these characteristics as reported in the literature, rather than on their motivation or significance. 
\end{abstract}

\hypersetup{pageanchor=false}
\maketitle
\hypersetup{pageanchor=true}

\section{Scope, format, and aim of this review}
Spin qubits are among platforms pursued to serve as quantum computing hardware. In this Technical Review we focus on spin qubits hosted in semiconducting nanostructures controlled and probed electrically. Their prospect for scalability stems from their compatibility with modern silicon industrial fabrication. Even restricted to gated nanostructures, the field of spin qubits is vast. There is a host of variants on the sample material and structure, device design, or qubit encoding. Although this versatility in the qubit types is beneficial for overcoming possible roadblocks, it also makes comparison of different spin qubits difficult. The main motivation for this Technical Review is to provide a basis for such a comparison and for an assessment of the progress of various spin-qubit types over time. We believe that for such tasks, a reliable database of figures of merit normalized to common definitions is of primary importance, and this is what we provide here.

The restriction to gated nanostructures suggests what is not covered in this Technical Review. We do not include other than solid-state qubits; within the solid-state, we do not include superconducting qubits and qubits based on optically-accessed impurities and self-assembled dots. We also skip qubits based on the spin of atomic nuclei, that is, hyperfine-spin qubits. Even though spin-related, we sacrifice these possible extensions to keep the review manageable in length and in the time of preparation. However, we include some characteristics of charge qubits, that is qubits with states encoded into the charge degree of freedom of a confined particle. One reason is that often the experimental devices are identical for both spin and charge qubit experiments. Another reason is that there are configurations where the spin and charge degrees of freedom are hybridized and tunable. With the character continuously tunable from fully spin- to fully charge-like, it would be difficult to decide objectively which cases to include and which ones not. Finally, we point out that this Technical Review is not meant as an overview of the physics of spin qubits, such as the principles of their operation and measurements, the decoherence channels, and so on. The reader interested in these aspects can consult, for example, Refs.~\cite{oosterkamp_photon_1999, van_der_wiel_electron_2002, schliemann_electron_2003, hanson_spins_2007, kloeffel_prospects_2013, schreiber_quantum_2014, tahan_opinion_2021, chatterjee_semiconductor_2021, gonzalez-zalba_scaling_2021, scappucci_germanium_2021, oiwa_conversion_2017, vandersypen_quantum_2019, kuemmeth_roadmap_2020, chatterjee_semiconductor_2021}. In particular, we point out the recent extensive review in Ref.~\cite{burkard_semiconductor_2023} as an excellent complement, covering the aspects intentionally omitted here.

The core of this Technical Review are the plots and tables of selected qubit characteristics on the qubit coherence, operation speed, operation fidelity, quality factors, and the size of multi-qubit arrays. The related quantities are defined, and their values collected from the literature given, in Sections \ref{sec:coherence}-\ref{sec:size}, respectively.

We hope this Technical Review will become useful and used as a database for spin-qubit characteristics. To this end, it is crucial that the database is up-to-date, error-free, and contains relevant quantities. These goals can hardly be met without an active participation of the spin-qubit community. We encourage the members of the community to provide us with feedback on any errors, omissions, or suggestions for changes.

\subsection{We present a database of values}

Though not necessarily of a concern for the reader, we note that the presented collection reflects a formally defined database. This fact might be useful to understand certain nomenclature, details of the presentation, and requirements on the included values and possible extensions. Let us briefly explain these aspects.

Every \myKey{value} given in this Technical Review belongs to a certain \myKey{attribute}. These two basic elements define the database and constitute keywords with precise meaning. The \myKey{attributes} are used as headers in tables and axes labels in plots. \recheck{For example, the first line of Table I contains the value \myKey{LD/e} under the attribute \myKey{Qubit}.} It means that the corresponding experiment used a qubit encoded into the spin of an electron. Whereas most of the attributes are self-explanatory, they are additionally listed alphabetically, together with their definitions, in Appendix \ref{sec:vocabulary}. To draw attention to their special role, we mark the attributes with single quotation marks.
This distinction is made only in this (first) section and Appendix~\ref{sec:vocabulary}.

\subsection{Spin-qubit types, device geometries, material choices}

Before we discuss specific characteristics, we comment on their common aspects. These stem from the fact that each \myKey{value} inherits a set of characteristics from the publication it was reported in and the experimental device it was measured with. We describe these common aspects below.

Every value given in this Technical Review (such as a coherence time of 10 ns) is a result of a measurement with a specific device and qubit type, reported in a published reference.\footnote{\recheck{Note that whereas we count arXiv preprints as `published references', the majority of the entries are from peer-reviewed journal publications. The current bibliography for the database contains 8 arXiv preprints, only four of which were uploaded before the year 2024, Refs.~\cite{hayes_lifetime_2009,cerfontaine_feedback-tuned_2016,ruffino_integrated_2021}.}} This means, first of all, that we include only numbers explicitly stated and directly measured. We do not include extrapolations to other conditions or materials. Also, we do not usually derive the values ourselves even if it would be possible. For example, if the reference gives the operation time and the dephasing time, but does not state the quality factor, we do not evaluate the latter ourselves. However, if a quantity is discussed and presented as a figure (for example, the operation time implied from oscillations displayed by a resonantly driven qubit), we might include a value read-off from the figure. In such cases, the table entry contains a \myKey{Note} with keywords $\derived$\! or $\estimated$\!, which are explained in Appendix \ref{sec:vocabulary}. With this requirement, every value given in this Technical Review should be easy to find using the reference, given under the attribute \myKey{Reference}, and within the reference as described by the attribute \myKey{Source}. An example of the latter is say ``page 4 and Fig.~1b''. One possible difference between the value given here and in the original work is normalization. In that case, a \myKey{Note} explains how the value was converted. Any additional information, for example, alerting on an unusual configuration or a specific method used in the experiment, is also given as a \myKey{Note}.

The second group of common characteristics concerns the details of the qubit. We found it useful to categorize the following: the sample material, the geometry of the host, and the qubit type. They belong to the attributes \myKey{Material}, \myKey{Host}, and \myKey{Qubit}, respectively. Whereas the value for material, for example, \myKey{Si/SiGe}, is self-explanatory,\footnote{Qubits based on silicon-oxide structures is one case that needs a comment. A unique identification in this Technical Review for such structures is the value \myKey{Si/SiO$_2$} of the attribute \myKey{Material}. The attribute \myKey{Host} is typically also straightforward to assign, being either \myKey{2D}, for example for an epilayer, or \myKey{1D} for finFETs or structures denoted as nanowires by their authors. One can often find further specifications for such devices, such as [complementary-]metal-oxide-semiconductor([C]MOS), silicon-on-insulator(SOI), field-effect-transistor(FET), foundry-compatible, and similar, including their combinations. These specifications hint at the fabrication details and the degree of compatibility with the industrial silicon technology. However, they are sometimes used interchangeably, even within one laboratory. Since the fabrication details are not our focus, we do not include such additional specifications even if given in the original work.}
 in some figures we group several different materials under a common tag, such as \myKey{Si}. Concerning the host geometry, we discriminate (qubits based on gating) the structures which are (quasi-)\myKey{2D}, for example, a 2D electron gas (2DEG), (quasi-)\myKey{1D}, for example a nanowire, and quasi-zero-dimensional, denoted by \myKey{imp}, an example being an implanted impurity. Some of these are not clear-cut cases, for example the hut wires with a flat cross-section \cite{watzinger_heavy-hole_2016}, or some variants of CMOS devices \cite{voisin_electrical_2016}; nevertheless, we assign both to \myKey{1D}. Finally, perhaps the largest variation exists among the qubit types. We distinguish the charge carrier: conduction electron, valence hole, and atomic impurity; and the spin-encoding: spin-1/2 (\myKey{LD}), singlet-triplet (\myKey{ST}), and hybrid (\myKey{HY}) qubits. We have not found it beneficial to subdivide further the hybrid qubits: everything which is not a spin-1/2 or a singlet-triplet is assigned the value hybrid here. A note might give additional information on the qubit type. The reader would benefit from consulting Appendix \ref{sec:vocabulary} now, to understand the database organization through attributes and values.

\subsection{The choice of values to collect is subjective}

A disclaimer is in order here: assigning a single specific value for each characteristic is necessary for a meta-analysis such as done here. However, converting an experimental investigation into a single number is inevitably a drastic compression. The choice of the value to quote requires subjective judgment: in a typical experiment, the value of a given figure of merit is seldom a unique value, but rather spans a range, sometimes a very large range, such as several orders of magnitude. We tend to take the most beneficial values, but it does not mean we simply take the largest ones. Especially when an experiment presents a set of values for different characteristics, we try to choose a representative set measured at a common setting. For example, in an experiment with three qubits, where each is measured for relaxation, dephasing, and the echo coherence time, we do not simply take the largest value seen among all experiments. We choose a qubit and quote the three numbers for this particular qubit. We proceed similarly when several characteristics are measured under various conditions. Our overall approach is to adopt values that are mutually consistent (such as the coherence time and the operation speed) as much as possible. Nevertheless, we warn the reader that all such choices are largely subjective. The final authority to judge the value meaning and importance is the original reference itself. 

\subsection{What we do and what we do not}

We would like to reiterate our goals, since this work is not a standard review. Our primary target is to provide a database of figures of merit and make its content accessible. This includes downloadable tables and figures, including the interface to produce figures and tables according to the user's design, and a public repository including the environment for feedback and discussions. This content is accessible through a public data depository \cite{noauthor_notitle_nodate}. The repository includes detailed instructions on providing feedback, though writing an email directly to the corresponding author is also encouraged. However, giving a subjective view of the spin-qubit field in the form of qubit-suitability judgments, interpretations, outlooks, summaries, recommendations, predictions, and similar, is not our target and the reader will not find much of such type of content here.

\separate{
After these preliminaries, we will present the spin-qubit figures of merit published until \recheck{the end of May 2025}.
}

\section{Coherence times}

\label{sec:coherence}

\begin{figure}
\centering
  \includegraphics[width=0.49\linewidth]{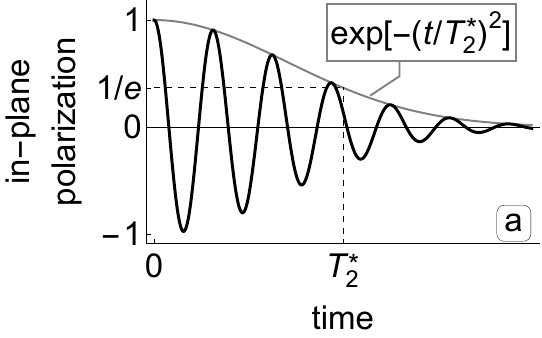} \hfill
  \includegraphics[width=0.49\linewidth]{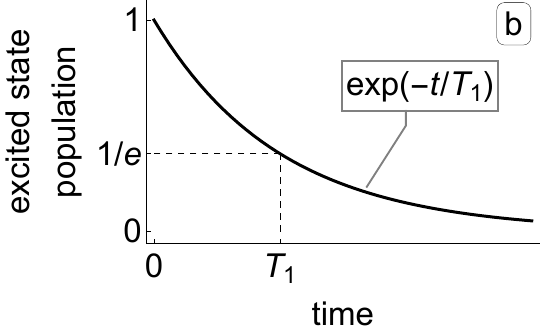} \hfill\\
  \includegraphics[width=0.99\linewidth]{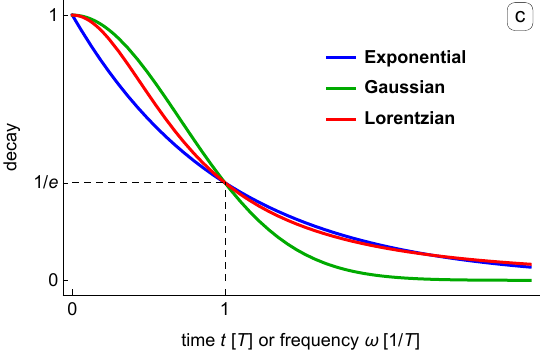} \hfill
  \caption{\label{fig:envelopes} Typical decay-curve envelopes. (a) Oscillations with a Gaussian decay envelope, typically appearing in Ramsey experiments measuring the inhomogeneous dephasing time $T_2^*$ in the time domain. (b) Exponential decay typical for relaxation processes, reflecting the relaxation time $T_1$. (c) Three curves are plotted for comparison. They are normalized to reach value 1 at the x-axis parameter equal to 0 and value $1/e$ at the x-axis parameter equal to 1. The Gaussian and Lorentzian, the Fourier transforms of the two envelopes given in (a) and (b), arise in probing the decay in the frequency domain. In that case, the horizontal axis on the figure is the frequency $\omega$, see Eq.~\eqref{eq:timeAndFreq}.}
\end{figure}

The largest amount of published data on spin qubits refers to their coherence times. Qualitatively, a coherence time extracted in an experiment has the meaning of a time during which state oscillations stemming from quantum mechanical superpositions can be observed. 

\subsection{Definition and meaning of experimentally-extracted coherence times}

Additional specifications of the conditions under which such a superposition decay is observed lead to several variants of the coherence time. The inhomogeneous dephasing time $T_2^*$ implies a Ramsey experiment, meaning the following sequence: the qubit is initialized to a state polarized within the equatorial plane of the Bloch sphere, for example along the $x$ axis, and is left to precess freely for time $t$ after which the in-plane polarization is measured; the evolution time $t$ is varied, and for each value of $t$ the sequence is repeated to gather enough statistics. A typical time-trace of the averaged signal fits a cosine with a Gaussian decay envelope $f(t)$,
\begin{equation}
P_x(t) = f(t)\frac{1+\cos \omega t}{2} = \exp\left[-(t / T_2^*)^2\right] \frac{1+\cos \omega t}{2},
\label{eq:dephasing}
\end{equation}
where $\omega$ is the precession frequency. This curve is plotted in Fig.~\ref{fig:envelopes}(a). 

The decay of a spin qubit in a Ramsey experiment described by Eq.~\eqref{eq:dephasing} is often due to fluctuating nuclear spins. The strong effect of nuclei on the spin coherence was predicted in Refs.~\cite{khaetskii_electron_2002, merkulov_electron_2002} and confirmed experimentally in Ref.~\cite{petta_coherent_2005}. Because the dynamics of nuclear spins is slow, one can protect the spin-qubit coherence using the spin-echo techniques developed in the field of Nuclear Magnetic Resonance \cite{slichter_principles_1996}. The simplest protecting protocol is the Hahn echo\cite{hahn_spin_1950}. It means that the spin is flipped (rotated around an in-plane axis by angle $\pi$) in the middle of the free evolution, at time $t/2$, of the described Ramsey sequence. The coherence time measured under a Hahn echo is denoted in this Technical Review as $ \TEcho$. Already in Ref.~\cite{petta_coherent_2005}, the Hahn echo prolonged the spin qubit coherence by a factor of a hundred. There are more elaborate protocols, applying more echo pulses, which prolong the coherence further. Although there are several different variants of such protocols \cite{alvarez_performance_2010}, here we assign any sequence containing more than a single echo under a common category, denoting the coherence time as $ \TDynD$. A typical member of this family is the Carr-Purcell-Meiboom-Gill (CPMG) protocol, with which the coherence of a singlet-triplet spin-qubit was prolonged to almost a millisecond in Ref.~\cite{malinowski_notch_2016}. 

All these protocols aim at prolonging the coherence of an idling spin qubit. The decay of a driven spin, meaning the decay of coherent Rabi oscillations, is another important time scale that is often reported. It is denoted here as $ \TRabi$. The understanding that the decay of coherence of driven and idling spins can strongly differ goes back to Alfred Redfield (Ref.~\cite{redfield_nuclear_1955}) and has been demonstrated with a spin qubit \cite{laucht_dressed_2016}. Finally, we include also the time $T_1$, called the relaxation time, to denote the decay of qubit energy. That it is a different type of process is denoted by its subscript `1' as opposed to `2' for the times describing the decay of phase. These subscripts refer to the notation usual in Bloch equations, where these two processes with distinct physical origins are called the `longitudinal' and `transverse' relaxation, respectively \cite{bloch_nuclear_1946}. See Appendix~\ref{app:notation} for the notation of various decay times.

Whereas we attributed the Gaussian decay in Eq.~\eqref{eq:dephasing} to nuclei as an example, the coherence-times nomenclature applies in the same way, irrespective of the noise source. Low-frequency charge and high-frequency phonon noise, influencing the spins through spin-orbit interactions, are most relevant. More importantly for the coherence-time measurement, the functional form of the envelope $f(t)$ in Eq.~\eqref{eq:dephasing} is often different from the Gaussian. Another typical case is an exponential,
\begin{equation}
f(t) = \exp(-t / T_1).
\label{eq:relaxation}
\end{equation}
We have suggestively used $T_1$ for the time scale, as the energy relaxation is often described by such an envelope. The function is plotted in Fig.~\ref{fig:envelopes}b. Because of the superimposed oscillations in Eq.~\eqref{eq:dephasing}, it is not easy to discriminate between the exponential and Gaussian decay envelopes. Although the functions differ strongly in their exponential tails, these tails are basically never resolvable, due to measurement errors and statistical fluctuations. If discrimination is possible, it is based on the different shapes of the two functions at short times: linear versus quadratic.

The discrimination becomes even more difficult in experiments where the qubit is probed in the frequency domain. A typical example is recording the amplitude and phase response of a resonant electrical circuit of which the qubit is a part. Both of these quantities are parametrized by the circuit reflection coefficient, a complex number. A standard result for it reads\footnote{
We copy Eq.~(A15) from Ref.~\cite{ezzouch_dispersively_2021}. Ref.~\cite{cottet_cavity_2017} gives an analogous result in its Eq.~(57).
}
\begin{equation}
r(\omega_\mathrm{p})=\frac{\omega_\mathrm{r}-\omega_\mathrm{p}+i\kappa/2+\chi(\omega_\mathrm{p})}{\omega_\mathrm{r}-\omega_\mathrm{p}-i\kappa/2+\chi(\omega_\mathrm{p})}.
\label{eq:circuitResponse}
\end{equation}
Here, $\omega_\mathrm{p}$ is the frequency of the signal probing the circuit, $\omega_\mathrm{r}$ is the circuit resonant frequency, 
$\kappa$ is the circuit-field decay rate, 
and $\chi$ is the qubit response function, proportional to the Fourier transform of the decay envelope $-i f(t)$. For a qubit described by an exponential decay, the latter becomes\cite{cottet_cavity_2017,ezzouch_dispersively_2021}
\begin{equation}
\chi(\omega_\mathrm{p}) = -\frac{g^2}{\omega_\mathrm{q}-\omega_\mathrm{p}-i\Gamma_2}D,
\label{eq:susceptibility}
\end{equation}
with $\hbar \omega_\mathrm{q}$ the energy difference of the qubit excited and ground states, $g$ the qubit--circuit coupling, $D$ the difference of the population probability of the qubit ground and excited state, and $\Gamma_2 = \Gamma_1/2 +\Gamma_\varphi$ is defined in Appendix~\ref{app:notation}. These formulas are also valid when the qubit itself is driven, the so-called two-tone spectroscopy\cite{scarlino_all-microwave_2019}. In that case, all parameters on the right hand side of Eq.~\eqref{eq:susceptibility} should be replaced by the corresponding quantities in the rotating reference frame \cite{ithier_decoherence_2005}.
From Eq.~\eqref{eq:circuitResponse}, one can express the fraction of the reflected power, $|r|^2$, through the following formula
\begin{equation}
\label{eq:r1}
|r(\omega_\mathrm{p})|^2 -1 = \frac{2\kappa\mathrm{Im}\{\chi\}}{\left(\frac{\kappa}{2} -\mathrm{Im}\{\chi\}\right)^2 + \left(\mathrm{Re}\{\chi \} -\omega_\mathrm{p} + \omega_\mathrm{r}\right)^2}.
\end{equation}
In the dispersive regime where $|\omega_\mathrm{p}-\omega_\mathrm{r}|$ is the largest frequency, the denominator can be approximated by a constant and Eq.~\eqref{eq:r1} reduces to 
a constant times $\mathrm{Im}\{\chi\}$. 
Scanning the probe frequency $\omega_\mathrm{p}$ around the resonance $\omega_\mathrm{p}=\omega_\mathrm{q}$, one observes a dip in the circuit steady-state response\footnote{Since normally $D>0$. However, the population inversion, $D<0$, is also possible, see Ref.~\cite{hauss_single-qubit_2008}. It would give a peak in $|r(\omega_\mathrm{p})|$, instead of a dip.}, and the width of the dip gives $2\Gamma_2$. The articles give the dip width as either the full or half width at half maximum (FWHM or HWHM) in frequency (and not angular frequency) units, or, in more general scenarios, $\Gamma_2$ as one of the fit parameters in fitting the data to Eq.~\eqref{eq:r1} or its analogs. In these cases, we evaluate the inhomogeneous dephasing time using
\begin{equation}
T_2^* = (\Gamma_2)^{-1}= \left( 2 \pi \Delta\! f_\mathrm{HWHM} \right)^{-1}.
\label{eq:HWHM}
\end{equation}

Returning to the possibility of discriminating between the decay envelopes, we now consider the Fourier transforms $\chi(\omega)$ for the above two examples. Whereas the Fourier transform of a Gaussian is a Gaussian, the exponential transforms into a complex function with the imaginary part a Lorentzian,
\begin{subequations}
\begin{align}
\textrm{Gaussian}\,f(t) &\overset{\mathrm{F.T.}}{\to}  \chi(\omega) = -i \exp( - \omega^2 T^2/4),\\
\nonumber\\
\textrm{Exponential}\,f(t) &\overset{\mathrm{F.T.}}{\to}   \chi(\omega) = \frac{- \omega T-i}{\omega^2 T^2 + 1},\\
\nonumber
\end{align}
\label{eq:timeAndFreq}
\end{subequations}
In  these equations, we Fourier transformed (F.T.) Eqs.~\eqref{eq:dephasing} and \eqref{eq:relaxation} multiplied by $-i$, normalized the results to the same value at zero frequency, and dropped the time-scale subscripts. To discriminate the two cases given in Eq.~\eqref{eq:timeAndFreq} in the frequency domain is even more difficult than in the time domain, since both Gaussian and Lorentzian are quadratic at small frequencies. Compared to the time domain, now the two functions differ more strongly at their tails (that is, for large $\omega$), with algebraic and exponential decay, respectively. However, as already stated, these tails are seldom accessible with the required precision. The three envelope functions discussed so far are plotted in Fig.~\ref{fig:envelopes}(c) for comparison.

The reason for discussing the discrimination between different functional forms of the decay envelopes is that it hints to the origin of the noise causing the decay. A minimal description of noise is to give its autocorrelation function, either in the time or frequency domain. If it is the latter, the function is called the noise spectrum. The form of the noise spectrum decides what will be the decay envelope. Noises with different physical origins, for example nuclear spins versus charge impurities, will have different spectra. The functional form of the decay envelope can then serve as an alternative to obtaining the noise spectrum, hinting at the possible origin of the dominant noise affecting the qubit. Finally, we note that the above three possibilities, Gaussian, exponential, and Lorentzian, are not the only ones. For example, going to the next order in the calculation reveals that algebraic tails in the decay exist \cite{ithier_decoherence_2005}. Algebraic tails were also obtained in calculations considering the backaction of the qubit on its environment giving rise to non-Markovian behavior \cite{fischer_dealing_2009}. 

To sum up the discussion, the coherence times are typically extracted from fits to simple functional forms. Some are given above, and there are more, such as the  `stretched exponential' used to fit data from dynamical-decoupling sequences.\footnote{The standard result is Ref.~\cite{cywinski_how_2008}, which derived $\log f(t) \sim (t/T_2)^{1+\alpha}$ as the envelope for decay under a generic dynamical-decoupling sequence assuming noise spectrum $1/f^\alpha$. In fitting the experimental data in GaAs, Ref.~\cite{medford_scaling_2012} found that a more robust way is to fit the observed decay time to $T_2 \propto n^\gamma$, where $n$ is the number of echos and, for the CPMG sequence, the noise-spectrum exponent is related to the fit parameter $\gamma$ by $\alpha=\gamma/(1-\gamma$).} The true decay envelope is a complicated function, which can hardly be parameterized by a single number. A typical fit returns the time scale over which the envelope decays to a fraction of its initial value, for example, $1/e$. This operational definition should be the first guess on the meaning of a coherence time in the tables we give. Not much is implied about the functional form itself, let alone the decay of long-time tails. 

\separate{The data on coherence times are listed in 
Table \ref{tab:spinCoherence}. 
We additionally present it here in figures, discussing them shortly. We split the figures into two groups, separating the charge qubits, the states of which do not rely on the spin degree of freedom in any way, from qubits that rely on the spin at least to some degree. We start with the latter group.}

\subsection{Measured coherence times of spin qubits}

\begin{figure}
  \includegraphics[width=\linewidth]{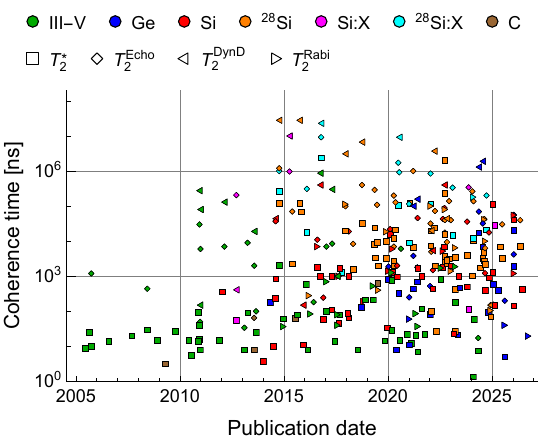}
  \caption{
  \label{fig:spinPhaseCoherenceAll}
  Spin coherence times according to the publication date. The color shows the device material and the symbol indicates the coherence type as given in the legend. The plotted data are values from Table \ref{tab:spinCoherence} excluding the data on the relaxation time $T_1$. 
  }
\end{figure}

The coherence times of spin qubits are given in Fig.~\ref{fig:spinPhaseCoherenceAll}. The values had started at around 10 ns inhomogeneous dephasing time in early experiments with qubits in GaAs. Echo techniques can extend the coherence by orders of magnitudes, as can a different material choice. \recheck{The coherence times published during the year 2024 span five orders of magnitude, depending on the qubit type, material, and protection measures.}

\begin{figure}
  \includegraphics[width=0.88\linewidth]{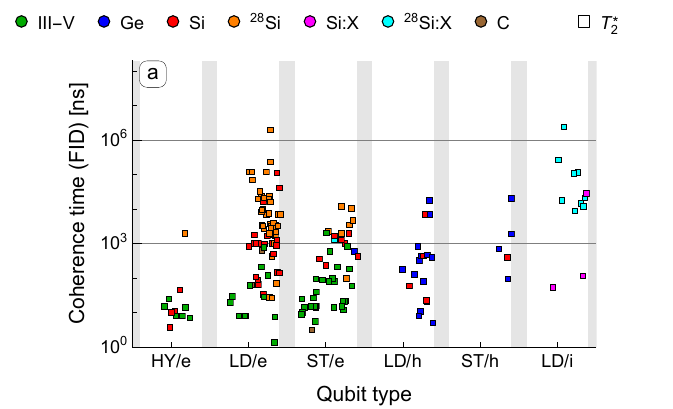}
  \includegraphics[width=0.88\linewidth]{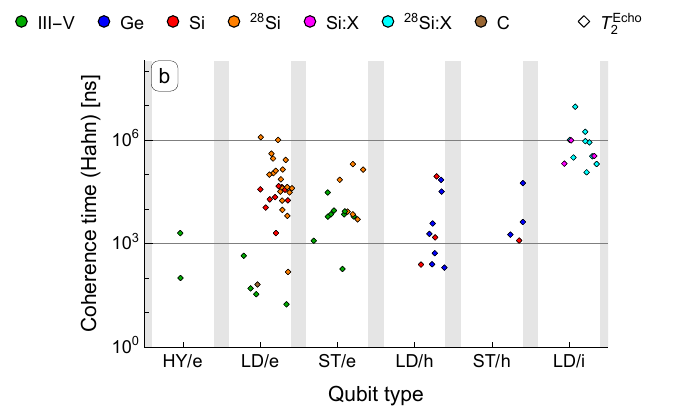}\hfill \\
  \includegraphics[width=0.88\linewidth]{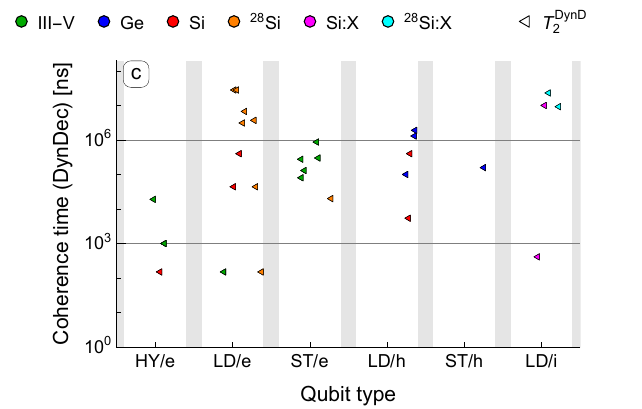} \hfill
  \includegraphics[width=0.88\linewidth]{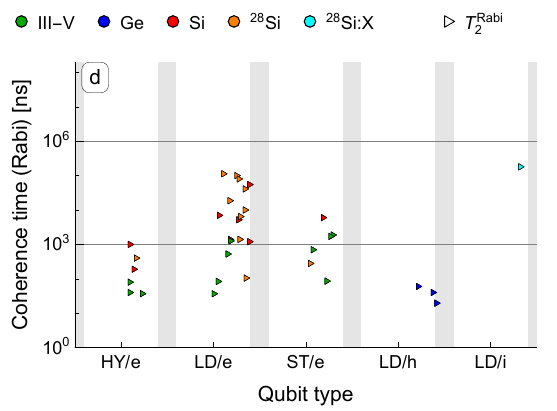}
  \caption{
  \label{fig:spinPhaseCoherenceSubplots}
  Spin coherence times. The panels display the data from Fig.~\ref{fig:spinPhaseCoherenceAll} split to the four panels according to the coherence type, denoted in the legend, from the set: inhomogeneous dephasing $T_2^*$, Hahn echo $\TEcho$, dynamical decoupling $\TDynD$, and Rabi decay $\TRabi$. In each panel, the horizontal axis uses the qubit type as a discrete category, displacing the points according to their publication date: more recent data are shifted to the right within the non-shaded area which is \recheck{normalized to the year span 2003-2025}. All vertical-axes ranges are the same and the point colors show the material.
  }
\end{figure}

To examine the influence of some of these factors, we plotted separately each type of coherence in Fig.~\ref{fig:spinPhaseCoherenceSubplots}. The separation allows us to group the values additionally according to the qubit type, being the discrete category on the horizontal axis. To reflect the publication date for an easier comparison, we displace the data within each category horizontally. \recheck{For example, Fig.~\ref{fig:spinPhaseCoherenceSubplots}(a) shows that the most recent electron spin-1/2 qubits implemented in purified silicon reach longer coherence times than singlet-triplet qubits. The coherences of holes have recently got on par with the best ones for singlet-triplet qubits in almost every category. Finally, impurity spins hold record $T_2^*$, $\TEcho$, and $\TRabi$ coherence times.}

\begin{figure}
  \includegraphics[width=0.99\linewidth]{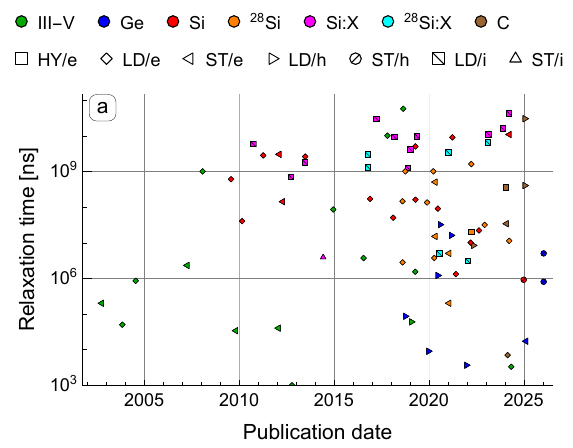} \hfill\\
  \includegraphics[width=0.99\linewidth]{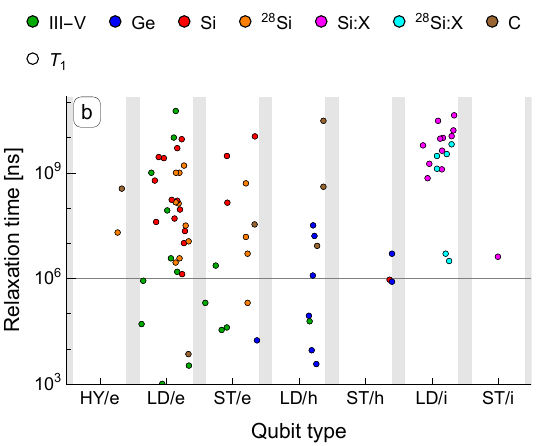} \hfill
  \caption{
  \label{fig:spinRelaxation}
  Relaxation times of spin qubits. Both panels show the same data, the relaxation times from Table \ref{tab:spinCoherence}. The datapoint color shows the material. In (a), the horizontal axis shows the publication date and the datapoint symbol shows the qubit type. In (b), the qubit type is on the horizontal axis as a discrete category, whereas the publication date is reflected by shifting the points horizontally, similarly as in Fig. ~\ref{fig:spinPhaseCoherenceSubplots}.}
\end{figure}

We next turn to the energy relaxation time. For a spin qubit, this time can be made very long by: isolating the qubit from reservoirs, so that the electron does not escape the dot; minimizing its dipole moment, so that the qubit couples weakly to phonons; and decreasing the transition energy, typically using a lower magnetic field, so that the available phase space for the process is reduced \cite{shrivastava_theory_1983,khaetskii_spin-flip_2001}. Under these conditions, relaxation times have reached seconds and can be considered not of concern for quantum computing. However, the relaxation times remain of concern if these conditions are not met: for example, a finite relaxation time of two-electron singlet and triplet states is the main limitation for the spin measurement fidelities \cite{barthel_rapid_2009,nakajima_robust_2017,broome_high-fidelity_2017}. 
Because of this crucial role, there are numerous values on the relaxation time for (mostly the triplet states of) singlet-triplet qubits, either explicitly reported or implicitly implied in many experiments using Pauli spin blockade for the spin measurement. As the relaxation time in this setting is a by-product of the maximization of the measurement fidelity, rather than a figure of merit maximized itself, we normally do not include singlet-triplet relaxation rates in this Technical Review. We do include a few values, from either early experiments, or when they are the article's main topic.

The reported relaxation times are shown in Fig.~\ref{fig:spinRelaxation}, plotting the same set of data in two different ways. Figure \ref{fig:spinRelaxation}(a) gives them according to the publication date. One can see how the longest-reached times developed: in the first decade, electron one-half spin-qubits were in the lead. Since 2010, impurities took over. \recheck{Currently, the record is back with a single electron in GaAs, with the relaxation time of one minute \cite{camenzind_hyperfine-phonon_2018}.} As already noted, while reaching such long times is not directly improving other figures of merit of the qubit, the increase of the record time illustrates the experimental progress with the given qubit platform. Figure \ref{fig:spinRelaxation}(b) groups the times according to qubit types, making it easier to judge the progress over the years within each group.

\subsection{Measured coherence times of charge qubits}

\begin{figure}
  \includegraphics[width=0.99\linewidth]{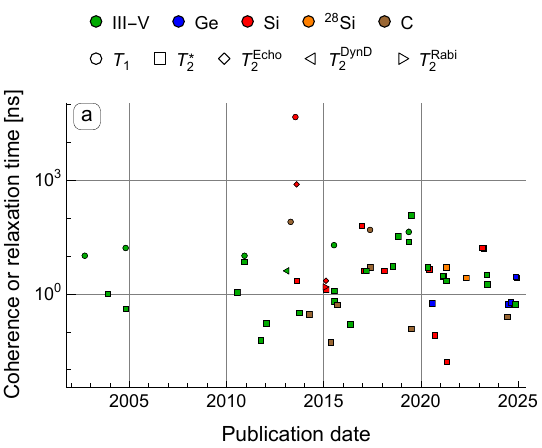} \hfill \\
  \includegraphics[width=0.99\linewidth]{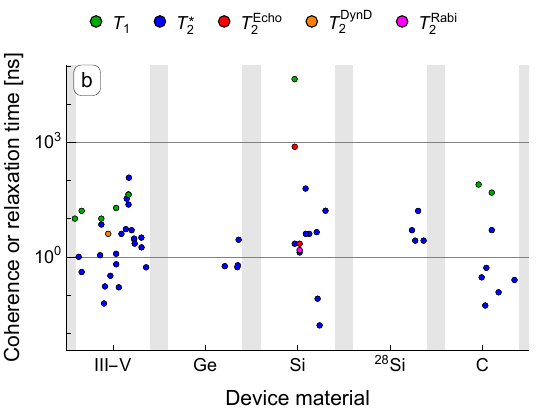} \hfill \\
  \includegraphics[width=0.99\linewidth]{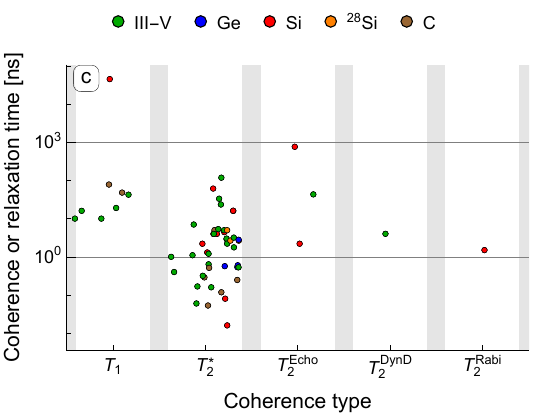}\\
  \caption{
 \label{fig:chargeCoherence}
  Charge coherence. Every panel plots the same data, the values from Table \ref{tab:chargeCoherence}. The panels differ by their horizontal axis, plotting the data according to the quantity given as the horizontal-axis label. Concerning the datapoint symbols and colors, (a) is analogous to Fig.~\ref{fig:spinPhaseCoherenceAll}. In (b) and (c), the role of the datapoint colors and the horizontal axis is swapped. 
  }
\end{figure}

We end the overview of the coherence times by looking at charge qubits. As already mentioned, we include them even though they implement qubits that do not rely on spin. The reason is a close relation of the devices where the two types of experiments are typically done and of used techniques: for example, the measurement of spin is done indirectly, converting different spin states to different charge states, which are then detected. 

All types of coherence times of charge qubits, including the relaxation time, are gathered in Fig.~\ref{fig:chargeCoherence}. In the three panels of the figure, the same data are shown according to the publication date, the device material, and the coherence type, respectively. One can see several differences compared to spin qubits. \recheck{
First, the relaxation times are often comparable to coherence times, unlike for most spin-qubits where they can be pushed to be by far the longest scale.} 
Therefore, the relaxation is a bigger issue for charge qubits even in single-qubit experiments. \recheck{Second, unlike for spin qubits, the echo techniques do not prolong coherence substantially.} \recheck{Third, there is much less variation among the data:} 
\recheck{Despite some upward trend in $T_2^*$ over time, the increase is not dramatic. Since the charge-qubit dephasing time $T_2^*$ is related to charge noise directly, a clear trend would indicate} an overall improvement of the available samples and devices concerning the level of charge noise.
\recheck{Finally, as one would expect based on the nuclear-spin noise being irrelevant for charge qubits, there is no apparent difference between devices made in Si and III-V materials concerning charge-qubit coherence.}

\vspace{-0.3cm}
\section{Operation times}

\label{sec:operations}

The second most abundant data exist on characteristics of spin-qubit operations. We consider the qubit gates, measurements, and initializations as three operation types, treated on equal footing.
We are motivated by their actual physical realizations which are similar: all three types of operations are typically implemented by pulsing the system to a specific configuration, or driving it resonantly, for a fixed time. Another reason is that whereas the algorithms considering logical qubits might assign initializations to the beginning and measurements to the end of the algorithm only, physical qubits will need error correction. In this case, the initializations and measurements are used heavily, interspersing the application of gates.

There is an additional attribute introduced in this section, the number of qubits that a given operation involves, \myKey{\#Qubits}. \recheck{The typical cases are one-qubit (1Q) and two-qubit (2Q) gates, even though the first three-qubit gate has already been used in an error-correction demonstration \cite{takeda_quantum_2022}. 
}

Finally, let us list the operation characteristics. In the next three sections, we discuss, respectively, operation times, a quantitative measure, and operation fidelities and quality factors, two related qualitative measures.

\subsection{The definition of experimentally extracted gate times}

The advantage of looking at gate times is that they reveal the natural time-scale for a given qubit and that they can be compared to the coherence times. However, they also have a shortcoming that arises if they are not compared to any coherence time: a fast qubit does not automatically mean a good qubit, since the judgment largely depends also on the coherence time. Conversely, a qubit with extremely long coherence becomes less appealing if the corresponding gates are also extremely slow. This shortcoming, namely a dimensionful quantity having a limited meaning without comparison to other dimensionful quantities, also applies to the coherence times presented in the previous section.

Let us specify the normalization for the operation times. For the measurements and initialization, the definition is straightforward, even though one should also count the preparation if it is a necessary part of the measurement or initialization sequence. More ambiguity exists for gates since gates are realized as unitary evolutions induced by certain Hamiltonians. A typical signal which is interpreted as a gate being applied looks as in Fig.~\ref{fig:envelopes}(a). Neglecting for now the decay and taking a single-qubit case, the oscillating signal is due to a unitary evolution such as
\begin{equation}
U(t) = \exp\left( -i \omega t s_z \right) \equiv \exp \left( -i 2\pi f t s_z \right).
\label{eq:gateU}
\end{equation}
Here, we used $f = \omega / 2\pi$ for the signal frequency and $s_z = \sigma_z/2$ for the rotation generator, with $\sigma_z$ being the Pauli matrix.
At various times, various gates are carried out on the qubit: for $f t=1/2$, one gets a $Z$ gate, for $f t=1$ an identity, whereas $f t=1/4$ gives a $\pi/2$ rotation, often used to prepare coherent superpositions such as $|0\rangle +|1\rangle$. If such a continuous signal is presented, we define the operation time to be one-half of the signal period,
\begin{equation}
T_\mathrm{op} \equiv \frac{1}{2f}.
\label{eq:gateTime}
\end{equation}
In the case of Eq.~\eqref{eq:gateU}, the value $t=1/2f$ would be taken as the operation time, and it would correspond to the $Z$ gate, that is a spin rotation by $\pi$. Note that this definition is different from the one adopted for the quality factor (see below).

\begin{figure}[t!]
  \includegraphics[width=0.89\linewidth]{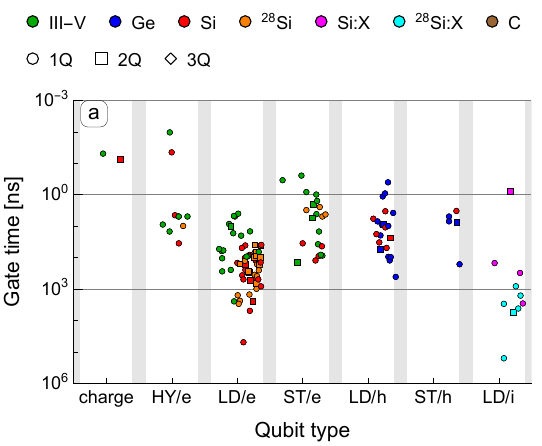} \hfill\\
  \includegraphics[width=0.89\linewidth]{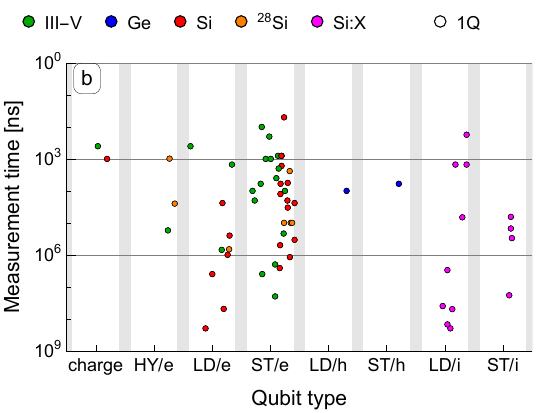} \hfill\\
  \includegraphics[width=0.89\linewidth]{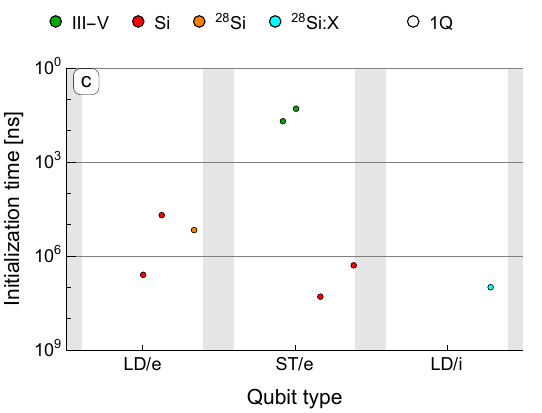}\\
  \caption{
  \label{fig:operationTime}
    Operation times: (a) gates, (b) measurements, (c) initializations. In all panels the horizontal axis shows the qubit type, restricting to those for which values exist. Within each type, the publication date is reflected by shifting the points horizontally, similarly as in Fig. ~\ref{fig:spinPhaseCoherenceSubplots}. The vertical axis shows the operation time in nanoseconds. The datapoint color indicates the material and its symbol the number of qubits involved in the operation, according to the panel legend. 
  }
\end{figure}

\subsection{Measured values of gate times}

Let us now look at the reported values plotted in Fig.~\ref{fig:operationTime}(a). Since the set of operations is diverse, we present the data with the qubit type as the primary category on the horizontal axis. The data are additionally tagged according to the device material and the number of qubits. \recheck{The shortest in the set are the times of single-qubit gates on charge qubits, being below $0.1$ ns.} One can deduce more instances of such short times in the literature than those shown in the figure. The reason is that these values are perhaps not claimed explicitly in the experiments: for charge qubits, the gate speed is limited by the experiment electronics, rather than the qubit itself \footnote{Instead of quoting the gate speed, the qubits are often judged by how strongly they couple to a microwave cavity field, using the charge-photon coupling strength. At the moment that figure of merit is not included in this Technical Review.}. \recheck{The hybrid qubits can reach similar speeds,} since they can be tuned into a configuration where they resemble a charge qubit. Arguably, this tunability is their biggest advantage. When they are tuned into a spin-like configuration, their gate (and coherence) times go up. Similarly, a strong direct coupling to the electric field can be exploited also for \recheck{holes, with gate speeds up to gigahertz seen}. Such electric-dipole spin resonance (EDSR) driving is less efficient for electron spin-1/2 qubits, for which \recheck{the highest speeds were reached using design-optimized micromagnets providing large magnetic-field gradients, or materials with strong spin-orbit interaction (for example, InAs).} As seen from the times in the singlet-triplet column of the figure, the singlet-triplet oscillations can reach gigahertz frequencies. Fast exchange-based gates were demonstrated for both one and two-qubit operations of spin-1/2 qubits and singlet-triplet qubits, and for two-qubit operations of hole and impurity qubits. Finally, we would like to note that the exchange-based gate for, first, a pair of one-half spins and, second, the singlet and triplet two-electron states, is an identical process. Whether such a process should be interpreted as a one-qubit gate or two-qubit gate depends on additional functionalities implemented, or implementable, in the given experiment. The boundary between the two cases is blurry. 

\subsection{The definition and values of measurement times}

We do not review the measurement times of charge qubits because the task of measuring a charge qubit is the task of detecting a charge---typically an elementary charge---in a nanodevice. First, this task is separable from, and not exclusive to, spin qubits, and thus not our focus. Second, the analysis of methods and results of this task is a topic rich enough for a review in its own right.\cite{gustavsson_electron_2009,clerk_introduction_2010,vigneau_probing_2022} For our purposes, it will be enough to give the following minimum. The charge detection is typically characterized through the detection sensitivity $s$ with a typical number being\footnote
{
Here are a few numbers that one can find in the literature: 
$10$ in Ref.~\cite{reilly_fast_2007},
$0.07$ in Ref.~\cite{angus_silicon_2008} %
$10$ in Ref.~\cite{wang_graphene_2010},
$30$ in Ref.~\cite{fringes_charge_2011}, 
$63$ in Ref.~\cite{colless_dispersive_2013},
$0.8$ in Ref.~\cite{stehlik_fast_2015},
$0.37$ in Ref.~\cite{gonzalez-zalba_probing_2015},
$8.2$ in Ref.~\cite{zajac_scalable_2016},
$4.1$ in Ref.~\cite{zheng_rapid_2019},
$3.3$ in Ref.~\cite{curry_single-shot_2019},
$21$ in Ref.~\cite{chanrion_charge_2020},
$0.60$ in Ref.~\cite{schupp_sensitive_2020},
$31$ in Ref.~\cite{park_single-shot_2024};
all in $10^{-4} \times e/\surd{\mathrm{Hz}}$.

} 
\begin{equation}
s = \mathrm{a\,few}\, \times 10^{-4}\mathrm{e}\, / \sqrt{\mathrm{Hz}}.  
\label{eq:chargeSensitivity}
\end{equation}
This parameter quantifies the reliability of the output of a charge-meter signal if integrated for time $T_\mathrm{M}$, through the variance
\begin{equation}
\textrm{var}[q] = s^2 / T_\mathrm{M}.
\end{equation}
Assuming that one aims to distinguish a signal of one elementary charge, $q_1=e$, from the signal of zero charge, $q_2=0$, where `to distinguish' means to make the signal error, $\sqrt{\textrm{var}[q]}$, as small as the signal magnitude, $|q_1-q_2|=e$, would give the required time
\begin{equation}
T_\mathrm{M} = s^2 / e^2 \sim 100\, \mathrm{ns}.
\label{eq:detectionLimit}
\end{equation}
Although higher sensitivity\cite{schaal_fast_2020} and shorter times\cite{keith_single-shot_2019} for elementary charge detection were reported, let us take this value for the measurement time $T_\mathrm{M}$ as a lower limit 
realistic for many experimental configurations. 

With 100 ns for the time required to detect an event involving elementary charge, we now look at the spin qubit measurement times shown in Fig.~\ref{fig:operationTime}(b). \recheck{We see that whereas the times start at about the elementary charge-detection limit of Eq.~\eqref{eq:detectionLimit}, they might become orders of magnitude longer.} The variance reflects the fact that the spin is detected indirectly, by first converting it to a charge event, which is in turn resolved by a charge sensor. The time required for the spin-to-charge conversion can vary a lot, depending on the process details. One example of a spin-to-charge conversion is spin-dependent tunneling: an electron with spin down can tunnel out a quantum dot, whereas the one with spin up can not, the difference originating in their Zeeman energies \cite{hanson_zeeman_2003}. Since the quantum dot has to be well isolated for the qubit to keep its coherence, the tunneling out of the dot is slow. As a second example, the spin singlet and triplet can be discriminated according to being allowed or not allowed to traverse a double dot (the Pauli Spin Blockade invented by Ono et al., see Ref.~\cite{ono_current_2002}). Although in this case the charge reconfiguration can be very fast, following the gate-voltage pulses on a nanosecond scale, the two charge configurations being discriminated differ only by a dot-size shift of an elementary charge. The discrimination of such states requires a longer time compared to those differing by the presence, rather than the displacement, of an elementary charge. A different class of measurements, called rf-based detection, relies on probing the above-described spin-dependent charge reconfigurations by an oscillatory (radio-frequency) electric field. \recheck{The shortest measurement times seen on Fig.~\ref{fig:operationTime}(b) were obtained using rf-sensors.}

\subsection{The values of initialization times}

The initialization times are shown in Fig.~\ref{fig:operationTime}(c). \recheck{Only a few values are present.} The reason might be that the most typical experimental scenario is to repeat a cycle: initialization -- operation -- measurement, where the measurement part plays the role of the initialization, and, therefore, the latter is not reported separately and explicitly. \recheck{There are more values published on the initialization fidelities, see below.}

\section{Operation fidelity}

The operation fidelity is a dimensionless figure of merit allowing the comparison of diverse qubits. Using randomized benchmarking \cite{epstein_investigating_2014}, one can extract the gate errors independently of the measurement errors, even if the former are orders of magnitude smaller than the latter. Although it is not strictly correct, the fidelity is used to judge the progress towards error-correction thresholds required for fault-tolerant quantum computing.\footnote{The problem is that the fidelity, as defined below, is not the error parameter entering the threshold theorem. The two parameters can differ by orders of magnitude, in the unfavorable way: whereas the fidelity extracted by the randomized benchmarking can be low, the error rate can remain much larger \cite{sanders_bounding_2015,blume-kohout_demonstration_2017}. } For all these reasons, evaluating the gate fidelities is popular and impressive values have been reached. 

\subsection{The definition and meaning of experimentally extracted fidelities}

\label{sec:fidelityDefinition}

The fidelity characterizes how close the actual operation is to the desired one. Although the fidelity of 1 means that the two operations are the same, the quantification of a measure when they are not the same is less straightforward. The usual definitions derive from the fidelity $\mathcal{F}$, in the sense of the distance between two quantum states described by their respective density matrices $\rho$ and $\rho^\prime$,
\begin{equation}
\mathcal{F} = \left( \mathrm{tr}\sqrt{\sqrt{\rho}\rho^\prime \sqrt{\rho}} \right)^2.
\label{eq:fidelity1}
\end{equation}
If one of the two states is pure, say $\rho^\prime = |\Psi\rangle \langle \Psi|$, the formula simplifies to
\begin{equation}
\mathcal{F} = \mathrm{tr} \left(  \rho |\Psi\rangle\langle \Psi| \right).
\label{eq:fidelity2}
\end{equation}
If the second state $\rho$ is also pure, $\mathcal{F}^{1/2}$ is closely related to unambiguous discrimination of the two pure states \cite{heinosaari_guide_2008}, whereas $\arccos \mathcal{F}^{1/2}$ is a measure of their statistical distinguishability \cite{wootters_statistical_1981}.\footnote{What is the best measure to quantify distance of two operations is discussed at length in Ref.~\cite{gilchrist_distance_2005}. We thank Andrea Morello for pointing this reference to us.}

We use the definition of Eq.~\eqref{eq:fidelity1} leading to Eq.~\eqref{eq:fidelity2} because of its connection to the randomized benchmarking. Namely, the essence of the latter is to prepare a pure state $|\Psi\rangle \langle \Psi|$, apply to it a sequence of gates, and then evaluate the overlap of the resulting density matrix $\rho$ with the original state $|\Psi\rangle$ using Eq.~\ref{eq:fidelity2}. In the sequence, the gates are randomly chosen from a discrete set, the Clifford group \cite{gottesman_theory_1998}, except for the last gate which is such that the whole sequence reduces to the identity if all gates are perfect. The resulting fidelity, also called the probability of the survival of the initial state, falls off exponentially with the sequence length $m$,
\begin{equation}
\mathcal{F}(m) \approx A p^m + B.
\label{eq:fidelityFit}
\end{equation}
Experimentally, the three coefficients $A$, $B$, and $p$ are fitted.\footnote{Even though Ref.~\cite{magesan_characterizing_2012} stresses that one should always use a more refined decay model, adding a term proportional to $(m-1)p^m$ to the right-hand side of Eq.~\eqref{eq:fidelityFit}, this piece of advice does not seem to be followed in practice.} The first two parameters absorb the errors of the state preparation and measurement, so that the infidelity $1-\mathcal{F}$ can be found from the parameter $p$ using the formula\footnote{Some comments are in order: The exponential decay reflected by Eq.~\eqref{eq:fidelityFit} happens under broad conditions investigated in detail in Refs.~\cite{magesan_characterizing_2012,epstein_investigating_2014}. The fidelity extracted in this way is the average of $\mathcal{F}$ in Eq.~\eqref{eq:fidelity2} over all pure input states $|\Psi\rangle\langle \Psi|$ and over the gates in the Clifford group. There is an extension of the procedure called interleaved randomized benchmarking \cite{magesan_efficient_2012} which can assign a fidelity---still averaged over all pure states---to a single specific gate from the Clifford group.}\footnote{Ref.~\cite{gilchrist_distance_2005} presents further arguments against using the average fidelity $\mathcal{F}$, as extracted from randomized benchmarking, Eq.~\eqref{eq:fidelityParameter}, and advocates entanglement fidelity $\mathcal{F}_e$ instead. The two measures are related by $(d+1)\mathcal{F}=d\mathcal{F}_e+1$. Since the majority of the spin-qubit publications give $\mathcal{F}$, we stick to $\mathcal{F}$ as the reported figure of merit.}
\begin{equation}
1-\mathcal{F} = (1-p) \frac{d-1}{d}.
\label{eq:fidelityParameter}
\end{equation}
Here, $d=2^n$ with $n$ the number of qubits; $d=2$ for a single-qubit gate benchmarking. Although the exponential decay form displayed by Eq.~\eqref{eq:fidelityFit} relies on the discrete set being the Clifford group, many articles convert $\mathcal{F}$ to fidelities of experiment-specific `elementary' or `primitive' gates, such as $\pi/2$ and $\pi$ rotations around various axes. Even though such a conversion is questionable \cite{magesan_characterizing_2012}, we follow the prevailing practice and in figures and tables we give the infidelity for the elementary-gate set (and not for the Clifford-gate set). For one-qubit or two-qubit gates, one Clifford gate typically requires a few elementary gates. Therefore, the infidelity of a Clifford gate would be around a factor of 2--3 larger than the infidelity of an elementary gate \footnote{Ref.~\cite{epstein_investigating_2014} gives one list of elementary gates for a single-qubit case, resulting in the ratio of 1.875. For the two-qubit case, Refs.~\cite{huang_fidelity_2019, petit_universal_2020} used elementary-gate sets with the ratio of 2.57. However, larger ratios also appear: for example, 9.75 in Ref.~\cite{xue_benchmarking_2019}.}, the value quoted in this Topical Review. 

\begin{figure}
  \includegraphics[width=0.99\linewidth]{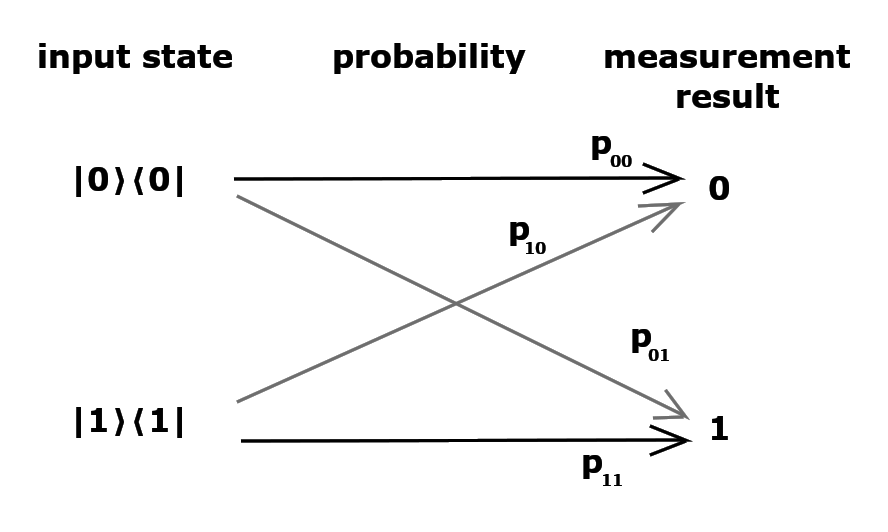} \hfill
  \caption{\label{fig:measurementScheme}
  Probabilities describing a two-outcome qubit measurement. The two mutually orthogonal pure states are on the left, the two measurement outcomes on the right. A perfect measurement would have unit probabilities $p_{ii}$ with $i=0,1$ denoted by horizontal lines. Owing to imperfections, diagonal lines appear with non-zero probabilities. The infidelity defined by Eq.~\eqref{eq:fidelityMeasurement} gives for this diagram $1-\mathcal{F} =\frac{1}{2} (p_{01}+p_{10})$.}
\end{figure}

The fidelities of initialization and measurement are less ambiguous to define as they can be based on Eq.~\eqref{eq:fidelity2} which is more intuitive than Eq.~\eqref{eq:fidelity1}. Let us start with the measurement. The probability to get an outcome $p$ in measuring the state described by a density matrix $\rho$ is given by Eq.~\eqref{eq:fidelity2} upon replacing the pure state $|\Psi\rangle\langle\Psi|$ by a positive semi-definite operator $A$. 
The most general measurement with possible outcomes labeled by index $i$ is specified by a set $\{A_i\}_i$ of such operators summing to identity, $\sum_i A_i = 1$. In experiments, these operators are approximations of a set of mutually orthogonal projectors spanning the qubit basis $A_i \approx |\Psi_i\rangle \langle \Psi_i|$. Owing to experimental imperfections, these approximations are not exact. The probability of the measurement outcome $j$ upon measuring the pure state $\rho=|\Psi_i\rangle\langle\Psi_i|$ follows as
\begin{equation}
p_{ij} = \mathrm{tr} \left(  |\Psi_i\rangle\langle \Psi_i| A_j \right).
\end{equation}
Owing to the normalization of $A_i$, the probabilities fulfill
\begin{equation}
\sum_j p_{ij}=1.
\label{eq:sumRule}
\end{equation}
Were the measurement perfect, all non-diagonal probabilities would be zero. The infidelity of the measurement can be quantified through the off-diagonal probabilities. For example,
\begin{equation}
1-\mathcal{F} = 1 - \frac{1}{d} \sum_i p_{ii} = \frac{1}{d} \sum_i \sum_{j\neq i} p_{ij}.
\label{eq:fidelityMeasurement}
\end{equation}
Here, the first equality sign is a definition, the second one follows from the sum rule, Eq.~\eqref{eq:sumRule}. Also, the number of outcomes is assumed to be equal to the size of the Hilbert space $d$: For example, $d=2$ for a two-outcome measurement of a two-level system (a qubit). In this most common case, the measurement probabilities are quantified by two error probabilities, $p_{01}$ and $p_{10}$, as depicted in Fig.~\ref{fig:measurementScheme}. The resulting measurement infidelity according to the definition in Eq.~\eqref{eq:fidelityMeasurement} is $(p_{01}+p_{10})/2$.

Finally, let us consider the fidelity of the initialization. Typically, one is interested in an initialization into a single pure state $|\Psi\rangle$. In this case, we can use Eq.~\eqref{eq:fidelity2} to define the initialization fidelity with $\rho$ the actual, perhaps imperfectly prepared, state.

\begin{figure}[ht!]
  \includegraphics[width=0.89\linewidth]{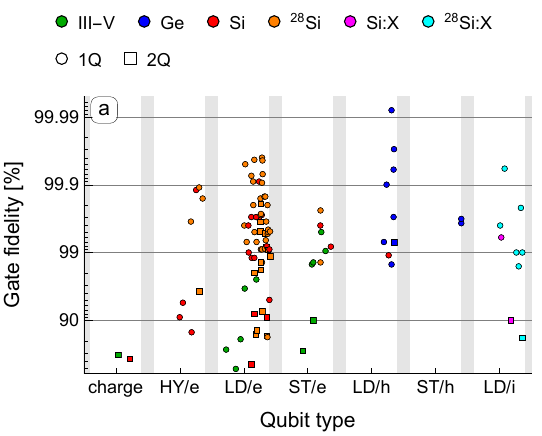} \hfill\\
  \includegraphics[width=0.89\linewidth]{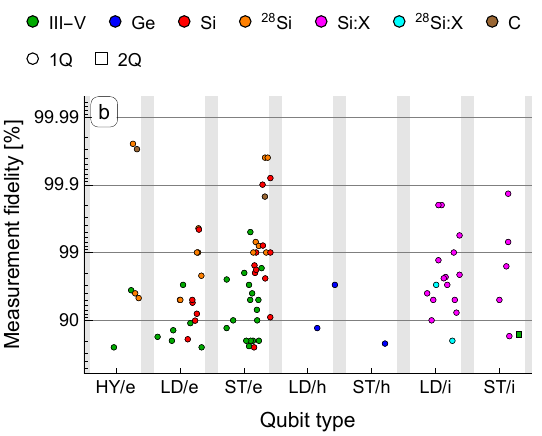} \hfill\\
  \includegraphics[width=0.89\linewidth]{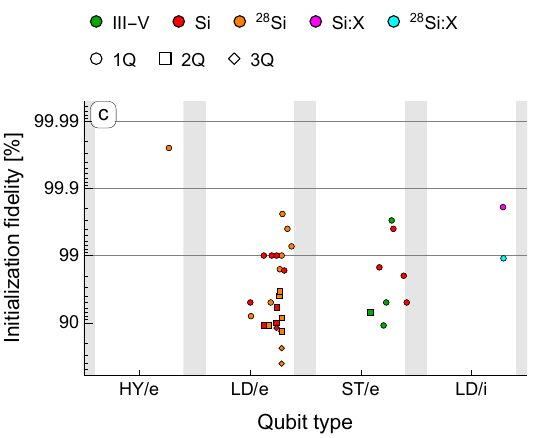} \hfill\\
  \caption{
    \label{fig:operationInfidelities}
  Operation fidelities: (a) gates, (b) measurements, (c) initializations. In all panels the horizontal axis shows the qubit type, restricting to those for which infidelity values exist. Within each type, the publication date is reflected by shifting the points horizontally, similarly as in Fig. ~\ref{fig:spinPhaseCoherenceSubplots}. The vertical axis shows the fidelity $\mathcal{F}$ in percent. The datapoint color indicates the material and its symbol the number of qubits involved in the operation, according to the individual panel legend.
  }
\end{figure}

\subsection{Measured values of fidelities}

The three panels of Fig.~\ref{fig:operationInfidelities} show the published fidelities of the gates, measurements, and initializations, respectively. For the gates, Fig.~\ref{fig:operationInfidelities}(a), \recheck{the electron spin-1/2 qubits previously reached the highest fidelities, well above 99.9 \%. Using silicon, both natural and isotopically purified, was crucial for this achievement. Recently, these values were overcome and a new record above 99.99 \% has been set by a hole in germanium \cite{lawrie_simultaneous_2023}.} \recheck{There is notable progress in almost every qubit category} and increasing the fidelity of single-qubit gates is one of the most impressive achievements within the whole spin-qubit field. 

Figure \ref{fig:operationInfidelities}(b) shows the fidelities of measurements. \recheck{Until recently, the infidelities remained above a few percent.} \recheck{Relying on a `latched' readout in the Pauli spin blockade \cite{nakajima_robust_2017,broome_high-fidelity_2017}, the fidelities above 99 \% were achieved with singlet-triplet qubits. Comparable high-fidelity results for impurity spins rely on their exceedingly long lifetimes. The current record fidelity is with Ref.~\cite{blumoff_fast_2022}, who almost reached 99.99\% for an exchange-only qubit in purified Si.}

We conclude with a remark on two-qubit fidelities. \recheck{In all categories, meaning gates, measurements, and initializations, their infidelities remain at least one order (usually more orders) of magnitude above the single-qubit ones. There is less data published for the two-qubit versions.} \recheck{Concerning initializations, the more-qubit infidelities that we list are initializations into a nontrivial state achieved through some simple quantum algorithm.} For example, initializing all individual qubits into single-qubit fiducial states, and then entangling them with gates into the desired entangled multi-qubit state, such as one of the two-qubit Bell states or the three-qubit Greenberger–Horne–Zeilinger (GHZ) state.

\section{Quality factor \texorpdfstring{$Q$}{Q}}

\label{sec:qualityFactor}

The quality factor is another dimensionless measure allowing for the comparison of diverse qubits, similar to the gate fidelity. The quality factor is a product of a gate frequency and a characteristic time scale. We denote the former by $f_\mathrm{R}$, with the subscript suggesting the most typical case where the gate is implemented by Rabi oscillations. There are two usual choices for the characteristic time, resulting in two groups of quality factors reported in the literature. 

The first choice is using the inhomogeneous dephasing time. We call the resulting metric the qubit quality-factor,
\begin{equation}
Q = f_\mathrm{R} T_2^*.
\label{eq:qualityFactorQubit}
\end{equation}
This $Q$ counts the number of operations which can be performed on a given qubit before other qubits, waiting idly, lose their coherence. With the possibility of applying dynamical decoupling, one might consider using other coherence times instead of $T_2^*$, as the time for which the other qubits can wait before losing coherence. However, we know only a single case where a different choice was made for this type of the quality factor, $ \TEcho$ in Ref.~\cite{dial_charge_2013}, and therefore stick to Eq.~\eqref{eq:qualityFactorQubit}. 

The second choice is using the gate-signal decay time. Though it is not exclusive to Rabi-induced gates, we denote it using the symbol $ \TRabi$ introduced in Sec.~\ref{sec:coherence}, again as the most typical scenario. We call the resulting quantity the gate quality-factor,
\begin{equation}
Q = f_\mathrm{R}  \TRabi.
\label{eq:qualityFactorGate}
\end{equation}
In this form, $Q$ gives the number of oscillations at frequency $f_\mathrm{R}$ during the decay time $ \TRabi$. Loosely speaking, it gives the number of discernible oscillations on a plot showing the Rabi oscillation signal.\footnote{Indeed, Ref.~\cite{veldhorst_two-qubit_2015} calls $Q$ defined in Eq.~\eqref{eq:qualityFactorGate} a `number of gate oscillations', which might have been a better name than a `quality factor'.} We note that many references reporting the gate quality-factors use Eq.~\eqref{eq:qualityFactorGate} with the right-hand side multiplied by a factor of two. Whereas such a factor is understandable looking at Eq.\eqref{eq:gateTime}, we adopt Eq.~\eqref{eq:qualityFactorGate} to make it more directly comparable to Eq.~\ref{eq:qualityFactorQubit}. As a consequence, we have to divide several of the reported values for the gate quality-factors by two. Finally, we note that the gate quality-factor is related to the gate fidelity, both being a qualitative measure of the gate imperfections. In Appendix~\ref{app:fidelity-quality} we derive $1-\mathcal{F} \approx 1/4Q$ valid for a large $Q$ in a toy model with exponential dephasing.

Figure \ref{fig:qualityFactor}(a) shows the published data on quality factors, both gate and qubit ones. \recheck{The majority of the values are within the range of a few to about a hundred. There are some exceptions that crawl toward a thousand.} Figure \ref{fig:qualityFactor}(b) makes it easier to compare different qubit types and discriminate the gate and qubit quality-factors. 
Of the two dephasing times, $T_2^*$ is typically smaller than $ \TRabi$, \recheck{so one expects that the gate quality-factors reach higher values than qubit quality-factors.
There is some evidence in favor of this expectation, but more data is needed to decide on its generality.}

\section{Size of qubit arrays}

\begin{figure}
  \includegraphics[width=0.99\linewidth]{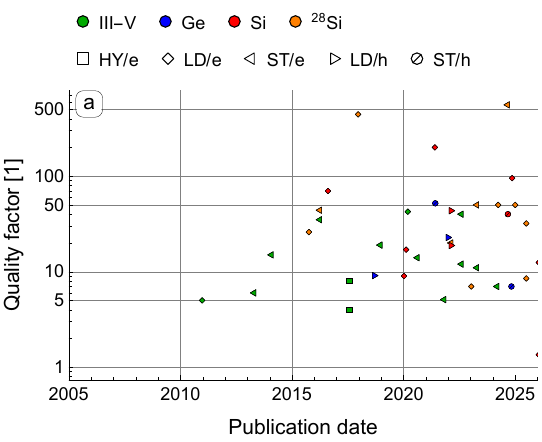}\\
    \includegraphics[width=0.99\linewidth]{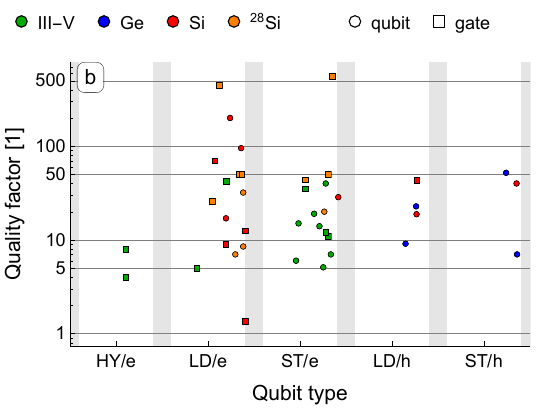}
 \caption{
  \label{fig:qualityFactor}
Quality factors. The two panels show the same data. In (a), according to the publication date, with the color indicating the material and the symbol the qubit type, according to the panel legend. In (b), the qubit type is on the horizontal axis, the publication date is reflected by shifting the points horizontally, similarly as in Fig. ~\ref{fig:spinPhaseCoherenceSubplots}. The datapoint color indicates the material and its symbol discriminates the qubit and the gate quality-factor as defined in Eq.~\eqref{eq:qualityFactorQubit} and Eq.~\eqref{eq:qualityFactorGate}, respectively.
  }
\end{figure}

\label{sec:size}

Scaling up devices to many qubits is the biggest current challenge of the field. The statement applies to all qubit platforms, not only to semiconducting qubits. For serious applications, being able to build large two-dimensional arrays seems necessary since the fault-tolerant thresholds for error correction in one-dimensional arrays remain too low (see Table VIII in Ref.~\cite{devitt_quantum_2013}). 

\subsection{The definition of qubit-array functionality levels}

\label{sec:arrayFunctionality}

Although it is possible to fabricate relatively large gated arrays with semiconductors, making them functional is a different story. Therefore, we discriminate several functionality levels. We note that assigning the levels as defined below to a specific experiment is often difficult. This assignment is perhaps the most subjective of all made in this Technical Review. The reader is advised to consult the original work to judge the details of the achievements.

We define the following values, in ascending order according to the device functionality:
\separate{
\textbf{\\$N$-qubit device.} A structure capable of hosting $N$ qubits has been fabricated. The gating and charge sensing work, so that all qubit hosts can be brought into the required charge configuration. On top of this minimal requirement, the articles assigned to this category report a large variety of additional features: single EDSR gates, tunable interdot tunneling or coupling, controllable charge shuffling, spin detection based on Pauli spin blockade, estimations of qubit-qubit interaction strength, and so on. 
\textbf{\\$N$-qubit simulator.} First of all, the device is stable and tunable enough to search large regions in the high-dimensional charge diagram. For example, aiming at single-electron LD qubits, the structure can be brought into the 1--1--1-- $\cdots$ --1 charge state, or the (11)--(11)-- $\cdots$ --(11) state for ST qubits. In such a configuration, qubits can interact pair-wisely and all $N$ qubits are connected by an interaction path. The interactions are tunable. Either one qubit can be measured and manipulated at the single-qubit level, or at least qubit-qubit (being spin-spin) correlations can be measured.
\textbf{\\$N$-qubit processor.} Every qubit (or most of them) has a two-axis control and can be measured. Qubits can interact pair-wisely through tunable interactions. The structure can perform $N$-qubit algorithms. 
}

To cast more light on the ambiguities that we met and the decisions we made to deal with them, let us make a few additional comments.

First, we do not include many-qubit structures which were fabricated, but no functionality was demonstrated, no matter what their size was. We do not include experiments where a multiple-dot structure was involved without any intention of using it for qubits (for example, it was used in a transport experiment).

Second, the vast majority of the experiments within the spin-qubit field until now were done with a single dot implementing a spin-1/2 qubit or a double dot implementing a singlet-triplet qubit. Therefore, we normally do not include these two cases in our tables and figures. One exception is if the experiment is somewhat outstanding, perhaps pioneering a spin qubit in a new material or platform. We included some of these. 

Third, the number of qubits is not the same as the number of dots. For example, a double-dot with two electrons can be viewed both as one singlet-triplet qubit as well as two qubits of spin-1/2 with limited functionality. In these cases, we follow the primary intention of the experiment as we understood it from the reference. For example, a triple dot implementing a resonant-exchange qubit is counted as a single-qubit structure implementing a fully functional resonant-exchange qubit (that is, a hybrid qubit in our nomenclature), and not as a structure implementing three spin-1/2 qubits with a limited functionality.

Finally, in Refs.~\cite{van_diepen_quantum_2021,dehollain_nagaoka_2020} there is no single-qubit gate nor measurement available. Still, we assign it to the quantum-simulator category, as in these experiments a simulation was the primary target and controllable spin-spin interactions played a critical role.

\subsection{Sizes of qubit arrays achieved experimentally}

\begin{figure}[t!]
  \includegraphics[width=0.89\linewidth]{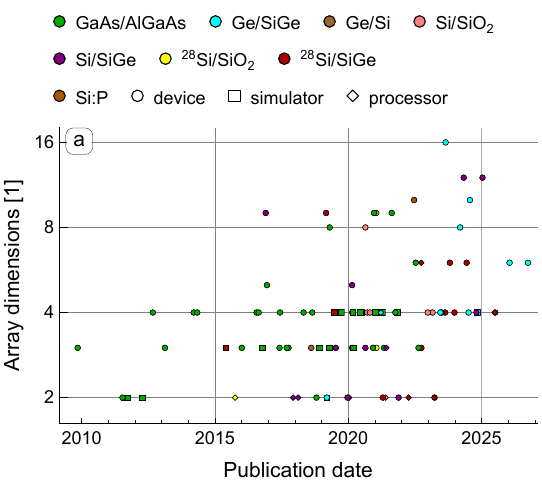}\\ 
    \includegraphics[width=0.89\linewidth]{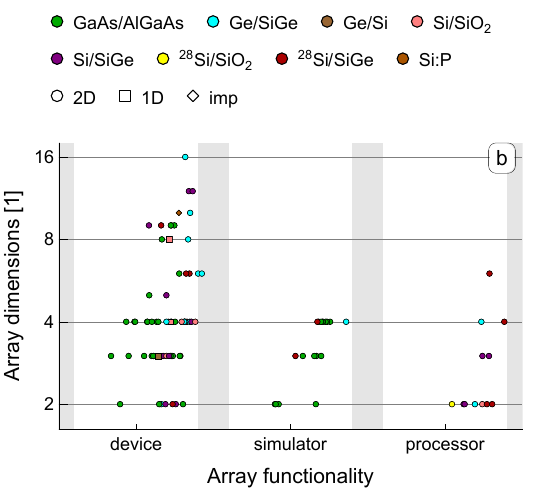}
  \caption{
  \label{fig:arraySize}
  Size of qubit arrays. The two panels show the same data: in (a) according to the publication date, with the color indicating the material and the symbol the array functionality (see the text for the definitions), according to the panel legend. In (b), the array functionality is on the horizontal axis, the publication date is reflected by shifting the points horizontally, similarly as in Fig.~\ref{fig:spinPhaseCoherenceSubplots}. The datapoint color indicates the material and its symbol the host geometry, according to the panel legend.
  }
\end{figure}

The sizes of qubit arrays appearing in publications on experiments are displayed in Fig.~\ref{fig:arraySize}. The panel (a) shows the progress over time. A lot of effort goes into scaling up the spin-qubit structures, with slow, but nevertheless steady progress.
It took about ten to fifteen years to bring the most basic structure of the spin-qubit field, the double dot, up to the functionality of a quantum processor, as defined in the above list: With the experiments starting \cite{petta_coherent_2005} around 2005, Refs.~\cite{veldhorst_two-qubit_2015,zajac_resonantly_2017,watson_programmable_2018} could be acknowledged to have reached the `processor' level, and only in the most recent one the fidelities where high enough to run the most elementary quantum circuits. We assign the accomplishment of the first fully-functional `processor' beyond the double dot to the year 2021, with Ref.~\cite{takeda_quantum_2021} using electrons in silicon and Ref.~\cite{hendrickx_four-qubit_2021} using holes in germanium. 
In Fig.~\ref{fig:arraySize}(b) one can see that the recent couple of years have brought a surge of results demonstrating progress in building spin-qubit arrays with advanced functionality. The fact that these recent breakthroughs come from many groups and happen in diverse materials and geometries gives excellent reasons for optimism on the scaling-up finally taking off.

\section{Acknowledgments}

We thank Takashi Nakajima, Akito Noiri, Kenta Takeda, and other members of Seigo Tarucha laboratory in RIKEN, for numerous discussions that were invaluable for preparing this review. We thank Leon Camenzind, Matthieu Delbecq, Georgios Katsaros, Benedikt Kratochwil, Ferdinand Kuemmeth, Tristan Meunier, Andrea Morello, Takashi Nakajima,  Akito Noiri, Matthew Reed, and Menno Veldhorst for providing feedback on early versions of the review. We also thank the anonymous and non-anonymous referees for many useful suggestions. Finally, we would like to acknowledge the financial support from CREST JST (JPMJCR1675) and Swiss National Science Foundation (SNSF) and NCCR SPIN.

\fi %

\appendix

\section{Relaxation time notations}

\label{app:notation}

To define the notation and give an example of the simplest dephasing form, we state here the result of Markovian (or exponential) decay displayed by the Bloch equations. Let us consider the following two parametrizations of the density matrix of a spin-1/2
\begin{equation}
\rho =\left( 
\begin{tabular}{cc}
 $\rho_{\uparrow\uparrow}$ &  $\rho_{\uparrow\downarrow}$\\
  $\rho_{\downarrow\uparrow}$&  $\rho_{\downarrow\downarrow}$
\end{tabular}
\right)
=\frac{1}{2} \left( 
\begin{tabular}{cc}
 $1+s_z$ & $s_x-i s_y$\\
 $s_x+i s_y$&$1-s_z$
\end{tabular}
\right) \equiv \frac{1}{2}\left( 1 + \mathbf{s} \cdot \boldsymbol{\sigma} \right).
\label{eq:densityMatrix}
\end{equation}
The defining and normalization conditions of a density matrix (trace one, Hermitivity, and positive semi-definiteness), require that the vector $\mathbf{s}$ is real and its length is not larger than one.

The Bloch equations \cite{bloch_nuclear_1946}, describing the exponential decay of the density-matrix components, read
\begin{subequations}
\begin{eqnarray}
\partial_t s_x &=&  -\frac{s_x}{T_2},\\
\partial_t s_y &=& -\frac{s_y}{T_2},\\
\partial_t s_z &=& -\frac{s_z-s_{0z}}{T_1},
\end{eqnarray}
\label{eq:Bloch}
\end{subequations}
with
$s_{0z}$ being the equilibrium spin polarization. The spin decay is described by two parameters, $T_1$ and $T_2$, called, respectively, the longitudinal and transverse relaxation time in the original work of Bloch. We denote the inverse of these times as rates, $\Gamma_1=1/T_1$ and $\Gamma_2=1/T_2$, that is, with the indexes corresponding.

We now use the alternative parameterization of the density matrix, where the exponential decay reads
\begin{subequations}
\begin{eqnarray}
\partial_t \rho_{\uparrow\uparrow} &=&-\Gamma_{ex} \rho_{\uparrow\uparrow} + \Gamma_{rel} \rho_{\downarrow\downarrow} ,\\
\partial_t \rho_{\uparrow\downarrow} &=&-\left( \frac{\Gamma_{ex}+\Gamma_{rel}}{2} + \Gamma_\varphi \right) \rho_{\uparrow\downarrow},
\end{eqnarray}
\label{eq:drhodt}
\end{subequations}
and the remaining matrix elements evolution is fixed by the requirements $\rho_{\uparrow\uparrow}+\rho_{\downarrow\downarrow} =1$ and $\rho_{\uparrow\downarrow} = \rho_{\downarrow\uparrow}^*$. Here, we use the notation of Ref.~\cite{ithier_decoherence_2005}, which originates in the assumption that the spin-up state is the energy ground state and the equations contain $\Gamma_{ex}$, the rate of excitations from the ground state, $\Gamma_{rel}$ the rate of relaxation into the ground state, and $\Gamma_\varphi$, the ``pure-dephasing'' rate. With the following relations,
\begin{subequations}
\begin{eqnarray}
\frac{1}{T_1} &=&\Gamma_{ex} + \Gamma_{rel},\\
\frac{1}{T_2} &=&\frac{\Gamma_{ex} + \Gamma_{rel}}{2} + \Gamma_\varphi,\\
s_{0z} &=& \frac{\Gamma_{rel} - \Gamma_{ex}}{\Gamma_{rel} + \Gamma_{ex}},
\end{eqnarray}
\end{subequations}
the two sets of equations, \eqref{eq:Bloch} and \eqref{eq:drhodt}, correspond to each other.

In the above equations for the density matrix evolution, we dropped the coherent part, which would read $\partial_t \mathbf{s} = (g \mu_B / \hbar) \mathbf{s} \times \mathbf{B}$, with $g$ being the electron g-factor and $\mathbf{B}$ the magnetic field. Equations \eqref{eq:Bloch} are derived in Ref.~\cite{fabian_semiconductor_2007} in a simple model on p.~707-712. The same reference also derives Eq.~\eqref{eq:drhodt} on p.~759-765, considering a spin qubit coupled to phonons as the decay source, following the standard methods of Ref.~\cite{blum_density_1996}.

\section{Relation of the gate fidelity and the gate quality factor}

\label{app:fidelity-quality}

The gate quality-factor is closely related to the gate fidelity, as we now illustrate. Let us consider a Rabi-oscillations experiment with a qubit. In the rotating frame, we describe it using a density matrix parameterized by vector $\mathbf{s}$. The Rabi oscillations induced by a resonant drive result in a time dependence of this vector. In an ideal scenario without errors or decoherence, the qubit rotates at a fixed frequency $f$ around a certain axis, set by the driving pulse phase. The time evolution is 
\begin{equation}
\rho_{ideal}(t) = \frac{1}{2}\Big(1+\mathbf{s}(t) \cdot \boldsymbol{\sigma}\Big).
\end{equation}
The ideal state at certain specific times corresponds to $|\Psi(t)\rangle\langle\Psi(t)|$. For example, with the initial state along $z$, $\mathbf{s}(0)=(0,0,1)$ and a rotation axis $x$, at time $t=1/2f$ an ideal $\pi$ rotation is performed. 

Let us now consider the coherence decay given by Eq.~\eqref{eq:Bloch}. Assuming, for simplicity, that either the rotation is in the plane of the Bloch sphere perpendicular to the energy-quantization axis, or taking $T_2=T_1$ and $s_{0z}=0$, the actual density matrix is described by $\mathbf{s}(t) \exp(-t/T_2)$. The amplitude of the oscillating polarization measured as a function of time will decay exponentially, as $\exp(-t/T_2)$. The fitted time-scale of the decay is called the ``gate decay time'' $ \TRabi$ entering Eq.~\eqref{eq:qualityFactorGate}. Evaluating the fidelity with Eq.~\eqref{eq:fidelity2} gives
\begin{equation}
\mathcal{F} = \frac{1}{2}\big(1+\exp(-t/T_2) \big).
\end{equation}
For a $\pi$-rotation gate, implemented imperfectly at time $t=1/2f$, it finally gives
\begin{equation}
\mathcal{F} = \frac{1}{2}\big(1+\exp(-1/2Q) \big) \approx 1 - \frac{1}{4Q},
\label{eq:fidelityQualityRelation}
\end{equation}
where the approximation is valid for small infidelity or large quality factor. 

In this toy model, the gate infidelity and the gate quality-factor are in a simple one-to-one correspondence. In reality, the relation is complicated by two facts. First, the dephasing might be more complicated than the Markovian decay described by the Bloch equations. More importantly, the signal decay and the fidelity are influenced differently by unitary gate-errors. The latter are deviations from the desired gate that do not originate in dephasing: for example, if the rotation axis is not exactly along $x$ or the rotation angle is below the target value $\pi$. By definition of the name, the unitary errors are systematic, that is they do not change in time. If known, they could be corrected by additional unitary rotations. Nevertheless, if they are not known, for example due to insufficient calibration, or they could not be corrected, they do diminish the gate fidelity. On the other hand, they do not lead to the gate-signal decay $ \TRabi$ as extracted experimentally. At most, they diminish the overall signal magnitude, called the visibility.  To conclude, because of the unitary errors diminishing the fidelity but not the gate quality-factor, the right-hand side in Eq.~\eqref{eq:fidelityQualityRelation} becomes only the upper limit for the left-hand side.

\begin{widetext}

\let\table\tablehere
\let\endtable\endtablehere

\rowcolors{2}{white}{gray!25}

\section{List of tables}

\label{app:tables}

\vspace{1cm}

\begin{table}[h!]
\Large
\begin{center}
\begin{tabular}{l@{\quad}ll}
\toprule
Content & Table & Page\\
\midrule
Spin coherence & Table \ref{tab:spinCoherence} & \pageref{spinCoherence1}-\pageref{spinCoherence8}\\
Charge coherence & Table \ref{tab:chargeCoherence} & \pageref{chargeCoherence1}--\pageref{chargeCoherence2}\\
Operation times & Table \ref{tab:operationTimes} & \pageref{operationTime1}-\pageref{operationTime5}\\
Operation fidelities & Table \ref{tab:operationFidelities} & \pageref{operationFidelity1}-\pageref{operationFidelity5}\\
Quality factors & Table \ref{tab:qualityFactors} & \pageref{qualityFactorMerged1}-\pageref{qualityFactorMerged2}\\
Qubit arrays & Table \ref{tab:qubitArrays} & \pageref{qubitArray1}-\pageref{qubitArray2}\\
\bottomrule
\end{tabular}
\end{center}
\end{table}

\vspace{1cm}

\newcounter{subtable}
\renewcommand{\theHtable}{\arabic{table}.\arabic{subtable}}

\setcounter{subtable}{0} \refstepcounter{table}\addtocounter{table}{-1}\label{tab:spinCoherence}
\renewcommand\thetable{\Roman{table}-\arabic{subtable}}
\setcounter{subtable}{1}

\input{table-spinCoherence-1.txt}

\addtocounter{table}{-1}
\addtocounter{subtable}{+1}

\ifFirstTableOnly \else
\input{table-spinCoherence-2.txt}

\addtocounter{table}{-1}
\addtocounter{subtable}{+1}

\input{table-spinCoherence-3.txt}

\addtocounter{table}{-1}
\addtocounter{subtable}{+1}

\input{table-spinCoherence-4.txt}

\addtocounter{table}{-1}
\addtocounter{subtable}{+1}

\input{table-spinCoherence-5.txt}

\addtocounter{table}{-1}
\addtocounter{subtable}{+1}

\input{table-spinCoherence-6.txt}

\addtocounter{table}{-1}
\addtocounter{subtable}{+1}

\input{table-spinCoherence-7.txt}

\addtocounter{table}{-1}
\addtocounter{subtable}{+1}

\input{table-spinCoherence-8.txt}

\addtocounter{table}{-1}
\addtocounter{subtable}{+1}
\fi

\addtocounter{table}{+1}

\setcounter{subtable}{0} \refstepcounter{table}\addtocounter{table}{-1}\label{tab:chargeCoherence}
\renewcommand\thetable{\Roman{table}-\arabic{subtable}}
\setcounter{subtable}{1}

\input{table-chargeCoherence-1.txt}

\addtocounter{table}{-1}
\addtocounter{subtable}{+1}

\ifFirstTableOnly \else
\input{table-chargeCoherence-2.txt}

\addtocounter{table}{-1}
\addtocounter{subtable}{+1}
\fi

\addtocounter{table}{+1}

\setcounter{subtable}{0} \refstepcounter{table}\addtocounter{table}{-1}\label{tab:operationTimes}
\renewcommand\thetable{\Roman{table}-\arabic{subtable}}
\setcounter{subtable}{1}

\input{table-operationTime-1.txt}

\addtocounter{table}{-1}
\addtocounter{subtable}{+1}

\ifFirstTableOnly \else
\input{table-operationTime-2.txt}

\addtocounter{table}{-1}
\addtocounter{subtable}{+1}

\input{table-operationTime-3.txt}

\addtocounter{table}{-1}
\addtocounter{subtable}{+1}

\input{table-operationTime-4.txt}

\addtocounter{table}{-1}
\addtocounter{subtable}{+1}

\input{table-operationTime-5.txt}

\addtocounter{table}{-1}
\addtocounter{subtable}{+1}
\fi

\addtocounter{table}{+1}

\renewcommand\thetable{\Roman{table}}
\setcounter{subtable}{0} \refstepcounter{table}\addtocounter{table}{-1}\label{tab:operationFidelities}

\renewcommand\thetable{\Roman{table}-\arabic{subtable}}
\setcounter{subtable}{1}

\input{table-operationFidelity-1.txt}

\addtocounter{table}{-1}
\addtocounter{subtable}{+1}

\ifFirstTableOnly \else
\input{table-operationFidelity-2.txt}

\addtocounter{table}{-1}
\addtocounter{subtable}{+1}

\input{table-operationFidelity-3.txt}

\addtocounter{table}{-1}
\addtocounter{subtable}{+1}

\input{table-operationFidelity-4.txt}

\addtocounter{table}{-1}
\addtocounter{subtable}{+1}

\input{table-operationFidelity-5.txt}

\addtocounter{table}{-1}
\addtocounter{subtable}{+1}
\fi

\addtocounter{table}{+1}

\renewcommand\thetable{\Roman{table}}
\setcounter{subtable}{0} \refstepcounter{table}\addtocounter{table}{-1}\label{tab:qualityFactors}

\renewcommand\thetable{\Roman{table}-\arabic{subtable}}
\setcounter{subtable}{1}

\input{table-qualityFactorMerged-1.txt}

\addtocounter{table}{-1}
\addtocounter{subtable}{+1}

\ifFirstTableOnly \else
\input{table-qualityFactorMerged-2.txt}

\addtocounter{table}{-1}
\addtocounter{subtable}{+1}

\fi

\addtocounter{table}{+1}

\renewcommand\thetable{\Roman{table}}
\setcounter{subtable}{0} \refstepcounter{table}\addtocounter{table}{-1}\label{tab:qubitArrays}

\renewcommand\thetable{\Roman{table}-\arabic{subtable}}
\setcounter{subtable}{1}

\input{table-qubitArrays-1.txt}

\addtocounter{table}{-1}
\addtocounter{subtable}{+1}

\ifFirstTableOnly \else
\input{table-qubitArrays-2.txt}

\addtocounter{table}{-1}
\addtocounter{subtable}{+1}
\fi

\addtocounter{table}{+1}

\renewcommand\thetable{\Roman{table}}

\clearpage

\end{widetext}

\renewcommand{\separate}[1]
{
 \vspace{0.5cm}
  \vspace{0.5cm}
  \noindent 
 #1
 \vspace{0.5cm}
  \vspace{0.5cm}
}

\section{Glossary}

\label{sec:vocabulary}

\newcommand{\keyword}[2]{\myKey{#1}. #2\\}

This section defines keywords with specific meaning used in plots and tables. Some are defined in the text, some are defined only here. 

\separate{First, we define the following special keywords.}

\begin{description}[leftmargin=5mm, topsep=0mm]
\item[\val{\derived}] The value is not stated in the reference.  We derived it as described in the \val{note}.
\item[\val{\estimated}] The value is not stated in the reference. We estimated it from available information, most often from a figure. The estimate is only rough, with a typical error of order one.
\item[\key{attribute}] A generic attribute. Several \val{value}s belong under it.
\item[\key{value}] A generic value. It belongs under a unique \val{attribute}.
\end{description}

\separate{The list of the \val{attribute}s and \val{value}s, the latter for the case when the allowed values are from a small set of enumerated alternatives.}

\begin{description}[leftmargin=5mm, topsep=0mm]
\input{vocabulary.txt}
\end{description}

\onecolumngrid

\bibliographystyle{plainnat}
\ifIncremental
\bibliography{../Added-bibtex-Western.bib}
\else
\bibliography{spinReview-bibtex-Western.1,spinReview-bibtex-Western.2}
\fi

\end{document}

%% file: figuresAndTables/table-spinCoherence-1.txt
\rowcolors{2}{white}{gray!25}
\setcounter{rowcount}{-1}
\begin{table}
\begin{tabular}{@{}S[table-format=2.2, round-mode=figures, round-pad=false, round-precision=2,  table-alignment-mode = none, table-number-alignment = center, table-column-width = 3em ]@{}lccccccc@{ \stepcounter{rowcount} \tiny \therowcount }}
\toprule
\multicolumn{2}{c}{Time} & {Coherence} & {Qubit} & {Material} & {Host} & {Date} & {Reference} & {Source}\\
\midrule
57. & \unit{\second} & $T_1$ & LD/e & GaAs/AlGaAs & 2D & 2018-08 & \onlinecite{camenzind_hyperfine-phonon_2018} & p3 and Fig. 4a\\
43. & \unit{\second} & $T_1$ & LD/i & Si:P & imp & 2024-03 & \onlinecite{hsueh_engineering_2024} & Tab. 2 and Fig. 4\\
30. & \unit{\second} & $T_1$ & LD/i & Si:P & imp & 2017-03 & \onlinecite{watson_atomically_2017} & Fig. 2b the lowest point\\
30. & \unit{\second}\textsuperscript{a} & $T_1$ & LD/h & BLG & 2D & 2025-02 & \onlinecite{denisov_spinvalley_2025} & p3\\
16. & \unit{\second} & $T_1$ & LD/i & Si:P & imp & 2023-11 & \onlinecite{kranz_exploiting_2023} & Tab. I\\
11. & \unit{\second}\textsuperscript{b} & $T_1$ & LD/i & Si:P & imp & 2023-02 & \onlinecite{hogg_single-shot_2023} & p6\\
10.8 & \unit{\second} & $T_1$ & ST/e & $\text{Si/}\text{SiO}_2$ & 2D & 2024-03 & \onlinecite{ma_singlet-triplet-state_2024} & p4\\
10. & \unit{\second} & $T_1$ & LD/e & GaAs/AlGaAs & 2D & 2017-10 & \onlinecite{hofmann_anisotropy_2017} & Fig. 2 the lowest green point\\
9.8 & \unit{\second} & $T_1$ & LD/i & Si:P & imp & 2019-05 & \onlinecite{tenberg_electron_2019} & Fig. 2c\\
9.3 & \unit{\second} & $T_1$ & LD/i & Si:P & imp & 2018-03 & \onlinecite{broome_two-electron_2018} & p3 and Fig. 1f\\
9. & \unit{\second} & $T_1$ & LD/e & $\text{Si/}\text{SiO}_2$ & 1D & 2021-03 & \onlinecite{ciriano-tejel_spin_2021} & p3 and Fig. 3a the leftmost blue point\\
6.45 & \unit{\second} & $T_1$ & LD/i & $\, ^{28}\text{Si:P}$ & imp & 2023-02 & \onlinecite{savytskyy_electrically_2023} & Fig. 3c\\
6. & \unit{\second} & $T_1$ & LD/i & Si:P & imp & 2010-09 & \onlinecite{morello_single-shot_2010} & p2\\
5. & \unit{\second}\textsuperscript{c} & $T_1$ & LD/e & Si/SiGe & 2D & 2019-04 & \onlinecite{borjans_single-spin_2019} & p4\\
4.23 & \unit{\second} & $T_1$ & LD/i & Si:P & imp & 2019-01 & \onlinecite{koch_spin_2019} & p3\\
3.4 & \unit{\second} & $T_1$ & LD/i & $\, ^{28}\text{Si:P}$ & imp & 2021-01 & \onlinecite{madzik_conditional_2021} & p6 and {SFig}. 3c\\
3. & \unit{\second} & $T_1$ & LD/i & $\, ^{28}\text{Si:P}$ & imp & 2016-10 & \onlinecite{dehollain_optimization_2016} & p3\\
3. & \unit{\second} & $T_1$ & ST/e & Si/SiGe & 2D & 2012-01 & \onlinecite{prance_single-shot_2012} & p4\\
2.8 & \unit{\second} & $T_1$ & LD/e & Si/SiGe & 2D & 2011-04 & \onlinecite{simmons_tunable_2011} & p3 and Fig. 3\\
2.6 & \unit{\second} & $T_1$ & LD/e & $\text{Si/}\text{SiO}_2$ & 2D & 2013-06 & \onlinecite{yang_spin-valley_2013} & p3\\
1.8 & \unit{\second} & $T_1$ & LD/i & Si:P & imp & 2013-06 & \onlinecite{buch_spin_2013} & Fig. 3\\
1.6 & \unit{\second} & $T_1$ & LD/e & $\, ^{28}\text{Si/}\text{SiO}_2$ & 2D & 2022-03 & \onlinecite{zwerver_qubits_2022} & p4 and Fig. 3c\\
1.3 & \unit{\second} & $T_1$ & LD/i\textsuperscript{d} & $\, ^{28}\text{Si:P}$ & imp & 2016-10 & \onlinecite{laucht_dressed_2016} & p4\\
1.25 & \unit{\second} & $T_1$ & LD/i & Si:P & imp & 2018-11 & \onlinecite{weber_spinorbit_2018} & p3 and Fig. 2b\\
1. & \unit{\second} & $T_1$ & LD/e & $\, ^{28}\text{Si/SiGe}$ & 2D & 2020-03 & \onlinecite{hollmann_large_2020} & p6 and Fig. 4a\\
1. & \unit{\second} & $T_1$ & LD/e & $\, ^{28}\text{Si/}\text{SiO}_2$ & 2D & 2018-10 & \onlinecite{chan_assessment_2018} & p2\\
1. & \unit{\second} & $T_1$ & LD/e & GaAs/AlGaAs & 2D & 2008-01 & \onlinecite{amasha_electrical_2008} & p4 and Fig. 3c the leftmost blue point\\
0.7 & \unit{\second} & $T_1$ & LD/i & Si:P & imp & 2012-09 & \onlinecite{pla_single-atom_2012} & p3\\
0.6 & \unit{\second}\textsuperscript{e} & $T_1$ & LD/e & Si/SiGe & 2D & 2009-08 & \onlinecite{hayes_lifetime_2009} & Fig. 5\\
0.5 & \unit{\second}\textsuperscript{f} & $T_1$ & ST/e & $\, ^{28}\text{Si/}\text{SiO}_2$ & 2D & 2020-04 & \onlinecite{yang_operation_2020} & Fig. 4 the leftmost black point\\
0.4 & \unit{\second} & $T_1$ & LD/h & BLG & 2D & 2025-02 & \onlinecite{denisov_spinvalley_2025} & p4\\
0.354 & \unit{\second}\textsuperscript{g} & $T_1$ & HY/e & BLG & 2D & 2024-01 & \onlinecite{garreis_long-lived_2024} & Fig. 3b\\
0.170 & \unit{\second} & $T_1$ & LD/e & Si/SiGe & 2D & 2016-11 & \onlinecite{zajac_scalable_2016} & Fig. 6\\
0.160 & \unit{\second}\textsuperscript{h} & $T_1$ & LD/e & Si/SiGe & 2D & 2019-04 & \onlinecite{borjans_single-spin_2019} & Fig. 2\\
0.145 & \unit{\second}\textsuperscript{i} & $T_1$ & LD/e & $\, ^{28}\text{Si/}\text{SiO}_2$ & 2D & 2018-08 & \onlinecite{petit_spin_2018} & p2 and p4\\
0.141 & \unit{\second} & $T_1$ & ST/e & Si/SiGe & 2D & 2012-04 & \onlinecite{shi_fast_2012} & Fig. 2d\\
0.134 & \unit{\second} & $T_1$ & LD/e & $\, ^{28}\text{Si/SiGe}$ & 2D & 2019-11 & \onlinecite{sigillito_coherent_2019} & p4\\
90. & \unit{\milli\second} & $T_1$ & LD/e & $\text{Si/}\text{SiO}_2$ & 2D & 2020-06 & \onlinecite{zhang_giant_2020} & Fig. 1c\\
85. & \unit{\milli\second} & $T_1$ & LD/e & GaAs/AlGaAs & 2D & 2014-12 & \onlinecite{scarlino_spin-relaxation_2014} & p2 and Fig. 3\\
50. & \unit{\milli\second} & $T_1$ & LD/e & Si/SiGe & 2D & 2018-02 & \onlinecite{watson_programmable_2018} & p1 and {ED} Fig. 3b\\
40. & \unit{\milli\second} & $T_1$ & LD/e & $\text{Si/}\text{SiO}_2$ & 2D & 2010-03 & \onlinecite{xiao_measurement_2010} & p4 and Fig. 4 the leftmost red point\\
34. & \unit{\milli\second} & $T_1$ & ST/e & BLG & 2D & 2024-01 & \onlinecite{garreis_long-lived_2024} & Fig. 3b\\
32. & \unit{\milli\second} & $T_1$ & LD/h & Ge/SiGe & 2D & 2020-08 & \onlinecite{lawrie_spin_2020} & p3\\
31.5 & \unit{\milli\second} & $T_1$ & LD/e & $\, ^{28}\text{Si/SiGe}$ & 2D & 2022-12 & \onlinecite{mills_high-fidelity_2022} & p3\\
22. & \unit{\milli\second}\textsuperscript{j} & $T_1$ & LD/e & Si/SiGe & 2D & 2022-08 & \onlinecite{takeda_quantum_2022} & p2 and {ED} Fig. 4b-d\\
20. & \unit{\milli\second} & $T_1$ & HY/e\textsuperscript{k} & $\, ^{28}\text{Si/SiGe}$ & 2D & 2022-03 & \onlinecite{blumoff_fast_2022} & p5\\
16. & \unit{\milli\second} & $T_1$ & LD/h & Ge/SiGe & 2D & 2021-03 & \onlinecite{hendrickx_four-qubit_2021} & Fig. S5 dot 3\\
15. & \unit{\milli\second}\textsuperscript{l} & $T_1$ & ST/e & $\, ^{28}\text{Si/}\text{SiO}_2$ & 2D & 2020-04 & \onlinecite{yang_operation_2020} & Fig. 4 the rightmost black point\\
11.23 & \unit{\milli\second}\textsuperscript{m} & $T_1$ & LD/e & $\, ^{28}\text{Si/}\text{SiO}_2$ & 2D & 2024-03 & \onlinecite{huang_high-fidelity_2024} & p3\\
10. & \unit{\milli\second} & $T_1$ & LD/e & $\text{Si/}\text{SiO}_2$ & 1D & 2022-03 & \onlinecite{spence_spin-valley_2022} & p2 and Fig. 2a\\
\bottomrule
\end{tabular}
\caption{Spin coherence times (part 1). Superscripts stand for the following. \textsuperscript{a}: Spin-valley relaxation. \textsuperscript{b}: Dot D3. \textsuperscript{c}: No micromagnet. \textsuperscript{d}: Qubit defined in the rotating frame. \textsuperscript{e}: \estimated Fig. 5 the lowest point. \textsuperscript{f}: At 0.04 K. \textsuperscript{g}: Valley degree of freedom. \textsuperscript{h}: With micromagnet. \textsuperscript{i}: At 0.1 K. \textsuperscript{j}: The average over the three qubits. \textsuperscript{k}: {EO} qubit. \textsuperscript{l}: At 1.5 K. \textsuperscript{m}: At 1 K.
\label{spinCoherence1}
}
\end{table}

%% file: figuresAndTables/table-spinCoherence-2.txt
\rowcolors{2}{white}{gray!25}
\setcounter{rowcount}{-1}
\begin{table}
\begin{tabular}{@{}S[table-format=2.2, round-mode=figures, round-pad=false, round-precision=2,  table-alignment-mode = none, table-number-alignment = center, table-column-width = 3em ]@{}lccccccc@{ \stepcounter{rowcount} \tiny \therowcount }}
\toprule
\multicolumn{2}{c}{Time} & {Coherence} & {Qubit} & {Material} & {Host} & {Date} & {Reference} & {Source}\\
\midrule
8.37 & \unit{\milli\second} & $T_1$ & LD/h & BLG & 2D & 2022-05 & \onlinecite{gachter_single-shot_2022} & p5 and Fig. 4\\
5. & \unit{\milli\second} & $T_1$ & LD/i & $\, ^{28}\text{Si:B}$ & imp & 2020-07 & \onlinecite{kobayashi_engineering_2020} & p3 and Fig. 3b\\
5. & \unit{\milli\second}\textsuperscript{a} & $T_1$ & ST/e & $\, ^{28}\text{Si/}\text{SiO}_2$ & 2D & 2021-01 & \onlinecite{seedhouse_pauli_2021} & p4 and Fig. 1d\\
5. & \unit{\milli\second}\textsuperscript{b} & $T_1$ & ST/h & Ge/SiGe & 2D & 2026-01 & \onlinecite{tsoukalas_dressed_2026} & p3\\
4.1 & \unit{\milli\second} & $T_1$ & ST/i & Si:P & imp & 2014-06 & \onlinecite{dehollain_single-shot_2014} & \\
3.7 & \unit{\milli\second} & $T_1$ & LD/e & GaAs/AlGaAs & 2D & 2016-07 & \onlinecite{baart_nanosecond-timescale_2016} & p3 and Fig. 2\\
3.7 & \unit{\milli\second} & $T_1$ & LD/e\textsuperscript{c} & $\, ^{28}\text{Si/}\text{SiO}_2$ & 2D & 2020-04 & \onlinecite{petit_universal_2020} & p2\\
3.1 & \unit{\milli\second} & $T_1$ & LD/i & $\, ^{28}\text{Si:P}$ & imp & 2022-01 & \onlinecite{madzik_precision_2022} & {ED} Fig. 3 first column\\
2.8 & \unit{\milli\second}\textsuperscript{d} & $T_1$ & LD/e & $\, ^{28}\text{Si/}\text{SiO}_2$ & 2D & 2018-08 & \onlinecite{petit_spin_2018} & p4 and Fig. 3a\\
2.3 & \unit{\milli\second}\textsuperscript{e} & $T_1$ & ST/e & GaAs/AlGaAs & 2D & 2007-03 & \onlinecite{meunier_experimental_2007} & p2\\
1.53 & \unit{\milli\second} & $T_1$ & LD/e & GaAs/AlGaAs & 2D & 2019-04 & \onlinecite{nakajima_quantum_2019} & Fig. 2\\
1.31 & \unit{\milli\second} & $T_1$ & LD/e & Si/SiGe & 2D & 2021-06 & \onlinecite{takeda_quantum_2021} & p1 for Q3\\
1.2 & \unit{\milli\second} & $T_1$ & LD/h & Ge/SiGe & 2D & 2020-07 & \onlinecite{hendrickx_single-hole_2020} & p4 and Fig. 3a\\
0.91 & \unit{\milli\second} & $T_1$ & ST/h & $\text{Si/}\text{SiO}_2$ & 2D & 2025-01 & \onlinecite{kondo_spin-blockade_2025} & p3\\
0.85 & \unit{\milli\second} & $T_1$ & LD/e & GaAs/AlGaAs & 2D & 2004-07 & \onlinecite{elzerman_single-shot_2004} & p4\\
0.8 & \unit{\milli\second} & $T_1$ & ST/h\textsuperscript{f} & Ge/SiGe & 2D & 2026-01 & \onlinecite{tsoukalas_dressed_2026} & Fig. 3c\\
0.200 & \unit{\milli\second} & $T_1$ & ST/e & InGaAs/AlGaAs & 2D & 2002-09 & \onlinecite{fujisawa_allowed_2002} & p3 and Fig. 3e\\
0.200 & \unit{\milli\second}\textsuperscript{g} & $T_1$ & ST/e & $\, ^{28}\text{Si/}\text{SiO}_2$ & 2D & 2021-01 & \onlinecite{seedhouse_pauli_2021} & p4 and Fig. 1d\\
86. & \unit{\micro\second} & $T_1$ & LD/h & Ge/Si & 1D & 2018-10 & \onlinecite{vukusic_single-shot_2018} & p3 and Fig. 3c the leftmost point\\
60. & \unit{\micro\second} & $T_1$ & LD/h & GaAs/AlGaAs & 2D & 2019-02 & \onlinecite{bogan_single_2019} & abstract and Fig. 4\\
50. & \unit{\micro\second} & $T_1$ & LD/e & GaAs/AlGaAs & 2D & 2003-11 & \onlinecite{hanson_zeeman_2003} & abstract\\
40. & \unit{\micro\second}\textsuperscript{h} & $T_1$ & ST/e & GaAs/AlGaAs & 2D & 2012-01 & \onlinecite{barthel_relaxation_2012} & p5\\
34. & \unit{\micro\second} & $T_1$ & ST/e & GaAs/AlGaAs & 2D & 2009-10 & \onlinecite{barthel_rapid_2009} & p3\\
17.2 & \unit{\micro\second} & $T_1$ & ST/e\textsuperscript{i} & Ge/SiGe & 2D & 2025-01 & \onlinecite{rooney_gate_2025} & p3 and Fig. 3b\\
9. & \unit{\micro\second} & $T_1$ & LD/h & Ge/SiGe & 2D & 2020-01 & \onlinecite{hendrickx_fast_2020} & p3 and Fig. 2f\\
7. & \unit{\micro\second}\textsuperscript{j} & $T_1$ & LD/e & BLG & 2D & 2024-02 & \onlinecite{banszerus_phonon-limited_2024} & p4 and Fig. 4\\
3.65 & \unit{\micro\second} & $T_1$ & LD/h & Ge/Si & 1D & 2022-01 & \onlinecite{wang_ultrafast_2022} & {SM} p13 and {SFig}. 7.1d\\
3.3 & \unit{\micro\second}\textsuperscript{k} & $T_1$ & LD/e & InAs & 1D & 2024-05 & \onlinecite{pita-vidal_strong_2024} & p3\\
1. & \unit{\micro\second} & $T_1$ & LD/e & InAs & 1D & 2012-10 & \onlinecite{petersson_circuit_2012} & Fig. 4d\\
5. & \unit{\nano\second} & $T_1$ & LD/h & GaAs/AlGaAs & 1D & 2016-11 & \onlinecite{wang_anisotropic_2016} & p4 and {SM} {pS}5\\
2.4 & \unit{\milli\second} & $T_2^*$ & LD/i\textsuperscript{f} & $\, ^{28}\text{Si:P}$ & imp & 2016-10 & \onlinecite{laucht_dressed_2016} & p4\\
2. & \unit{\milli\second}\textsuperscript{l} & $T_2^*$ & LD/e & $\, ^{28}\text{Si/}\text{SiO}_2$ & 2D & 2022-09 & \onlinecite{hansen_implementation_2022} & p4 and Fig. 3f\\
0.270 & \unit{\milli\second} & $T_2^*$ & LD/i & $\, ^{28}\text{Si:P}$ & imp & 2014-10 & \onlinecite{muhonen_storing_2014} & p1 and Fig. 2b\\
0.235 & \unit{\milli\second}\textsuperscript{m} & $T_2^*$ & LD/e & $\, ^{28}\text{Si/}\text{SiO}_2$ & 2D & 2022-09 & \onlinecite{hansen_implementation_2022} & p4 and Fig. 3e\\
0.120 & \unit{\milli\second} & $T_2^*$ & LD/e & $\, ^{28}\text{Si/}\text{SiO}_2$ & 2D & 2014-10 & \onlinecite{veldhorst_addressable_2014} & p2 and Fig. 3a\\
0.120 & \unit{\milli\second} & $T_2^*$ & LD/e & $\, ^{28}\text{Si/}\text{SiO}_2$ & 2D & 2015-10 & \onlinecite{veldhorst_two-qubit_2015} & p1\\
0.120 & \unit{\milli\second} & $T_2^*$ & LD/e & $\, ^{28}\text{Si/}\text{SiO}_2$ & 2D & 2021-01 & \onlinecite{chan_exchange_2021} & p13 and Fig. 4a\\
0.116 & \unit{\milli\second} & $T_2^*$ & LD/i & $\, ^{28}\text{Si:P}$ & imp & 2022-01 & \onlinecite{madzik_precision_2022} & {ED} Fig. 3 first column\\
0.112 & \unit{\milli\second} & $T_2^*$ & LD/e & Si/SiGe & 2D & 2025-01 & \onlinecite{george_12-spin-qubit_2025} & p5 and Fig. 5f\\
0.107 & \unit{\milli\second} & $T_2^*$ & LD/i & $\, ^{28}\text{Si:P}$ & imp & 2020-07 & \onlinecite{madzik_controllable_2020} & Fig. 2b and i\\
70. & \unit{\micro\second} & $T_2^*$ & LD/e\textsuperscript{n} & $\, ^{28}\text{Si/}\text{SiO}_2$ & 2D & 2015-11 & \onlinecite{veldhorst_spin-orbit_2015} & p2 and Fig. 2b\\
40.7 & \unit{\micro\second}\textsuperscript{o,p} & $T_2^*$ & LD/e & $\text{Si/}\text{SiO}_2$ & 2D & 2026-01 & \onlinecite{kuno_robust_2026} & p4\\
33. & \unit{\micro\second} & $T_2^*$ & LD/e & $\, ^{28}\text{Si/}\text{SiO}_2$ & 2D & 2018-10 & \onlinecite{chan_assessment_2018} & p2\\
28.1 & \unit{\micro\second} & $T_2^*$ & LD/i & Si:P & imp & 2025-02 & \onlinecite{thorvaldson_grovers_2025} & p3\\
24. & \unit{\micro\second} & $T_2^*$ & LD/e & $\, ^{28}\text{Si/}\text{SiO}_2$ & 2D & 2019-05 & \onlinecite{huang_fidelity_2019} & p3 and {ExtData} Fig. 1a\\
24. & \unit{\micro\second} & $T_2^*$ & LD/e & $\, ^{28}\text{Si/}\text{SiO}_2$ & 2D & 2022-03 & \onlinecite{zwerver_qubits_2022} & p5 and {SM} Fig. 14\\
21. & \unit{\micro\second} & $T_2^*$ & LD/e & $\, ^{28}\text{Si/SiGe}$ & 2D & 2020-05 & \onlinecite{struck_low-frequency_2020} & Fig. 3b\\
21. & \unit{\micro\second} & $T_2^*$ & LD/i & $\, ^{28}\text{Si:P}$ & imp & 2024-09 & \onlinecite{stemp_tomography_2024} & Fig. 2a\\
20.3 & \unit{\micro\second} & $T_2^*$ & ST/h\textsuperscript{f} & Ge/SiGe & 2D & 2026-01 & \onlinecite{tsoukalas_dressed_2026} & Fig. 2f\\
20. & \unit{\micro\second} & $T_2^*$ & LD/e & $\, ^{28}\text{Si/SiGe}$ & 2D & 2017-12 & \onlinecite{yoneda_quantum-dot_2017} & p3 and Fig. 3a\\
\bottomrule
\end{tabular}
\caption{Spin coherence times (part 2). Superscripts stand for the following. \textsuperscript{a}: Lifetime of $T_-$ state. \textsuperscript{b}: Lower limit. \textsuperscript{c}: At 1 K. \textsuperscript{d}: At 1.1 K. \textsuperscript{e}: Triplet-singlet relaxation in a single dot. \textsuperscript{f}: Qubit defined in the rotating frame. \textsuperscript{g}: Lifetime of $T_0$ state. \textsuperscript{h}: From several relaxation times defined, we take the inverse of the "bare triplet relaxation rate" $1/\Gamma_T=40$ \unit{\micro\second}. \textsuperscript{i}: $S$-$T_-$ qubit. \textsuperscript{j}: Valley lifetime. \textsuperscript{k}: Q1. \textsuperscript{l}: Qubit dressed with an oscillatory drive-amplitude. \textsuperscript{m}: Qubit dressed with a constant drive-amplitude. \textsuperscript{n}: A three-electron qubit. \textsuperscript{o}: Driven qubit. \textsuperscript{p}: The reference denotes it $T_\mathrm{ccd-Ramsey}$.
\label{spinCoherence2}
}
\end{table}

%% file: figuresAndTables/table-spinCoherence-3.txt
\rowcolors{2}{white}{gray!25}
\setcounter{rowcount}{-1}
\begin{table}
\begin{tabular}{@{}S[table-format=2.2, round-mode=figures, round-pad=false, round-precision=2,  table-alignment-mode = none, table-number-alignment = center, table-column-width = 3em ]@{}lccccccc@{ \stepcounter{rowcount} \tiny \therowcount }}
\toprule
\multicolumn{2}{c}{Time} & {Coherence} & {Qubit} & {Material} & {Host} & {Date} & {Reference} & {Source}\\
\midrule
20. & \unit{\micro\second} & $T_2^*$ & LD/e & $\, ^{28}\text{Si/SiGe}$ & 2D & 2022-01 & \onlinecite{xue_quantum_2022} & Fig. 3c\\
18. & \unit{\micro\second} & $T_2^*$ & LD/i & $\, ^{28}\text{Si:P}$ & imp & 2016-02 & \onlinecite{tracy_single_2016} & Fig. 4c\\
17.6 & \unit{\micro\second} & $T_2^*$ & LD/h & Ge/SiGe & 2D & 2024-05 & \onlinecite{hendrickx_sweet-spot_2024} & p5 and Fig. 6a\\
16. & \unit{\micro\second} & $T_2^*$ & LD/e\textsuperscript{a} & $\text{Si/}\text{SiO}_2$ & 2D & 2020-02 & \onlinecite{leon_coherent_2020} & {SM} Tab. I\\
16. & \unit{\micro\second} & $T_2^*$ & LD/e & $\, ^{28}\text{Si/}\text{SiO}_2$ & 2D & 2022-09 & \onlinecite{hansen_implementation_2022} & p4 and Fig. 3d\\
14.9 & \unit{\micro\second} & $T_2^*$ & LD/i & $\, ^{28}\text{Si:P}$ & imp & 2023-02 & \onlinecite{savytskyy_electrically_2023} & Fig. 3c\\
12. & \unit{\micro\second}\textsuperscript{b} & $T_2^*$ & ST/e & $\, ^{28}\text{Si/}\text{SiO}_2$ & 2D & 2020-04 & \onlinecite{yang_operation_2020} & Fig. 3c\\
12. & \unit{\micro\second} & $T_2^*$ & LD/i & $\, ^{28}\text{Si:P}$ & imp & 2024-02 & \onlinecite{reiner_high-fidelity_2024} & Fig. 5c\\
10.6 & \unit{\micro\second} & $T_2^*$ & ST/e & $\, ^{28}\text{Si/SiGe}$ & 2D & 2024-02 & \onlinecite{takeda_rapid_2024} & Fig. 4b\\
10. & \unit{\micro\second} & $T_2^*$ & LD/e & $\, ^{28}\text{Si/SiGe}$ & 2D & 2019-11 & \onlinecite{sigillito_coherent_2019} & p4\\
9.4 & \unit{\micro\second} & $T_2^*$ & LD/e & $\, ^{28}\text{Si/SiGe}$ & 2D & 2019-06 & \onlinecite{sigillito_site-selective_2019} & Tab.1\\
9. & \unit{\micro\second} & $T_2^*$ & LD/i & $\, ^{28}\text{Si:P}$ & imp & 2021-01 & \onlinecite{madzik_conditional_2021} & p7 and {SFig}. 3a\\
8.2 & \unit{\micro\second} & $T_2^*$ & LD/e & $\, ^{28}\text{Si/}\text{SiO}_2$ & 2D & 2019-05 & \onlinecite{tanttu_controlling_2019} & p5 and Fig. 3b\\
7.4 & \unit{\micro\second} & $T_2^*$ & LD/e & $\, ^{28}\text{Si/SiGe}$ & 2D & 2022-01 & \onlinecite{noiri_fast_2022} & {ED} Fig.2f the topmost trace\\
7.14 & \unit{\micro\second}\textsuperscript{c} & $T_2^*$ & LD/e & $\, ^{28}\text{Si/SiGe}$ & 2D & 2026-05 & \onlinecite{matsumoto_two-qubit_2026} & Fig. S1b\\
7. & \unit{\micro\second} & $T_2^*$ & LD/h & $\text{Si/}\text{SiO}_2$ & 1D & 2022-09 & \onlinecite{piot_single_2022} & p4 and Fig. 4c\\
7. & \unit{\micro\second}\textsuperscript{d} & $T_2^*$ & LD/h & Ge/SiGe & 2D & 2024-07 & \onlinecite{wang_operating_2024} & Fig. 1f\\
7. & \unit{\micro\second} & $T_2^*$ & LD/e & $\, ^{28}\text{Si/SiGe}$ & 2D & 2025-05 & \onlinecite{kobayashi_charge-induced_2025} & p4\\
6.8 & \unit{\micro\second} & $T_2^*$ & LD/e & $\, ^{28}\text{Si/SiGe}$ & 2D & 2021-03 & \onlinecite{kerckhoff_magnetic_2021} & Fig. 6 the blue point\\
4.8 & \unit{\micro\second} & $T_2^*$ & ST/e & $\, ^{28}\text{Si/SiGe}$ & 2D & 2024-08 & \onlinecite{song_coherence_2024} & p3 and Fig. 2b\\
4. & \unit{\micro\second}\textsuperscript{e} & $T_2^*$ & LD/e & $\, ^{28}\text{Si/SiGe}$ & 2D & 2023-10 & \onlinecite{undseth_hotter_2023} & Fig. 11\\
3.7 & \unit{\micro\second}\textsuperscript{f} & $T_2^*$ & LD/e & $\, ^{28}\text{Si/SiGe}$ & 2D & 2022-09 & \onlinecite{philips_universal_2022} & Fig. 2e\\
3.5 & \unit{\micro\second} & $T_2^*$ & ST/e & $\, ^{28}\text{Si/SiGe}$ & 2D & 2023-03 & \onlinecite{weinstein_universal_2023} & p7 and {ED} Fig. 1\\
3.4 & \unit{\micro\second} & $T_2^*$ & LD/e & $\, ^{28}\text{Si/}\text{SiO}_2$ & 2D & 2019-05 & \onlinecite{harvey-collard_spin-orbit_2019} & p2\\
3.32 & \unit{\micro\second}\textsuperscript{g} & $T_2^*$ & LD/e & $\, ^{28}\text{Si/}\text{SiO}_2$ & 2D & 2024-07 & \onlinecite{bartee_spin_2024} & Fig. 3d\\
3.31 & \unit{\micro\second}\textsuperscript{h} & $T_2^*$ & LD/e & $\, ^{28}\text{Si/SiGe}$ & 2D & 2025-07 & \onlinecite{unseld_baseband_2025} & p2\\
3.3 & \unit{\micro\second} & $T_2^*$ & LD/e & $\, ^{28}\text{Si/}\text{SiO}_2$ & 2D & 2019-12 & \onlinecite{zhao_single-spin_2019} & Tab. 1\\
3.2 & \unit{\micro\second} & $T_2^*$ & LD/e & $\, ^{28}\text{Si/SiGe}$ & 2D & 2022-12 & \onlinecite{mills_high-fidelity_2022} & p5\\
2.9 & \unit{\micro\second}\textsuperscript{i,j} & $T_2^*$ & LD/e & $\, ^{28}\text{Si/SiGe}$ & 2D & 2022-11 & \onlinecite{petit_design_2022} & p3\\
2.8 & \unit{\micro\second}\textsuperscript{k} & $T_2^*$ & LD/e & $\, ^{28}\text{Si/}\text{SiO}_2$ & 2D & 2024-08 & \onlinecite{tanttu_assessment_2024} & {ED} Tab. {II}\\
2.7 & \unit{\micro\second} & $T_2^*$ & LD/e\textsuperscript{i} & $\, ^{28}\text{Si/}\text{SiO}_2$ & 2D & 2020-04 & \onlinecite{petit_universal_2020} & p2 and Fig. 2e\\
2.59 & \unit{\micro\second}\textsuperscript{l} & $T_2^*$ & LD/e & $\, ^{28}\text{Si/SiGe}$ & 2D & 2022-09 & \onlinecite{noiri_shuttling-based_2022} & p3 and {ED} Fig. 2a\\
2.33 & \unit{\micro\second}\textsuperscript{m} & $T_2^*$ & LD/e & $\, ^{28}\text{Si/}\text{SiO}_2$ & 2D & 2022-11 & \onlinecite{vahapoglu_coherent_2022} & p3 and Fig. 3a\\
2.3 & \unit{\micro\second} & $T_2^*$ & ST/e & $\, ^{28}\text{Si/SiGe}$ & 2D & 2015-05 & \onlinecite{eng_isotopically_2015} & p3\\
2.2 & \unit{\micro\second} & $T_2^*$ & LD/e\textsuperscript{n} & $\, ^{28}\text{Si/}\text{SiO}_2$ & 2D & 2024-09 & \onlinecite{hansen_entangling_2024} & Fig. 3a and p5\\
2.066 & \unit{\micro\second}\textsuperscript{o} & $T_2^*$ & ST/e & GaAs/AlGaAs & 2D & 2014-10 & \onlinecite{shulman_suppressing_2014} & p4 and Fig. 3b\\
2.011 & \unit{\micro\second} & $T_2^*$ & ST/e\textsuperscript{p} & Si/SiGe & 2D & 2023-01 & \onlinecite{cai_coherent_2023} & Fig. 6b\\
2. & \unit{\micro\second}\textsuperscript{q} & $T_2^*$ & ST/e & $\, ^{28}\text{Si/}\text{SiO}_2$ & 2D & 2020-04 & \onlinecite{yang_operation_2020} & Fig. 3g\\
2. & \unit{\micro\second}\textsuperscript{r} & $T_2^*$ & HY/e\textsuperscript{s} & $\, ^{28}\text{Si/SiGe}$ & 2D & 2019-07 & \onlinecite{andrews_quantifying_2019} & Fig. S2a\\
2. & \unit{\micro\second}\textsuperscript{t} & $T_2^*$ & LD/e & $\, ^{28}\text{Si/SiGe}$ & 2D & 2022-04 & \onlinecite{mills_two-qubit_2022} & p1\\
1.9 & \unit{\micro\second} & $T_2^*$ & ST/h & Ge/SiGe & 2D & 2026-01 & \onlinecite{tsoukalas_dressed_2026} & Fig. 2e\\
1.8 & \unit{\micro\second}\textsuperscript{u} & $T_2^*$ & LD/e & Si/SiGe & 2D & 2022-08 & \onlinecite{takeda_quantum_2022} & p2 and {ED} Fig. 4f-h\\
1.8 & \unit{\micro\second} & $T_2^*$ & LD/e & Si/SiGe & 2D & 2016-08 & \onlinecite{takeda_fault-tolerant_2016} & p3 and Fig. 2c\\
1.75 & \unit{\micro\second} & $T_2^*$ & LD/e & $\, ^{28}\text{Si/SiGe}$ & 2D & 2024-06 & \onlinecite{de_smet_high-fidelity_2024} & p3\\
1.7 & \unit{\micro\second}\textsuperscript{v} & $T_2^*$ & ST/e & Si/SiGe & 2D & 2017-09 & \onlinecite{qi_effects_2017} & p2 and Fig. 4\\
1.69 & \unit{\micro\second} & $T_2^*$ & LD/e & Si/SiGe & 2D & 2021-06 & \onlinecite{takeda_quantum_2021} & p1 for Q3\\
1.6 & \unit{\micro\second} & $T_2^*$ & ST/e & $\, ^{28}\text{Si/}\text{SiO}_2$ & 2D & 2018-05 & \onlinecite{jock_silicon_2018} & p6\\
1.3 & \unit{\micro\second} & $T_2^*$ & ST/e & Si/SiGe & 2D & 2020-03 & \onlinecite{takeda_resonantly_2020} & Fig. 3c\\
1.3 & \unit{\micro\second} & $T_2^*$ & ST/e\textsuperscript{w} & $\, ^{28}\text{Si:P}$ & imp & 2017-10 & \onlinecite{harvey-collard_coherent_2017} & p4\\
1.3 & \unit{\micro\second} & $T_2^*$ & LD/e & Si/SiGe & 2D & 2025-01 & \onlinecite{george_12-spin-qubit_2025} & p5\\
\bottomrule
\end{tabular}
\caption{Spin coherence times (part 3). Superscripts stand for the following. \textsuperscript{a}: In a five-electron dot configuration. \textsuperscript{b}: At 0.04 K. \textsuperscript{c}: Q5. \textsuperscript{d}: $Q_A$. \textsuperscript{e}: We take a representative value from Fig. 11. \textsuperscript{f}: Q3. \textsuperscript{g}: With cryo-{CMOS} control electronics on chip. \textsuperscript{h}: Q1. \textsuperscript{i}: At 1 K. \textsuperscript{j}: Q2. \textsuperscript{k}: Device B. \textsuperscript{l}: Left qubit. \textsuperscript{m}: Qubit 1. \textsuperscript{n}: Qubit dressed by constant driving. \textsuperscript{o}: With feedback. \textsuperscript{p}: $S$-$T_-$ qubit. \textsuperscript{q}: At 1.5 K. \textsuperscript{r}: \estimated from Fig. S2a as the signal decay by factor $1/e$. \textsuperscript{s}: {EO} qubit. \textsuperscript{t}: The average figure of merit for the two qubits measured. \textsuperscript{u}: The average over the three qubits. \textsuperscript{v}: From {LZSM} oscillations. \textsuperscript{w}: Impurity-gated dot hybrid qubit.
\label{spinCoherence3}
}
\end{table}

%% file: figuresAndTables/table-spinCoherence-4.txt
\rowcolors{2}{white}{gray!25}
\setcounter{rowcount}{-1}
\begin{table}
\begin{tabular}{@{}S[table-format=2.2, round-mode=figures, round-pad=false, round-precision=2,  table-alignment-mode = none, table-number-alignment = center, table-column-width = 3em ]@{}lccccccc@{ \stepcounter{rowcount} \tiny \therowcount }}
\toprule
\multicolumn{2}{c}{Time} & {Coherence} & {Qubit} & {Material} & {Host} & {Date} & {Reference} & {Source}\\
\midrule
1.025 & \unit{\micro\second} & $T_2^*$ & LD/e & $\, ^{28}\text{Si/}\text{SiO}_2$ & 2D & 2024-01 & \onlinecite{ma_single-spin-qubit_2024} & p2 and Fig. 2c\\
1. & \unit{\micro\second} & $T_2^*$ & LD/e & Si/SiGe & 2D & 2018-02 & \onlinecite{watson_programmable_2018} & p1 and {ED} Fig. 3d\\
1. & \unit{\micro\second}\textsuperscript{a} & $T_2^*$ & LD/e & Si/SiGe & 2D & 2016-10 & \onlinecite{kawakami_gate_2016} & p3\\
1. & \unit{\micro\second} & $T_2^*$ & ST/e & Si/SiGe & 2D & 2021-08 & \onlinecite{liu_magnetic-gradient-free_2021} & p3\\
1. & \unit{\micro\second}\textsuperscript{a} & $T_2^*$ & LD/e & Si/SiGe & 2D & 2017-04 & \onlinecite{scarlino_dressed_2017} & p2\\
1. & \unit{\micro\second} & $T_2^*$ & LD/e & Si/SiGe & 2D & 2023-06 & \onlinecite{kobayashi_feedback-based_2023} & p7\\
0.97 & \unit{\micro\second} & $T_2^*$ & LD/e & Si/SiGe & 2D & 2020-06 & \onlinecite{boter_spatial_2020} & p3\\
0.9 & \unit{\micro\second}\textsuperscript{b} & $T_2^*$ & LD/e & Si/SiGe & 2D & 2024-11 & \onlinecite{wang_pursuing_2024} & p3\\
0.840 & \unit{\micro\second}\textsuperscript{a} & $T_2^*$ & LD/e & Si/SiGe & 2D & 2014-08 & \onlinecite{kawakami_electrical_2014} & p2\\
0.835 & \unit{\micro\second}\textsuperscript{c} & $T_2^*$ & ST/e & GaAs/AlGaAs & 2D & 2022-07 & \onlinecite{kim_approaching_2022} & p3\\
0.833 & \unit{\micro\second} & $T_2^*$ & LD/h & Ge/SiGe & 2D & 2020-01 & \onlinecite{hendrickx_fast_2020} & p2 and Fig. 2f\\
0.7667 & \unit{\micro\second}\textsuperscript{d} & $T_2^*$ & LD/e & GaAs/AlGaAs & 2D & 2020-03 & \onlinecite{nakajima_coherence_2020} & p3 and Fig. 3b\\
0.697 & \unit{\micro\second} & $T_2^*$ & ST/h & Ge/SiGe & 2D & 2021-06 & \onlinecite{jirovec_singlet-triplet_2021} & Fig. 3f 1.0 {mT} trace\\
0.639 & \unit{\micro\second}\textsuperscript{e} & $T_2^*$ & LD/e & CNT & 1D & 2019-07 & \onlinecite{cubaynes_highly_2019} & p2 and Fig. 3d\\
0.600 & \unit{\micro\second}\textsuperscript{f} & $T_2^*$ & ST/e & GaAs/AlGaAs & 2D & 2016-01 & \onlinecite{delbecq_quantum_2016} & p3\\
0.600 & \unit{\micro\second} & $T_2^*$ & ST/e\textsuperscript{g} & Ge/SiGe & 2D & 2025-01 & \onlinecite{rooney_gate_2025} & Fig. 3a the topmost point\\
0.500 & \unit{\micro\second} & $T_2^*$ & LD/e & $\text{Si/}\text{SiO}_2$ & 1D & 2023-10 & \onlinecite{klemt_electrical_2023} & p3 and Fig. 3d\\
0.472 & \unit{\micro\second}\textsuperscript{h} & $T_2^*$ & LD/h & Ge/SiGe & 2D & 2023-06 & \onlinecite{lawrie_simultaneous_2023} & Fig. 1c caption\\
0.445 & \unit{\micro\second} & $T_2^*$ & LD/h & Ge/SiGe & 2D & 2021-03 & \onlinecite{hendrickx_four-qubit_2021} & Fig. S6 dot 3\\
0.440 & \unit{\micro\second}\textsuperscript{i} & $T_2^*$ & LD/h\textsuperscript{i} & $\text{Si/}\text{SiO}_2$ & 1D & 2022-03 & \onlinecite{camenzind_spin_2022} & p4 and Fig. S9\\
0.42 & \unit{\micro\second} & $T_2^*$ & LD/e & $\, ^{28}\text{Si/}\text{SiO}_2$ & 2D & 2023-03 & \onlinecite{hu_flopping-mode_2023} & Fig. 4b\\
0.419 & \unit{\micro\second} & $T_2^*$ & ST/e & $\text{Si/}\text{SiO}_2$ & 2D & 2026-06 & \onlinecite{hamonic_single-shot_2026} & Fig. 3c\\
0.400 & \unit{\micro\second} & $T_2^*$ & ST/h & $\text{Si/}\text{SiO}_2$ & 2D & 2024-09 & \onlinecite{liles_singlet-triplet_2024} & Fig 3g the leftmost black point\\
0.400 & \unit{\micro\second} & $T_2^*$ & LD/h & Ge/SiGe & 2D & 2025-04 & \onlinecite{saez-mollejo_exchange_2025} & p6 and Fig. 4cd\\
0.360 & \unit{\micro\second} & $T_2^*$ & ST/e & Si/SiGe & 2D & 2012-01 & \onlinecite{maune_coherent_2012} & p3\\
0.330 & \unit{\micro\second} & $T_2^*$ & LD/h & Ge/SiGe & 2D & 2020-07 & \onlinecite{hendrickx_single-hole_2020} & p4 and Fig. 4b\\
0.236 & \unit{\micro\second} & $T_2^*$ & ST/e & Si/SiGe & 2D & 2014-08 & \onlinecite{wu_two-axis_2014} & Fig. 2d the blue trace\\
0.212 & \unit{\micro\second} & $T_2^*$ & LD/e & GaAs/AlGaAs & 2D & 2019-04 & \onlinecite{nakajima_quantum_2019} & Fig. 2\\
0.211 & \unit{\micro\second} & $T_2^*$ & ST/e & GaAs/AlGaAs & 2D & 2018-11 & \onlinecite{noiri_fast_2018} & p4\\
0.183 & \unit{\micro\second}\textsuperscript{j} & $T_2^*$ & ST/e & GaAs/AlGaAs & 2D & 2023-03 & \onlinecite{yun_probing_2023} & Fig. 2d\\
0.18 & \unit{\micro\second} & $T_2^*$ & LD/h & Ge/Si & 1D & 2014-05 & \onlinecite{higginbotham_hole_2014} & p3\\
0.147 & \unit{\micro\second} & $T_2^*$ & LD/e\textsuperscript{k} & $\text{Si/}\text{SiO}_2$ & 1D & 2025-02 & \onlinecite{eggli_coupling_2025} & p3 and Fig. 2c\\
0.143 & \unit{\micro\second}\textsuperscript{l} & $T_2^*$ & LD/e & $\text{Si/}\text{SiO}_2$ & 2D & 2026-01 & \onlinecite{kuno_robust_2026} & p2 and Fig. S1d\\
0.130 & \unit{\micro\second} & $T_2^*$ & LD/h & Ge/Si & 1D & 2018-09 & \onlinecite{watzinger_germanium_2018} & p4 and Fig. 6d\\
0.118 & \unit{\micro\second}\textsuperscript{m} & $T_2^*$ & LD/e & GaAs/AlGaAs & 2D & 2021-08 & \onlinecite{mortemousque_enhanced_2021} & Fig. 5f the topmost point\\
0.116 & \unit{\micro\second} & $T_2^*$ & LD/i & Si:P & imp & 2023-11 & \onlinecite{kranz_exploiting_2023} & Fig. 4g and Tab. I\\
0.110 & \unit{\micro\second}\textsuperscript{a,n} & $T_2^*$ & LD/e & Si/SiGe & 2D & 2017-04 & \onlinecite{scarlino_dressed_2017} & Fig. 3a\\
100. & \unit{\nano\second}\textsuperscript{o} & $T_2^*$ & ST/e & $\, ^{28}\text{Si/}\text{SiO}_2$ & 2D & 2022-02 & \onlinecite{jock_silicon_2022} & p4 and Fig. 2e-g\\
0.100 & \unit{\micro\second}\textsuperscript{d} & $T_2^*$ & ST/e & GaAs/AlGaAs & 2D & 2017-01 & \onlinecite{nichol_high-fidelity_2017} & p2\\
95. & \unit{\nano\second}\textsuperscript{p} & $T_2^*$ & ST/h & Ge/SiGe & 2D & 2024-10 & \onlinecite{zhang_universal_2024} & p5\\
94. & \unit{\nano\second}\textsuperscript{d} & $T_2^*$ & ST/e & GaAs/AlGaAs & 2D & 2010-11 & \onlinecite{bluhm_enhancing_2010} & Fig. 3b\\
90. & \unit{\nano\second} & $T_2^*$ & ST/e & GaAs/AlGaAs & 2D & 2013-04 & \onlinecite{dial_charge_2013} & p2\\
88. & \unit{\nano\second}\textsuperscript{q} & $T_2^*$ & LD/e & Si/SiGe & 2D & 2018-01 & \onlinecite{samkharadze_strong_2018} & p1\\
82. & \unit{\nano\second} & $T_2^*$ & LD/h & Ge/Si & 1D & 2022-01 & \onlinecite{wang_ultrafast_2022} & p3 and Fig. 3d\\
80. & \unit{\nano\second}\textsuperscript{r} & $T_2^*$ & ST/e & GaAs/AlGaAs & 2D & 2017-09 & \onlinecite{flentje_coherent_2017} & p3\\
80. & \unit{\nano\second}\textsuperscript{s} & $T_2^*$ & ST/e & GaAs/AlGaAs & 2D & 2015-08 & \onlinecite{bertrand_quantum_2015} & Fig. 4d\\
70. & \unit{\nano\second}\textsuperscript{t} & $T_2^*$ & LD/e & $\, ^{28}\text{Si/SiGe}$ & 2D & 2024-12 & \onlinecite{dijkema_cavity-mediated_2024} & Tab. I\\
66. & \unit{\nano\second}\textsuperscript{u} & $T_2^*$ & LD/e & Si/SiGe & 2D & 2018-02 & \onlinecite{mi_coherent_2018} & p3 and Fig. 3b\\
64. & \unit{\nano\second}\textsuperscript{v} & $T_2^*$ & LD/e & CNT & 1D & 2015-07 & \onlinecite{viennot_coherent_2015} & p4\\
61. & \unit{\nano\second}\textsuperscript{a} & $T_2^*$ & LD/e & GaAs/AlGaAs & 2D & 2014-12 & \onlinecite{yoneda_fast_2014} & p3\\
\bottomrule
\end{tabular}
\caption{Spin coherence times (part 4). Superscripts stand for the following. \textsuperscript{a}: From {EDSR} linewidth. \textsuperscript{b}: Q1. \textsuperscript{c}: With estimation and/or feedback. \textsuperscript{d}: With feedback. \textsuperscript{e}: From $f_\mathrm{FWHM}= 498$ {kHz}. \textsuperscript{f}: In non-ergodic regime. \textsuperscript{g}: $S$-$T_-$ qubit. \textsuperscript{h}: We take the value for qubit 3 at 0.65 T. \textsuperscript{i}: At 1.5 K. \textsuperscript{j}: $Q_R$. \textsuperscript{k}: At 1 K. \textsuperscript{l}: Bare qubit. \textsuperscript{m}: The value of the point, 118 ns, is from T. Meunier, private communication. \textsuperscript{n}: Valley and spin flip. \textsuperscript{o}: \derived from $T_2^* = Q/f$ with $f=200$ {MHz} and $Q=20$. \textsuperscript{p}: Average of the given range 42-147 ns. \textsuperscript{q}: From $\Gamma_2/2\pi = 1.8$ {MHz}. \textsuperscript{r}: Dephasing of transported spin. \textsuperscript{s}: \estimated from Fig. 4d. \textsuperscript{t}: Qubit 1 representative value. \textsuperscript{u}: From $\Gamma_2/2\pi = 2.4$ {MHz}. \textsuperscript{v}: From $\Gamma_2/2\pi = 2.5$ {MHz}.
\label{spinCoherence4}
}
\end{table}

%% file: figuresAndTables/table-spinCoherence-5.txt
\rowcolors{2}{white}{gray!25}
\setcounter{rowcount}{-1}
\begin{table}
\begin{tabular}{@{}S[table-format=2.2, round-mode=figures, round-pad=false, round-precision=2,  table-alignment-mode = none, table-number-alignment = center, table-column-width = 3em ]@{}lccccccc@{ \stepcounter{rowcount} \tiny \therowcount }}
\toprule
\multicolumn{2}{c}{Time} & {Coherence} & {Qubit} & {Material} & {Host} & {Date} & {Reference} & {Source}\\
\midrule
60. & \unit{\nano\second} & $T_2^*$ & ST/e & GaAs/AlGaAs & 2D & 2024-02 & \onlinecite{berritta_real-time_2024} & p5\\
59. & \unit{\nano\second} & $T_2^*$ & LD/h & $\text{Si/}\text{SiO}_2$ & 1D & 2016-11 & \onlinecite{maurand_cmos_2016} & p4 and Fig. 3a\\
55. & \unit{\nano\second} & $T_2^*$ & LD/i & Si:P & imp & 2012-09 & \onlinecite{pla_single-atom_2012} & p2\\
45.3 & \unit{\nano\second} & $T_2^*$ & HY/e\textsuperscript{a} & Si/SiGe & 2D & 2017-08 & \onlinecite{thorgrimsson_extending_2017} & Fig. 2i the blue trace\\
40. & \unit{\nano\second}\textsuperscript{b} & $T_2^*$ & ST/e & GaAs/AlGaAs & 2D & 2010-12 & \onlinecite{bluhm_dephasing_2010} & {SM} {SFig}. 1\\
34. & \unit{\nano\second}\textsuperscript{c} & $T_2^*$ & LD/e & Si/SiGe & 2D & 2019-12 & \onlinecite{borjans_resonant_2019} & p1\\
30. & \unit{\nano\second} & $T_2^*$ & LD/e & GaAs/AlGaAs & 2D & 2008-06 & \onlinecite{koppens_spin_2008} & p2\\
28.4 & \unit{\nano\second} & $T_2^*$ & LD/e & GaAs/AlGaAs & 2D & 2020-03 & \onlinecite{nakajima_coherence_2020} & p2 and Fig. 2c\\
27.2 & \unit{\nano\second}\textsuperscript{d} & $T_2^*$ & LD/e & $\, ^{28}\text{Si/SiGe}$ & 2D & 2022-05 & \onlinecite{harvey-collard_coherent_2022} & p4\\
27. & \unit{\nano\second} & $T_2^*$ & ST/e & GaAs/AlGaAs & 2D & 2009-10 & \onlinecite{barthel_rapid_2009} & p4\\
30. & \unit{\nano\second}\textsuperscript{e} & $T_2^*$ & LD/e & $\, ^{28}\text{Si/SiGe}$ & 2D & 2023-03 & \onlinecite{bonsen_probing_2023} & p5\\
25. & \unit{\nano\second} & $T_2^*$ & ST/e & GaAs/AlGaAs & 2D & 2005-08 & \onlinecite{koppens_control_2005} & p5\\
25. & \unit{\nano\second} & $T_2^*$ & HY/e\textsuperscript{f} & GaAs/AlGaAs & 2D & 2013-09 & \onlinecite{medford_self-consistent_2013} & p3\\
20. & \unit{\nano\second}\textsuperscript{g} & $T_2^*$ & LD/h & $\text{Si/}\text{SiO}_2$ & 1D & 2023-03 & \onlinecite{yu_strong_2023} & p3\\
22.1 & \unit{\nano\second} & $T_2^*$ & ST/e & GaAs/AlGaAs & 2D & 2021-10 & \onlinecite{fedele_simultaneous_2021} & Fig. 2c Qubit 2\\
22. & \unit{\nano\second} & $T_2^*$ & ST/e & GaAs/AlGaAs & 2D & 2020-12 & \onlinecite{jang_individual_2020} & Fig. 2a the green trace\\
21.5 & \unit{\nano\second} & $T_2^*$ & LD/h & Ge/Si & 1D & 2023-05 & \onlinecite{liu_ultrafast_2023} & p6\\
21. & \unit{\nano\second} & $T_2^*$ & ST/e & GaAs/AlGaAs & 2D & 2021-05 & \onlinecite{kojima_probabilistic_2021} & p5\\
20. & \unit{\nano\second} & $T_2^*$ & LD/e & GaAs/AlGaAs & 2D & 2007-09 & \onlinecite{koppens_universal_2007} & p1\\
15.8 & \unit{\nano\second} & $T_2^*$ & ST/e & GaAs/AlGaAs & 2D & 2010-12 & \onlinecite{bluhm_dephasing_2010} & {SM} {SFig}. 1\\
15. & \unit{\nano\second} & $T_2^*$ & ST/e & GaAs/AlGaAs & 2D & 2020-07 & \onlinecite{jang_robust_2020} & p4\\
15. & \unit{\nano\second} & $T_2^*$ & ST/e & GaAs/AlGaAs & 2D & 2010-06 & \onlinecite{reilly_exchange_2010} & p2 and Fig. 1\\
15. & \unit{\nano\second} & $T_2^*$ & ST/e & GaAs/AlGaAs & 2D & 2008-12 & \onlinecite{reilly_measurement_2008} & p2 and Fig. 1e\\
15. & \unit{\nano\second}\textsuperscript{h} & $T_2^*$ & HY/e\textsuperscript{f} & GaAs/AlGaAs & 2D & 2011-11 & \onlinecite{gaudreau_coherent_2011} & p2 and Fig. 3d\\
14. & \unit{\nano\second}\textsuperscript{i} & $T_2^*$ & HY/e\textsuperscript{j} & GaAs/AlGaAs & 2D & 2019-11 & \onlinecite{landig_virtual-photon-mediated_2019} & p3\\
14. & \unit{\nano\second} & $T_2^*$ & ST/e & GaAs/AlGaAs & 2D & 2017-06 & \onlinecite{fujita_coherent_2017} & p2\\
14. & \unit{\nano\second}\textsuperscript{k} & $T_2^*$ & ST/e & GaAs/AlGaAs & 2D & 2010-11 & \onlinecite{bluhm_enhancing_2010} & p2 and Fig. 3a\\
14. & \unit{\nano\second} & $T_2^*$ & ST/e & GaAs/AlGaAs & 2D & 2006-07 & \onlinecite{laird_effect_2006} & p1\\
12.2 & \unit{\nano\second} & $T_2^*$ & ST/e & GaAs/AlGaAs & 2D & 2021-02 & \onlinecite{jadot_distant_2021} & p1 and Fig. 2b\\
11. & \unit{\nano\second} & $T_2^*$ & HY/e & Si/SiGe & 2D & 2015-12 & \onlinecite{kim_high-fidelity_2015} & Fig. 2c\\
11. & \unit{\nano\second} & $T_2^*$ & LD/h & Ge/Si & 1D & 2021-01 & \onlinecite{froning_ultrafast_2021} & Fig. 2e\\
10. & \unit{\nano\second} & $T_2^*$ & ST/e & GaAs/AlGaAs & 2D & 2005-09 & \onlinecite{petta_coherent_2005} & p3 and Fig. 3b\\
10. & \unit{\nano\second} & $T_2^*$ & HY/e\textsuperscript{l} & Si/SiGe & 2D & 2014-07 & \onlinecite{kim_quantum_2014} & p1\\
9. & \unit{\nano\second} & $T_2^*$ & ST/e & GaAs/AlGaAs & 2D & 2005-06 & \onlinecite{johnson_tripletsinglet_2005} & p3\\
8.12 & \unit{\nano\second}\textsuperscript{m} & $T_2^*$ & HY/e\textsuperscript{j} & GaAs/AlGaAs & 2D & 2018-07 & \onlinecite{landig_coherent_2018} & p2\\
8.1 & \unit{\nano\second} & $T_2^*$ & HY/e & GaAs/AlGaAs & 2D & 2016-02 & \onlinecite{cao_tunable_2016} & p4 and Fig. 4d\\
8. & \unit{\nano\second} & $T_2^*$ & LD/e & InAs & 1D & 2010-12 & \onlinecite{nadj-perge_spinorbit_2010} & p3\\
8. & \unit{\nano\second} & $T_2^*$ & LD/e & CNT & 1D & 2013-07 & \onlinecite{laird_valleyspin_2013} & p4 and Fig. 4\\
8. & \unit{\nano\second} & $T_2^*$ & LD/e & InSb & 1D & 2013-02 & \onlinecite{van_den_berg_fast_2013} & p3\\
8. & \unit{\nano\second}\textsuperscript{n} & $T_2^*$ & LD/h & Ge/Si & 1D & 2020-06 & \onlinecite{xu_hole_2020} & p4 and Fig. 4b\\
7.6 & \unit{\nano\second}\textsuperscript{o} & $T_2^*$ & LD/e & InAs & 1D & 2024-05 & \onlinecite{pita-vidal_strong_2024} & p3\\
7. & \unit{\nano\second} & $T_2^*$ & HY/e & GaAs/AlGaAs & 2D & 2021-06 & \onlinecite{jang_single-shot_2021} & p4 and Fig. 3c\\
5.6 & \unit{\nano\second} & $T_2^*$ & ST/e & GaAs/AlGaAs & 2D & 2010-07 & \onlinecite{petersson_charge_2010} & p4\\
5. & \unit{\nano\second}\textsuperscript{p} & $T_2^*$ & LD/h & Ge/Si & 1D & 2025-08 & \onlinecite{carballido_compromise-free_2025} & Fig. S5c\\
3.7 & \unit{\nano\second} & $T_2^*$ & HY/e & Si/SiGe & 2D & 2014-01 & \onlinecite{shi_fast_2014} & p2\\
3.2 & \unit{\nano\second}\textsuperscript{q} & $T_2^*$ & ST/e & CNT & 1D & 2009-04 & \onlinecite{churchill_relaxation_2009} & p3\\
1.37 & \unit{\nano\second} & $T_2^*$ & LD/e & InAs & 1D & 2024-02 & \onlinecite{ungerer_strong_2024} & p3\\
0.446 & \unit{\nano\second}\textsuperscript{r} & $T_2^*$ & ST/e & GaAs/AlGaAs & 2D & 2019-05 & \onlinecite{landig_microwave-cavity-detected_2019} & p2\\
0.120 & \unit{\nano\second}\textsuperscript{s} & $T_2^*$ & LD/h & GaAs/AlGaAs & 2D & 2018-05 & \onlinecite{bogan_landau-zener-stuckelberg-majorana_2018} & p4\\
9.2 & \unit{\milli\second} & $T_2^{\text{Echo}}$ & LD/i\textsuperscript{t} & $\, ^{28}\text{Si:P}$ & imp & 2016-10 & \onlinecite{laucht_dressed_2016} & p4\\
\bottomrule
\end{tabular}
\caption{Spin coherence times (part 5). Superscripts stand for the following. \textsuperscript{a}: Tunable between spin- and charge- qubit. \textsuperscript{b}: With feedback. \textsuperscript{c}: From $\Gamma_2/2\pi = 4.7$ {MHz}. \textsuperscript{d}: From $f_\mathrm{FWHM} = 11.7$ {MHz}. \textsuperscript{e}: From $\Gamma_2/2\pi = 6$ {MHz}. \textsuperscript{f}: {EO} qubit. \textsuperscript{g}: From $\Gamma_2/2\pi=7$ {MHz}. \textsuperscript{h}: From interference of Landau-Zener transitions. \textsuperscript{i}: From $\Gamma_2/2\pi = 11$ {MHz}. \textsuperscript{j}: {RX} qubit. \textsuperscript{k}: Without feedback. \textsuperscript{l}: Hybrid qubit with three electrons in two dots. \textsuperscript{m}: From $\Gamma_2/2\pi = 19.6$ {MHz}. \textsuperscript{n}: From {EDSR} linewidth. \textsuperscript{o}: Q1. \textsuperscript{p}: Representative value. \textsuperscript{q}: Carbon nanotube made from $^{13}$C isotope. \textsuperscript{r}: From $\Gamma_2/2\pi = 357$ {MHz}. \textsuperscript{s}: From Landau-Zener interference. \textsuperscript{t}: Qubit defined in the rotating frame.
\label{spinCoherence5}
}
\end{table}

%% file: figuresAndTables/table-spinCoherence-6.txt
\rowcolors{2}{white}{gray!25}
\setcounter{rowcount}{-1}
\begin{table}
\begin{tabular}{@{}S[table-format=2.2, round-mode=figures, round-pad=false, round-precision=2,  table-alignment-mode = none, table-number-alignment = center, table-column-width = 3em ]@{}lccccccc@{ \stepcounter{rowcount} \tiny \therowcount }}
\toprule
\multicolumn{2}{c}{Time} & {Coherence} & {Qubit} & {Material} & {Host} & {Date} & {Reference} & {Source}\\
\midrule
1.73 & \unit{\milli\second} & $T_2^{\text{Echo}}$ & LD/i & $\, ^{28}\text{Si:P}$ & imp & 2020-07 & \onlinecite{madzik_controllable_2020} & Fig. 3\\
1.2 & \unit{\milli\second} & $T_2^{\text{Echo}}$ & LD/e & $\, ^{28}\text{Si/}\text{SiO}_2$ & 2D & 2014-10 & \onlinecite{veldhorst_addressable_2014} & p2 and Fig. 3b\\
1. & \unit{\milli\second} & $T_2^{\text{Echo}}$ & LD/i & $\, ^{28}\text{Si:P}$ & imp & 2014-10 & \onlinecite{muhonen_storing_2014} & p2 and Fig. 2d\\
1. & \unit{\milli\second} & $T_2^{\text{Echo}}$ & LD/e & $\, ^{28}\text{Si/SiGe}$ & 2D & 2021-03 & \onlinecite{kerckhoff_magnetic_2021} & p6 and Fig. 5\\
0.97 & \unit{\milli\second} & $T_2^{\text{Echo}}$ & LD/i & Si:P & imp & 2015-04 & \onlinecite{laucht_electrically_2015} & p4 and Fig. 3h\\
0.92 & \unit{\milli\second} & $T_2^{\text{Echo}}$ & LD/i & $\, ^{28}\text{Si:B}$ & imp & 2020-07 & \onlinecite{kobayashi_engineering_2020} & p3 and Fig. 3a\\
0.850 & \unit{\milli\second} & $T_2^{\text{Echo}}$ & LD/i & $\, ^{28}\text{Si:P}$ & imp & 2022-01 & \onlinecite{madzik_precision_2022} & {ED} Fig. 3 first column\\
0.401 & \unit{\milli\second} & $T_2^{\text{Echo}}$ & LD/e & $\, ^{28}\text{Si/}\text{SiO}_2$ & 2D & 2018-10 & \onlinecite{chan_assessment_2018} & p2\\
0.343 & \unit{\milli\second} & $T_2^{\text{Echo}}$ & LD/i & Si:P & imp & 2023-11 & \onlinecite{kranz_exploiting_2023} & Fig. 4e and Tab. I\\
0.336 & \unit{\milli\second} & $T_2^{\text{Echo}}$ & LD/i & $\, ^{28}\text{Si:P}$ & imp & 2023-02 & \onlinecite{savytskyy_electrically_2023} & Fig. 3c\\
0.310 & \unit{\milli\second} & $T_2^{\text{Echo}}$ & LD/i & $\, ^{28}\text{Si:P}$ & imp & 2016-02 & \onlinecite{tracy_single_2016} & Fig. 4d\\
0.290 & \unit{\milli\second} & $T_2^{\text{Echo}}$ & LD/e & $\, ^{28}\text{Si/}\text{SiO}_2$ & 2D & 2019-05 & \onlinecite{huang_fidelity_2019} & p3 and {ExtData} Fig. 1b\\
0.264 & \unit{\milli\second} & $T_2^{\text{Echo}}$ & LD/e & $\, ^{28}\text{Si/}\text{SiO}_2$ & 2D & 2024-01 & \onlinecite{ma_single-spin-qubit_2024} & p2 and Fig. 2d\\
0.206 & \unit{\milli\second} & $T_2^{\text{Echo}}$ & LD/i & Si:P & imp & 2012-09 & \onlinecite{pla_single-atom_2012} & p3\\
0.201 & \unit{\milli\second} & $T_2^{\text{Echo}}$ & LD/i & $\, ^{28}\text{Si:P}$ & imp & 2024-09 & \onlinecite{stemp_tomography_2024} & Fig. 2b\\
0.2 & \unit{\milli\second}\textsuperscript{a} & $T_2^{\text{Echo}}$ & ST/e & $\, ^{28}\text{Si/}\text{SiO}_2$ & 2D & 2020-04 & \onlinecite{yang_operation_2020} & Fig. 4 the leftmost blue point\\
0.139 & \unit{\milli\second} & $T_2^{\text{Echo}}$ & LD/e & $\, ^{28}\text{Si/SiGe}$ & 2D & 2022-12 & \onlinecite{mills_high-fidelity_2022} & p5\\
0.138 & \unit{\milli\second} & $T_2^{\text{Echo}}$ & ST/e & $\, ^{28}\text{Si/SiGe}$ & 2D & 2024-02 & \onlinecite{takeda_rapid_2024} & Fig. 4b\\
0.128 & \unit{\milli\second} & $T_2^{\text{Echo}}$ & LD/e & $\, ^{28}\text{Si/SiGe}$ & 2D & 2020-05 & \onlinecite{struck_low-frequency_2020} & Fig. 2d\\
0.1157 & \unit{\milli\second} & $T_2^{\text{Echo}}$ & LD/i & $\, ^{28}\text{Si:P}$ & imp & 2021-01 & \onlinecite{madzik_conditional_2021} & {SFig}. 3b\\
0.109 & \unit{\milli\second} & $T_2^{\text{Echo}}$ & LD/e & $\, ^{28}\text{Si/SiGe}$ & 2D & 2019-06 & \onlinecite{sigillito_site-selective_2019} & Tab. 1\\
99. & \unit{\micro\second} & $T_2^{\text{Echo}}$ & LD/e & $\, ^{28}\text{Si/SiGe}$ & 2D & 2017-12 & \onlinecite{yoneda_quantum-dot_2017} & p3 and Fig. 3c\\
88. & \unit{\micro\second} & $T_2^{\text{Echo}}$ & LD/h & $\text{Si/}\text{SiO}_2$ & 1D & 2022-09 & \onlinecite{piot_single_2022} & p3 and Fig. 3b\\
72.5 & \unit{\micro\second}\textsuperscript{b} & $T_2^{\text{Echo}}$ & LD/e & $\, ^{28}\text{Si/SiGe}$ & 2D & 2022-04 & \onlinecite{mills_two-qubit_2022} & p1\\
70. & \unit{\micro\second} & $T_2^{\text{Echo}}$ & ST/e & $\, ^{28}\text{Si/SiGe}$ & 2D & 2015-05 & \onlinecite{eng_isotopically_2015} & Fig. S2 the topmost red point\\
70. & \unit{\micro\second} & $T_2^{\text{Echo}}$ & LD/h & Ge/SiGe & 2D & 2024-05 & \onlinecite{hendrickx_sweet-spot_2024} & Fig. {ED}3\\
56.3 & \unit{\micro\second} & $T_2^{\text{Echo}}$ & ST/h\textsuperscript{c} & Ge/SiGe & 2D & 2026-01 & \onlinecite{tsoukalas_dressed_2026} & Fig. 3c\\
45.8 & \unit{\micro\second} & $T_2^{\text{Echo}}$ & LD/e & Si/SiGe & 2D & 2021-06 & \onlinecite{takeda_quantum_2021} & p1 for Q3\\
43. & \unit{\micro\second}\textsuperscript{d} & $T_2^{\text{Echo}}$ & LD/e & Si/SiGe & 2D & 2022-08 & \onlinecite{takeda_quantum_2022} & p2 and {ED} Fig. 4j-l\\
43. & \unit{\micro\second}\textsuperscript{e} & $T_2^{\text{Echo}}$ & LD/e & $\, ^{28}\text{Si/}\text{SiO}_2$ & 2D & 2024-08 & \onlinecite{tanttu_assessment_2024} & {ED} Tab. {II}\\
40.1 & \unit{\micro\second}\textsuperscript{f} & $T_2^{\text{Echo}}$ & LD/e & $\, ^{28}\text{Si/SiGe}$ & 2D & 2022-09 & \onlinecite{philips_universal_2022} & Fig. 2e\\
40. & \unit{\micro\second}\textsuperscript{g} & $T_2^{\text{Echo}}$ & LD/e & $\, ^{28}\text{Si/}\text{SiO}_2$ & 2D & 2023-01 & \onlinecite{gilbert_-demand_2023} & Fig. 3d, green point\\
39.8 & \unit{\micro\second}\textsuperscript{h} & $T_2^{\text{Echo}}$ & LD/e & $\, ^{28}\text{Si/SiGe}$ & 2D & 2026-05 & \onlinecite{matsumoto_two-qubit_2026} & Fig. S1d\\
37. & \unit{\micro\second} & $T_2^{\text{Echo}}$ & LD/e & Si/SiGe & 2D & 2014-08 & \onlinecite{kawakami_electrical_2014} & p4 and Fig. 4a\\
36. & \unit{\micro\second} & $T_2^{\text{Echo}}$ & LD/e & $\text{Si/}\text{SiO}_2$ & 1D & 2023-10 & \onlinecite{klemt_electrical_2023} & p4\\
32. & \unit{\micro\second} & $T_2^{\text{Echo}}$ & LD/e & $\, ^{28}\text{Si/SiGe}$ & 2D & 2022-01 & \onlinecite{noiri_fast_2022} & {ED} Fig. 2i the topmost trace\\
32. & \unit{\micro\second}\textsuperscript{i} & $T_2^{\text{Echo}}$ & LD/h & Ge/SiGe & 2D & 2024-07 & \onlinecite{wang_operating_2024} & Fig. 1f\\
30.21 & \unit{\micro\second}\textsuperscript{j} & $T_2^{\text{Echo}}$ & LD/e & $\, ^{28}\text{Si/SiGe}$ & 2D & 2025-07 & \onlinecite{unseld_baseband_2025} & p2\\
30. & \unit{\micro\second} & $T_2^{\text{Echo}}$ & ST/e & GaAs/AlGaAs & 2D & 2010-12 & \onlinecite{bluhm_dephasing_2010} & abstract and {SM} p9\\
22. & \unit{\micro\second} & $T_2^{\text{Echo}}$ & LD/e\textsuperscript{k} & $\text{Si/}\text{SiO}_2$ & 2D & 2020-02 & \onlinecite{leon_coherent_2020} & {SM} Tab. I\\
19. & \unit{\micro\second} & $T_2^{\text{Echo}}$ & LD/e & Si/SiGe & 2D & 2018-02 & \onlinecite{watson_programmable_2018} & p1 and {ED} Fig. 3e\\
17.7 & \unit{\micro\second}\textsuperscript{j} & $T_2^{\text{Echo}}$ & LD/e & Si/SiGe & 2D & 2024-11 & \onlinecite{wang_pursuing_2024} & p3\\
17.5 & \unit{\micro\second}\textsuperscript{l} & $T_2^{\text{Echo}}$ & LD/e & $\, ^{28}\text{Si/SiGe}$ & 2D & 2022-09 & \onlinecite{noiri_shuttling-based_2022} & p3 and {ED} Fig. 2c\\
11. & \unit{\micro\second}\textsuperscript{m} & $T_2^{\text{Echo}}$ & LD/e & Si/SiGe & 2D & 2016-08 & \onlinecite{takeda_fault-tolerant_2016} & p3 and Fig. 2c\\
9.4 & \unit{\micro\second}\textsuperscript{n} & $T_2^{\text{Echo}}$ & LD/e & $\, ^{28}\text{Si/}\text{SiO}_2$ & 2D & 2022-11 & \onlinecite{vahapoglu_coherent_2022} & p3 and Fig. 3a\\
9. & \unit{\micro\second} & $T_2^{\text{Echo}}$ & ST/e & GaAs/AlGaAs & 2D & 2013-04 & \onlinecite{dial_charge_2013} & p3\\
8.5 & \unit{\micro\second} & $T_2^{\text{Echo}}$ & ST/e & GaAs/AlGaAs & 2D & 2017-04 & \onlinecite{malinowski_spectrum_2017} & Fig. 4c the point n=1\\
8.4 & \unit{\micro\second} & $T_2^{\text{Echo}}$ & ST/e & $\, ^{28}\text{Si/}\text{SiO}_2$ & 2D & 2018-05 & \onlinecite{jock_silicon_2018} & p6 and Fig. 4f\\
7. & \unit{\micro\second}\textsuperscript{o} & $T_2^{\text{Echo}}$ & ST/e & $\, ^{28}\text{Si/}\text{SiO}_2$ & 2D & 2020-04 & \onlinecite{yang_operation_2020} & Fig. 4 the rightmost blue point\\
7. & \unit{\micro\second}\textsuperscript{p} & $T_2^{\text{Echo}}$ & ST/e & GaAs/AlGaAs & 2D & 2012-02 & \onlinecite{medford_scaling_2012} & Fig. 4\\
\bottomrule
\end{tabular}
\caption{Spin coherence times (part 6). Superscripts stand for the following. \textsuperscript{a}: At 0.04 K. \textsuperscript{b}: The average figure of merit for the two qubits measured. \textsuperscript{c}: Qubit defined in the rotating frame. \textsuperscript{d}: The average over the three qubits. \textsuperscript{e}: Device B. \textsuperscript{f}: Q3. \textsuperscript{g}: \estimated from Fig. 3d at $\Delta V_J=10$ {mV}. \textsuperscript{h}: Q5. \textsuperscript{i}: $Q_A$. \textsuperscript{j}: Q1. \textsuperscript{k}: In a five-electron dot configuration. \textsuperscript{l}: Left qubit. \textsuperscript{m}: Measured in Ref.~\cite{takeda_optimized_2018}. \textsuperscript{n}: Qubit 1. \textsuperscript{o}: At 1.5 K. \textsuperscript{p}: \estimated Fig. 4 the leftmost point.
\label{spinCoherence6}
}
\end{table}

%% file: figuresAndTables/table-spinCoherence-7.txt
\rowcolors{2}{white}{gray!25}
\setcounter{rowcount}{-1}
\begin{table}
\begin{tabular}{@{}S[table-format=2.2, round-mode=figures, round-pad=false, round-precision=2,  table-alignment-mode = none, table-number-alignment = center, table-column-width = 3em ]@{}lccccccc@{ \stepcounter{rowcount} \tiny \therowcount }}
\toprule
\multicolumn{2}{c}{Time} & {Coherence} & {Qubit} & {Material} & {Host} & {Date} & {Reference} & {Source}\\
\midrule
7.000 & \unit{\micro\second} & $T_2^{\text{Echo}}$ & ST/e & GaAs/AlGaAs & 2D & 2017-01 & \onlinecite{nichol_high-fidelity_2017} & Fig. 2b\\
6.3 & \unit{\micro\second} & $T_2^{\text{Echo}}$ & LD/e\textsuperscript{a} & $\, ^{28}\text{Si/}\text{SiO}_2$ & 2D & 2024-09 & \onlinecite{hansen_entangling_2024} & Fig. 3b and p5\\
6.1 & \unit{\micro\second} & $T_2^{\text{Echo}}$ & ST/e & GaAs/AlGaAs & 2D & 2020-08 & \onlinecite{cerfontaine_closed-loop_2020} & p4\\
6. & \unit{\micro\second} & $T_2^{\text{Echo}}$ & ST/e & GaAs/AlGaAs & 2D & 2010-12 & \onlinecite{barthel_interlaced_2010} & p3\\
5. & \unit{\micro\second}\textsuperscript{b} & $T_2^{\text{Echo}}$ & ST/e & $\, ^{28}\text{Si/}\text{SiO}_2$ & 2D & 2022-02 & \onlinecite{jock_silicon_2022} & Fig. 4c\\
4.2 & \unit{\micro\second} & $T_2^{\text{Echo}}$ & ST/h & Ge/SiGe & 2D & 2026-01 & \onlinecite{tsoukalas_dressed_2026} & p3 and Fig. 2e\\
3.8 & \unit{\micro\second} & $T_2^{\text{Echo}}$ & LD/h & Ge/SiGe & 2D & 2021-03 & \onlinecite{hendrickx_four-qubit_2021} & Fig. S6 dot 3\\
2. & \unit{\micro\second} & $T_2^{\text{Echo}}$ & LD/e & Si/SiGe & 2D & 2020-06 & \onlinecite{boter_spatial_2020} & Fig. 4\\
2. & \unit{\micro\second}\textsuperscript{c} & $T_2^{\text{Echo}}$ & HY/e\textsuperscript{d} & GaAs/AlGaAs & 2D & 2013-07 & \onlinecite{medford_quantum-dot-based_2013} & Fig. 5\\
1.9 & \unit{\micro\second} & $T_2^{\text{Echo}}$ & LD/h & Ge/SiGe & 2D & 2020-01 & \onlinecite{hendrickx_fast_2020} & p2 and Fig. 2f\\
1.8 & \unit{\micro\second}\textsuperscript{e} & $T_2^{\text{Echo}}$ & ST/h & Ge/SiGe & 2D & 2021-06 & \onlinecite{jirovec_singlet-triplet_2021} & Fig. 5 of {arXiv}-v3\\
1.5 & \unit{\micro\second}\textsuperscript{f,g} & $T_2^{\text{Echo}}$ & LD/h\textsuperscript{g} & $\text{Si/}\text{SiO}_2$ & 1D & 2022-03 & \onlinecite{camenzind_spin_2022} & Fig. 4a\\
1.220 & \unit{\micro\second} & $T_2^{\text{Echo}}$ & ST/h & $\text{Si/}\text{SiO}_2$ & 2D & 2024-09 & \onlinecite{liles_singlet-triplet_2024} & p7\\
1.2 & \unit{\micro\second} & $T_2^{\text{Echo}}$ & ST/e & GaAs/AlGaAs & 2D & 2005-09 & \onlinecite{petta_coherent_2005} & p4 and Fig. 5c\\
0.523 & \unit{\micro\second} & $T_2^{\text{Echo}}$ & LD/h & Ge/Si & 1D & 2022-01 & \onlinecite{wang_ultrafast_2022} & p3 and Fig. 3f\\
0.44 & \unit{\micro\second} & $T_2^{\text{Echo}}$ & LD/e & GaAs/AlGaAs & 2D & 2008-06 & \onlinecite{koppens_spin_2008} & p4 and Fig. 4b\\
0.250 & \unit{\micro\second} & $T_2^{\text{Echo}}$ & LD/h & Ge/Si & 1D & 2021-01 & \onlinecite{froning_ultrafast_2021} & Fig. 2e\\
0.245 & \unit{\micro\second} & $T_2^{\text{Echo}}$ & LD/h & $\text{Si/}\text{SiO}_2$ & 1D & 2016-11 & \onlinecite{maurand_cmos_2016} & p5 and Fig. 4b\\
0.200 & \unit{\micro\second} & $T_2^{\text{Echo}}$ & LD/h & Ge/Si & 1D & 2025-08 & \onlinecite{carballido_compromise-free_2025} & Fig. 3a at $V_L=1330$ {mV}\\
0.183 & \unit{\micro\second} & $T_2^{\text{Echo}}$ & ST/e & GaAs/AlGaAs & 2D & 2016-06 & \onlinecite{cerfontaine_feedback-tuned_2016} & p4\\
0.150 & \unit{\micro\second}\textsuperscript{h} & $T_2^{\text{Echo}}$ & LD/e & $\, ^{28}\text{Si/SiGe}$ & 2D & 2024-12 & \onlinecite{dijkema_cavity-mediated_2024} & Tab. I\\
0.100 & \unit{\micro\second} & $T_2^{\text{Echo}}$ & HY/e\textsuperscript{i} & GaAs/AlGaAs & 2D & 2013-09 & \onlinecite{medford_self-consistent_2013} & p3\\
65. & \unit{\nano\second} & $T_2^{\text{Echo}}$ & LD/e & CNT & 1D & 2013-07 & \onlinecite{laird_valleyspin_2013} & p4 and Fig. 4\\
50. & \unit{\nano\second} & $T_2^{\text{Echo}}$ & LD/e & InAs & 1D & 2010-12 & \onlinecite{nadj-perge_spinorbit_2010} & p3\\
34. & \unit{\nano\second} & $T_2^{\text{Echo}}$ & LD/e & InSb & 1D & 2013-02 & \onlinecite{van_den_berg_fast_2013} & p3\\
17.3 & \unit{\nano\second}\textsuperscript{j} & $T_2^{\text{Echo}}$ & LD/e & InAs & 1D & 2024-05 & \onlinecite{pita-vidal_strong_2024} & p3\\
0.56 & \unit{\second} & $T_2^{\text{DynD}}$ & LD/i & $\, ^{28}\text{Si:P}$ & imp & 2014-10 & \onlinecite{muhonen_storing_2014} & p2 and Fig. 2e\\
28. & \unit{\milli\second} & $T_2^{\text{DynD}}$ & LD/e & $\, ^{28}\text{Si/}\text{SiO}_2$ & 2D & 2014-10 & \onlinecite{veldhorst_addressable_2014} & p2 and Fig. 3c\\
28. & \unit{\milli\second} & $T_2^{\text{DynD}}$ & LD/e & $\, ^{28}\text{Si/}\text{SiO}_2$ & 2D & 2015-10 & \onlinecite{veldhorst_two-qubit_2015} & p1\\
23. & \unit{\milli\second} & $T_2^{\text{DynD}}$ & LD/i\textsuperscript{k} & $\, ^{28}\text{Si:P}$ & imp & 2016-10 & \onlinecite{laucht_dressed_2016} & p4\\
10. & \unit{\milli\second} & $T_2^{\text{DynD}}$ & LD/i & Si:P & imp & 2015-04 & \onlinecite{laucht_electrically_2015} & p4 and Fig. 3j\\
9.2 & \unit{\milli\second} & $T_2^{\text{DynD}}$ & LD/i & $\, ^{28}\text{Si:B}$ & imp & 2020-07 & \onlinecite{kobayashi_engineering_2020} & p3 and Fig. 4\\
6.7 & \unit{\milli\second} & $T_2^{\text{DynD}}$ & LD/e & $\, ^{28}\text{Si/}\text{SiO}_2$ & 2D & 2018-10 & \onlinecite{chan_assessment_2018} & p2\\
3.7 & \unit{\milli\second} & $T_2^{\text{DynD}}$ & LD/e & $\, ^{28}\text{Si/}\text{SiO}_2$ & 2D & 2022-03 & \onlinecite{zwerver_qubits_2022} & Fig. 4d\\
3.1 & \unit{\milli\second} & $T_2^{\text{DynD}}$ & LD/e & $\, ^{28}\text{Si/SiGe}$ & 2D & 2017-12 & \onlinecite{yoneda_quantum-dot_2017} & Fig. 4a\\
1.9 & \unit{\milli\second}\textsuperscript{l} & $T_2^{\text{DynD}}$ & LD/h & Ge/SiGe & 2D & 2024-07 & \onlinecite{wang_operating_2024} & Fig. 1f\\
1.3 & \unit{\milli\second} & $T_2^{\text{DynD}}$ & LD/h & Ge/SiGe & 2D & 2024-05 & \onlinecite{hendrickx_sweet-spot_2024} & Fig. 6b\\
0.87 & \unit{\milli\second} & $T_2^{\text{DynD}}$ & ST/e & GaAs/AlGaAs & 2D & 2016-10 & \onlinecite{malinowski_notch_2016} & Fig. 5\\
0.400 & \unit{\milli\second} & $T_2^{\text{DynD}}$ & LD/e & Si/SiGe & 2D & 2016-10 & \onlinecite{kawakami_gate_2016} & p3\\
0.4 & \unit{\milli\second} & $T_2^{\text{DynD}}$ & LD/h & $\text{Si/}\text{SiO}_2$ & 1D & 2022-09 & \onlinecite{piot_single_2022} & p4 and Fig. 3d\\
0.300 & \unit{\milli\second} & $T_2^{\text{DynD}}$ & ST/e & GaAs/AlGaAs & 2D & 2017-04 & \onlinecite{malinowski_spectrum_2017} & Fig. 4c the point n=256\\
0.276 & \unit{\milli\second} & $T_2^{\text{DynD}}$ & ST/e & GaAs/AlGaAs & 2D & 2010-12 & \onlinecite{bluhm_dephasing_2010} & Fig. 3\\
0.158 & \unit{\milli\second} & $T_2^{\text{DynD}}$ & ST/h & Ge/SiGe & 2D & 2021-06 & \onlinecite{jirovec_singlet-triplet_2021} & p6 and Fig. 5\\
0.130 & \unit{\milli\second}\textsuperscript{m} & $T_2^{\text{DynD}}$ & ST/e & GaAs/AlGaAs & 2D & 2012-02 & \onlinecite{medford_scaling_2012} & Fig. 3b\\
0.10 & \unit{\milli\second} & $T_2^{\text{DynD}}$ & LD/h & Ge/SiGe & 2D & 2021-03 & \onlinecite{hendrickx_four-qubit_2021} & Fig. S6 dot 3\\
80. & \unit{\micro\second} & $T_2^{\text{DynD}}$ & ST/e & GaAs/AlGaAs & 2D & 2010-12 & \onlinecite{barthel_interlaced_2010} & p3\\
44. & \unit{\micro\second} & $T_2^{\text{DynD}}$ & LD/e & Si/SiGe & 2D & 2014-08 & \onlinecite{kawakami_electrical_2014} & Fig. 4b\\
44. & \unit{\micro\second}\textsuperscript{n,o} & $T_2^{\text{DynD}}$ & LD/e & $\, ^{28}\text{Si/SiGe}$ & 2D & 2022-11 & \onlinecite{petit_design_2022} & p3\\
20. & \unit{\micro\second}\textsuperscript{p} & $T_2^{\text{DynD}}$ & ST/e & $\, ^{28}\text{Si/}\text{SiO}_2$ & 2D & 2022-02 & \onlinecite{jock_silicon_2022} & Fig. 4c the rightmost blue point\\
19. & \unit{\micro\second} & $T_2^{\text{DynD}}$ & HY/e\textsuperscript{d} & GaAs/AlGaAs & 2D & 2013-07 & \onlinecite{medford_quantum-dot-based_2013} & p3 and Fig. 5 the rightmost point\\
\bottomrule
\end{tabular}
\caption{Spin coherence times (part 7). Superscripts stand for the following. \textsuperscript{a}: Qubit dressed by constant driving. \textsuperscript{b}: \estimated from Fig. 4c the blue points extrapolation to $N_\pi=1$. \textsuperscript{c}: \estimated Fig. 5 the leftmost point. \textsuperscript{d}: {RX} qubit. \textsuperscript{e}: {arXiv}:2011.13755v3. \textsuperscript{f}: \estimated we take the value for $n=2$ {CPMG} sequence, which seems comparable to the Hahn echo data according to Fig. S12. \textsuperscript{g}: At 1.5 K. \textsuperscript{h}: Qubit 1 representative value. \textsuperscript{i}: {EO} qubit. \textsuperscript{j}: Q1. \textsuperscript{k}: Qubit defined in the rotating frame. \textsuperscript{l}: $Q_A$. \textsuperscript{m}: \estimated Fig. 3b the rightmost point. \textsuperscript{n}: At 1 K. \textsuperscript{o}: Q2. \textsuperscript{p}: \estimated from Fig. 4c.
\label{spinCoherence7}
}
\end{table}

%% file: figuresAndTables/table-spinCoherence-8.txt
\rowcolors{2}{white}{gray!25}
\setcounter{rowcount}{-1}
\begin{table}
\begin{tabular}{@{}S[table-format=2.2, round-mode=figures, round-pad=false, round-precision=2,  table-alignment-mode = none, table-number-alignment = center, table-column-width = 3em ]@{}lccccccc@{ \stepcounter{rowcount} \tiny \therowcount }}
\toprule
\multicolumn{2}{c}{Time} & {Coherence} & {Qubit} & {Material} & {Host} & {Date} & {Reference} & {Source}\\
\midrule
5.4 & \unit{\micro\second}\textsuperscript{a} & $T_2^{\text{DynD}}$ & LD/h\textsuperscript{a} & $\text{Si/}\text{SiO}_2$ & 1D & 2022-03 & \onlinecite{camenzind_spin_2022} & p3 and Fig. 4a\\
1. & \unit{\micro\second} & $T_2^{\text{DynD}}$ & HY/e\textsuperscript{b,c} & GaAs/AlGaAs & 2D & 2017-07 & \onlinecite{malinowski_symmetric_2017} & Fig. 5b\\
1. & \unit{\micro\second} & $T_2^{\text{DynD}}$ & HY/e\textsuperscript{d,c} & GaAs/AlGaAs & 2D & 2017-07 & \onlinecite{malinowski_symmetric_2017} & Fig. 5b\\
0.410 & \unit{\micro\second} & $T_2^{\text{DynD}}$ & LD/i & Si:P & imp & 2012-09 & \onlinecite{pla_single-atom_2012} & p3\\
0.150 & \unit{\micro\second} & $T_2^{\text{DynD}}$ & LD/e & InAs & 1D & 2010-12 & \onlinecite{nadj-perge_spinorbit_2010} & Fig. 4a inset the highest point\\
0.150 & \unit{\micro\second} & $T_2^{\text{DynD}}$ & HY/e & Si/SiGe & 2D & 2015-12 & \onlinecite{kim_high-fidelity_2015} & abstract and Fig. 3d\\
0.150 & \unit{\micro\second}\textsuperscript{e} & $T_2^{\text{DynD}}$ & LD/e & $\, ^{28}\text{Si/SiGe}$ & 2D & 2024-12 & \onlinecite{dijkema_cavity-mediated_2024} & Tab. I\\
0.180 & \unit{\milli\second} & $T_2^{\text{Rabi}}$ & LD/i & $\, ^{28}\text{Si:P}$ & imp & 2024-02 & \onlinecite{reiner_high-fidelity_2024} & p5\\
0.113 & \unit{\milli\second} & $T_2^{\text{Rabi}}$ & LD/e & $\, ^{28}\text{Si/SiGe}$ & 2D & 2017-12 & \onlinecite{yoneda_quantum-dot_2017} & p2\\
0.10 & \unit{\milli\second}\textsuperscript{f} & $T_2^{\text{Rabi}}$ & LD/e & $\, ^{28}\text{Si/SiGe}$ & 2D & 2022-01 & \onlinecite{noiri_fast_2022} & {ED} Fig. 2l the topmost trace\\
80. & \unit{\micro\second}\textsuperscript{g} & $T_2^{\text{Rabi}}$ & LD/e & $\, ^{28}\text{Si/SiGe}$ & 2D & 2022-09 & \onlinecite{noiri_shuttling-based_2022} & {ED} Fig. 2e\\
54.9 & \unit{\micro\second}\textsuperscript{h} & $T_2^{\text{Rabi}}$ & LD/e & $\text{Si/}\text{SiO}_2$ & 2D & 2026-01 & \onlinecite{kuno_robust_2026} & p4\\
41. & \unit{\micro\second}\textsuperscript{i} & $T_2^{\text{Rabi}}$ & LD/e & $\, ^{28}\text{Si/}\text{SiO}_2$ & 2D & 2024-08 & \onlinecite{tanttu_assessment_2024} & {ED} Tab. {II}\\
18.6 & \unit{\micro\second} & $T_2^{\text{Rabi}}$ & LD/e & $\, ^{28}\text{Si/}\text{SiO}_2$ & 2D & 2019-12 & \onlinecite{zhao_single-spin_2019} & p3 and Fig. 2d\\
10. & \unit{\micro\second}\textsuperscript{j} & $T_2^{\text{Rabi}}$ & LD/e\textsuperscript{k} & $\, ^{28}\text{Si/}\text{SiO}_2$ & 2D & 2024-09 & \onlinecite{hansen_entangling_2024} & Fig. 3d\\
7. & \unit{\micro\second}\textsuperscript{l} & $T_2^{\text{Rabi}}$ & LD/e & Si/SiGe & 2D & 2016-08 & \onlinecite{takeda_fault-tolerant_2016} & p3\\
6.46 & \unit{\micro\second} & $T_2^{\text{Rabi}}$ & LD/e & $\, ^{28}\text{Si/}\text{SiO}_2$ & 2D & 2023-03 & \onlinecite{hu_flopping-mode_2023} & Fig. 3c\\
6. & \unit{\micro\second} & $T_2^{\text{Rabi}}$ & ST/e & Si/SiGe & 2D & 2020-03 & \onlinecite{takeda_resonantly_2020} & p3\\
5.2 & \unit{\micro\second}\textsuperscript{m} & $T_2^{\text{Rabi}}$ & LD/e & Si/SiGe & 2D & 2022-08 & \onlinecite{takeda_quantum_2022} & {ED} Fig. 3i\\
1.88 & \unit{\micro\second}\textsuperscript{n} & $T_2^{\text{Rabi}}$ & ST/e & GaAs/AlGaAs & 2D & 2023-03 & \onlinecite{yun_probing_2023} & p4\\
1.71 & \unit{\micro\second}\textsuperscript{o} & $T_2^{\text{Rabi}}$ & ST/e & GaAs/AlGaAs & 2D & 2022-07 & \onlinecite{kim_approaching_2022} & p4\\
1.4 & \unit{\micro\second}\textsuperscript{p} & $T_2^{\text{Rabi}}$ & LD/e & $\, ^{28}\text{Si/}\text{SiO}_2$ & 2D & 2023-01 & \onlinecite{gilbert_-demand_2023} & Fig. 3d, blue point\\
1.4 & \unit{\micro\second} & $T_2^{\text{Rabi}}$ & LD/e & Si/SiGe & 2D & 2020-01 & \onlinecite{croot_flopping-mode_2020} & p4\\
1.26 & \unit{\micro\second} & $T_2^{\text{Rabi}}$ & LD/e & GaAs/AlGaAs & 2D & 2020-03 & \onlinecite{nakajima_coherence_2020} & Fig. 6\\
1.2 & \unit{\micro\second}\textsuperscript{q} & $T_2^{\text{Rabi}}$ & LD/e & $\text{Si/}\text{SiO}_2$ & 2D & 2026-01 & \onlinecite{kuno_robust_2026} & p2 and Fig. S1b\\
1. & \unit{\micro\second} & $T_2^{\text{Rabi}}$ & HY/e\textsuperscript{r} & Si/SiGe & 2D & 2017-08 & \onlinecite{thorgrimsson_extending_2017} & Fig. 3g\\
0.700 & \unit{\micro\second} & $T_2^{\text{Rabi}}$ & ST/e & GaAs/AlGaAs & 2D & 2017-01 & \onlinecite{nichol_high-fidelity_2017} & p3 and Fig. 2b\\
0.526 & \unit{\micro\second} & $T_2^{\text{Rabi}}$ & LD/e & GaAs/AlGaAs & 2D & 2019-04 & \onlinecite{nakajima_quantum_2019} & Fig. 2\\
0.400 & \unit{\micro\second}\textsuperscript{s} & $T_2^{\text{Rabi}}$ & HY/e\textsuperscript{t} & $\, ^{28}\text{Si/SiGe}$ & 2D & 2019-07 & \onlinecite{andrews_quantifying_2019} & Fig. S2c\\
0.28 & \unit{\micro\second}\textsuperscript{u} & $T_2^{\text{Rabi}}$ & ST/e & $\, ^{28}\text{Si/SiGe}$ & 2D & 2016-03 & \onlinecite{reed_reduced_2016} & p3, Fig. 4a and b\\
0.19 & \unit{\micro\second}\textsuperscript{v} & $T_2^{\text{Rabi}}$ & HY/e & Si/SiGe & 2D & 2018-10 & \onlinecite{abadillo-uriel_signatures_2018} & p3\\
0.105 & \unit{\micro\second}\textsuperscript{e} & $T_2^{\text{Rabi}}$ & LD/e & $\, ^{28}\text{Si/SiGe}$ & 2D & 2024-12 & \onlinecite{dijkema_cavity-mediated_2024} & Tab. I\\
86. & \unit{\nano\second} & $T_2^{\text{Rabi}}$ & ST/e & GaAs/AlGaAs & 2D & 2021-05 & \onlinecite{kojima_probabilistic_2021} & p3\\
84. & \unit{\nano\second} & $T_2^{\text{Rabi}}$ & LD/e & GaAs/AlGaAs & 2D & 2016-04 & \onlinecite{noiri_coherent_2016} & Fig. 3\\
80. & \unit{\nano\second}\textsuperscript{w} & $T_2^{\text{Rabi}}$ & HY/e\textsuperscript{d,c} & GaAs/AlGaAs & 2D & 2017-07 & \onlinecite{malinowski_symmetric_2017} & Fig. 4c\\
60. & \unit{\nano\second} & $T_2^{\text{Rabi}}$ & LD/h & Ge/Si & 1D & 2021-01 & \onlinecite{froning_ultrafast_2021} & Fig. 3d\\
40. & \unit{\nano\second}\textsuperscript{w} & $T_2^{\text{Rabi}}$ & HY/e\textsuperscript{b,c} & GaAs/AlGaAs & 2D & 2017-07 & \onlinecite{malinowski_symmetric_2017} & Fig. 4c\\
40. & \unit{\nano\second} & $T_2^{\text{Rabi}}$ & LD/h & Ge/Si & 1D & 2025-08 & \onlinecite{carballido_compromise-free_2025} & Fig. 1 f\\
37. & \unit{\nano\second} & $T_2^{\text{Rabi}}$ & HY/e & GaAs/AlGaAs & 2D & 2021-06 & \onlinecite{jang_single-shot_2021} & Fig. 3b\\
36.7 & \unit{\nano\second}\textsuperscript{x} & $T_2^{\text{Rabi}}$ & LD/e & GaAs/AlGaAs & 2D & 2014-12 & \onlinecite{yoneda_fast_2014} & p3\\
19.6 & \unit{\nano\second}\textsuperscript{y} & $T_2^{\text{Rabi}}$ & LD/h & Ge/SiGe & 2D & 2026-09 & \onlinecite{kelly_identification_2026} & p6\\
\bottomrule
\end{tabular}
\caption{Spin coherence times (part 8). Superscripts stand for the following. \textsuperscript{a}: At 1.5 K. \textsuperscript{b}: At a partial sweet spot. \textsuperscript{c}: Resonant exchange qubit. \textsuperscript{d}: At a full sweet spot. \textsuperscript{e}: Qubit 1 representative value. \textsuperscript{f}: The Rabi decay time $\TRabi = 100$ \unit{\micro\second} is obtained by fitting the Rabi decay data to a function with two parameters, $\TRabi$ and $T_2^*$. \textsuperscript{g}: Left qubit. \textsuperscript{h}: Driven qubit. \textsuperscript{i}: Device B. \textsuperscript{j}: The lower limit. \textsuperscript{k}: Qubit dressed by constant driving. \textsuperscript{l}: From $Q=140$ and $f_\mathrm{R}=10$ {MHz}. \textsuperscript{m}: Decay of {CZ} oscillations. \textsuperscript{n}: $Q_R$. \textsuperscript{o}: With estimation and/or feedback. \textsuperscript{p}: \estimated from Fig. 3d at $\Delta V_J=10$ {mV}. \textsuperscript{q}: Bare qubit. \textsuperscript{r}: Tunable between spin- and charge- qubit. \textsuperscript{s}: \estimated from Fig. S2c as the signal decay by factor $1/e$. \textsuperscript{t}: {EO} qubit. \textsuperscript{u}: \derived from $T = h N_\mathrm{R} / J$ with $J/h=160$ {MHz} and $N_\mathrm{R}=44$. \textsuperscript{v}: Reference gives Rabi decay rate $\Gamma=5.4$ {MHz}. We convert it to Rabi decay time by $\TRabi=1/\Gamma$. \textsuperscript{w}: Approximate value at $f_\mathrm{R}=100$ {MHz}. \textsuperscript{x}: At Rabi frequency $f_\mathrm{R}=85.9$ {MHz}. \textsuperscript{y}: Q2.
\label{spinCoherence8}
}
\end{table}

%% file: figuresAndTables/table-chargeCoherence-1.txt
\rowcolors{2}{white}{gray!25}
\setcounter{rowcount}{-1}
\begin{table}
\begin{tabular}{@{}S[table-format=2.2, round-mode=figures, round-pad=false, round-precision=2,  table-alignment-mode = none, table-number-alignment = center, table-column-width = 3em ]@{}lccccccc@{ \stepcounter{rowcount} \tiny \therowcount }}
\toprule
\multicolumn{2}{c}{Time} & {Coherence} & {Qubit} & {Material} & {Host} & {Date} & {Reference} & {Source}\\
\midrule
45. & \unit{\micro\second} & $T_1$ & charge & Si/SiGe & 2D & 2013-07 & \onlinecite{wang_charge_2013} & p3 and Fig. 3c\\
78. & \unit{\nano\second} & $T_1$ & charge & SLG & 2D & 2013-04 & \onlinecite{volk_probing_2013} & Fig. 4b\\
48. & \unit{\nano\second} & $T_1$ & charge & CNT & 1D & 2017-05 & \onlinecite{penfold-fitch_microwave_2017} & p4 and Fig. 3c\\
42.3 & \unit{\nano\second} & $T_1$ & charge & GaAs/AlGaAs & 2D & 2019-05 & \onlinecite{scarlino_all-microwave_2019} & p4 and Fig. 5d\\
19. & \unit{\nano\second} & $T_1$ & charge & GaAs/AlGaAs & 2D & 2015-07 & \onlinecite{li_conditional_2015} & p4\\
16. & \unit{\nano\second} & $T_1$ & charge & GaAs/AlGaAs & 2D & 2004-10 & \onlinecite{petta_manipulation_2004} & p3 and Fig. 4\\
10. & \unit{\nano\second} & $T_1$ & charge & GaAs/AlGaAs & 2D & 2010-12 & \onlinecite{petersson_quantum_2010} & p2\\
10. & \unit{\nano\second} & $T_1$ & charge & InGaAs/AlGaAs & 2D & 2002-09 & \onlinecite{fujisawa_allowed_2002} & p1 and Fig. 2d\\
0.118 & \unit{\micro\second}\textsuperscript{a} & $T_2^*$ & charge & GaAs/AlGaAs & 2D & 2019-07 & \onlinecite{scarlino_coherent_2019} & p2\\
61. & \unit{\nano\second}\textsuperscript{b} & $T_2^*$ & charge & Si/SiGe & 2D & 2016-12 & \onlinecite{mi_strong_2016} & p2\\
33. & \unit{\nano\second}\textsuperscript{c} & $T_2^*$ & charge & GaAs/AlGaAs & 2D & 2018-10 & \onlinecite{van_woerkom_microwave_2018} & p2\\
23.4 & \unit{\nano\second} & $T_2^*$ & charge & GaAs/AlGaAs & 2D & 2019-05 & \onlinecite{scarlino_all-microwave_2019} & p4 and Fig. 5c\\
16. & \unit{\nano\second}\textsuperscript{d} & $T_2^*$ & charge & $\text{Si/}\text{SiO}_2$ & 1D & 2023-03 & \onlinecite{yu_strong_2023} & p3\\
16. & \unit{\nano\second}\textsuperscript{e} & $T_2^*$ & charge & $\, ^{28}\text{Si/SiGe}$ & 2D & 2023-03 & \onlinecite{bonsen_probing_2023} & Tab. S1\\
7. & \unit{\nano\second} & $T_2^*$ & charge & GaAs/AlGaAs & 2D & 2010-12 & \onlinecite{petersson_quantum_2010} & p4\\
5.3 & \unit{\nano\second}\textsuperscript{f} & $T_2^*$ & charge & GaAs/AlGaAs & 2D & 2018-07 & \onlinecite{landig_coherent_2018} & p4\\
5. & \unit{\nano\second} & $T_2^*$ & charge & CNT & 1D & 2017-05 & \onlinecite{penfold-fitch_microwave_2017} & p4 and Fig. 3b\\
5.0 & \unit{\nano\second}\textsuperscript{g} & $T_2^*$ & charge & GaAs/AlGaAs & 2D & 2020-05 & \onlinecite{koski_strong_2020} & p3\\
5.0 & \unit{\nano\second}\textsuperscript{g} & $T_2^*$ & charge & $\, ^{28}\text{Si/SiGe}$ & 2D & 2021-04 & \onlinecite{borjans_probing_2021} & p6\\
4.4 & \unit{\nano\second}\textsuperscript{h} & $T_2^*$ & charge & Si/SiGe & 2D & 2020-06 & \onlinecite{borjans_split-gate_2020} & Fig. 3b\\
4.0 & \unit{\nano\second}\textsuperscript{i} & $T_2^*$ & charge & Si/SiGe & 2D & 2018-02 & \onlinecite{mi_coherent_2018} & p2 and Fig. 1d\\
4.0 & \unit{\nano\second}\textsuperscript{i} & $T_2^*$ & charge & Si/SiGe & 2D & 2017-01 & \onlinecite{mi_circuit_2017} & p5\\
4.0 & \unit{\nano\second}\textsuperscript{j} & $T_2^*$ & charge & GaAs/AlGaAs & 2D & 2017-03 & \onlinecite{stockklauser_strong_2017} & p5\\
3.2 & \unit{\nano\second}\textsuperscript{k,l} & $T_2^*$ & charge & GaAs/AlGaAs & 2D & 2023-05 & \onlinecite{lin_collective_2023} & Tab. 1\\
3.0 & \unit{\nano\second}\textsuperscript{m} & $T_2^*$ & charge & GaAs/AlGaAs & 2D & 2021-02 & \onlinecite{kratochwil_charge_2021} & p5 and Fig. 4a\\
2.9 & \unit{\nano\second}\textsuperscript{n} & $T_2^*$ & charge & GaAs/AlGaAs & 2D & 2021-02 & \onlinecite{wang_correlated_2021} & p4\\
2.8 & \unit{\nano\second} & $T_2^*$ & charge & Ge/SiGe & 2D & 2024-11 & \onlinecite{de_palma_strong_2024} & Fig. 3b\\
2.7 & \unit{\nano\second}\textsuperscript{o} & $T_2^*$ & charge & $\, ^{28}\text{Si/SiGe}$ & 2D & 2022-05 & \onlinecite{harvey-collard_coherent_2022} & p6\\
2.7 & \unit{\nano\second} & $T_2^*$ & charge & $\, ^{28}\text{Si/SiGe}$ & 2D & 2024-12 & \onlinecite{dijkema_cavity-mediated_2024} & p6\\
2.2 & \unit{\nano\second}\textsuperscript{p} & $T_2^*$ & charge & GaAs/AlGaAs & 2D & 2021-04 & \onlinecite{chen_microwave-resonator-detected_2021} & p3\\
2.2 & \unit{\nano\second} & $T_2^*$ & charge & Si/SiGe & 2D & 2013-08 & \onlinecite{shi_coherent_2013} & abstract and Fig. 1\\
1.8 & \unit{\nano\second}\textsuperscript{q} & $T_2^*$ & charge & GaAs/AlGaAs & 2D & 2023-06 & \onlinecite{gu_probing_2023} & p2\\
1.3 & \unit{\nano\second} & $T_2^*$ & charge & Si/SiGe & 2D & 2015-02 & \onlinecite{kim_microwave-driven_2015} & Fig. 2c\\
1.2 & \unit{\nano\second} & $T_2^*$ & charge & GaAs/AlGaAs & 2D & 2015-07 & \onlinecite{li_conditional_2015} & p4\\
1.1 & \unit{\nano\second} & $T_2^*$ & charge & GaAs/AlGaAs & 2D & 2010-07 & \onlinecite{petersson_charge_2010} & p3\\
1. & \unit{\nano\second} & $T_2^*$ & charge & GaAs/AlGaAs & 2D & 2003-11 & \onlinecite{hayashi_coherent_2003} & p3\\
0.637 & \unit{\nano\second}\textsuperscript{r} & $T_2^*$ & charge & GaAs/AlGaAs & 2D & 2015-07 & \onlinecite{stockklauser_microwave_2015} & Fig. 2b\\
0.601 & \unit{\nano\second} & $T_2^*$ & charge & Ge/SiGe & 2D & 2024-08 & \onlinecite{kang_coupling_2024} & p4\\
0.57 & \unit{\nano\second}\textsuperscript{s} & $T_2^*$ & charge & Ge/SiGe & 1D & 2020-08 & \onlinecite{xu_dipole_2020} & p5 and Fig.3\\
0.536 & \unit{\nano\second} & $T_2^*$ & charge & Ge/SiGe & 2D & 2024-07 & \onlinecite{janik_strong_2024} & p5\\
0.531 & \unit{\nano\second}\textsuperscript{t} & $T_2^*$ & charge & InAs & 1D & 2024-11 & \onlinecite{ranni_decoherence_2024} & p5\\
0.51 & \unit{\nano\second}\textsuperscript{u} & $T_2^*$ & charge & SLG & 2D & 2015-09 & \onlinecite{deng_charge_2015} & p3\\
0.400 & \unit{\nano\second} & $T_2^*$ & charge & GaAs/AlGaAs & 2D & 2004-10 & \onlinecite{petta_manipulation_2004} & p4 and Fig. 4\\
0.3 & \unit{\nano\second}\textsuperscript{v} & $T_2^*$ & charge & GaAs/AlGaAs & 2D & 2013-09 & \onlinecite{basset_single-electron_2013} & Fig. 3b\\
0.289 & \unit{\nano\second}\textsuperscript{w} & $T_2^*$ & charge & CNT & 1D & 2014-04 & \onlinecite{viennot_out--equilibrium_2014} & p2\\
0.248 & \unit{\nano\second} & $T_2^*$ & charge & BLG & 2D & 2024-06 & \onlinecite{ruckriegel_electric_2024} & p5 and Fig. 4\\
0.17 & \unit{\nano\second}\textsuperscript{x} & $T_2^*$ & charge & GaAs/AlGaAs & 2D & 2012-01 & \onlinecite{frey_dipole_2012} & p4\\
0.2 & \unit{\nano\second}\textsuperscript{y} & $T_2^*$ & charge & InSb & 1D & 2016-05 & \onlinecite{wang_insb_2016} & p4\\
0.118 & \unit{\nano\second}\textsuperscript{z} & $T_2^*$ & charge & CNT & 1D & 2019-07 & \onlinecite{cubaynes_highly_2019} & p2\\
80. & \unit{\pico\second} & $T_2^*$ & charge & Si/SiGe & 2D & 2020-09 & \onlinecite{macquarrie_progress_2020} & p3\\
\bottomrule
\end{tabular}
\caption{Charge coherence times (part 1). Superscripts stand for the following. \textsuperscript{a}: From $f_\textrm{FWHM}/2\pi = 2.7$ {MHz}. \textsuperscript{b}: From $\Gamma_2/2\pi = 2.6$ {MHz}. \textsuperscript{c}: From $\Gamma_2/2\pi = 4.8$ {MHz}. \textsuperscript{d}: From $\Gamma_2/2\pi=9.9$ {MHz}. \textsuperscript{e}: From $\Gamma_2/2\pi = 6$ {MHz}. \textsuperscript{f}: From $\Gamma_2/2\pi = 30$ {MHz}. \textsuperscript{g}: From $\Gamma_2/2\pi = 32$ {MHz}. \textsuperscript{h}: From $\Gamma_2/2\pi = 36$ {MHz}. \textsuperscript{i}: From $\Gamma_2/2\pi = 40$ {MHz}. \textsuperscript{j}: From $f_\mathrm{FWHM}/2\pi = 80$ {MHz}. \textsuperscript{k}: {DQD}1. \textsuperscript{l}: From $\Gamma_2/2\pi =50$ {MHz}. \textsuperscript{m}: From $\Gamma_2/2\pi = 53$ {MHz}. \textsuperscript{n}: From $\Gamma_2/2\pi = 55$ {MHz}. \textsuperscript{o}: From $\Gamma_2/2\pi = 60$ {MHz}. \textsuperscript{p}: From $\Gamma_2/2\pi=72$ {MHz}. \textsuperscript{q}: From $\Gamma_2 /2\pi = 90$ {MHz}. \textsuperscript{r}: From $\Gamma_2/2\pi = 250$ {MHz}. \textsuperscript{s}: From $\Gamma_2/2\pi = 0.28$ {GHz}. \textsuperscript{t}: From $\Gamma/2\pi=300$ {MHz}. \textsuperscript{u}: From $\Gamma_\phi/2\pi = 0.31$ {GHz}. \textsuperscript{v}: From $\Gamma_\phi /2pi = 0.5$ {GHz}. \textsuperscript{w}: From $\Gamma_2/2\pi = 550$ {MHz}. \textsuperscript{x}: From $\Gamma_\phi/2\pi = 0.9$ {GHz} and $\Gamma_1=100$ {MHz}. \textsuperscript{y}: From $\Gamma_\phi/2\pi$; we take the lower limit from the range 1-4 {GHz} given in the reference. \textsuperscript{z}: From $\Gamma_2/2\pi = 1.35$ {GHz}.
\label{chargeCoherence1}
}
\end{table}

%% file: figuresAndTables/table-chargeCoherence-2.txt
\rowcolors{2}{white}{gray!25}
\setcounter{rowcount}{-1}
\begin{table}
\begin{tabular}{@{}S[table-format=2.2, round-mode=figures, round-pad=false, round-precision=2,  table-alignment-mode = none, table-number-alignment = center, table-column-width = 3em ]@{}lccccccc@{ \stepcounter{rowcount} \tiny \therowcount }}
\toprule
\multicolumn{2}{c}{Time} & {Coherence} & {Qubit} & {Material} & {Host} & {Date} & {Reference} & {Source}\\
\midrule
60. & \unit{\pico\second} & $T_2^*$ & charge & GaAs/AlGaAs & 2D & 2011-10 & \onlinecite{dovzhenko_nonadiabatic_2011} & p2\\
50. & \unit{\pico\second}\textsuperscript{a} & $T_2^*$ & charge & CNT & 1D & 2015-05 & \onlinecite{ranjan_clean_2015} & Fig. 4f the lowest black point\\
16.0 & \unit{\pico\second}\textsuperscript{b} & $T_2^*$ & charge & $\text{Si/}\text{SiO}_2$ & 1D & 2021-05 & \onlinecite{ibberson_large_2021} & p4\\
0.760 & \unit{\micro\second} & $T_2^{\text{Echo}}$ & charge & Si/SiGe & 2D & 2013-08 & \onlinecite{shi_coherent_2013} & Fg. 2\\
43.1 & \unit{\nano\second} & $T_2^{\text{Echo}}$ & charge & GaAs/AlGaAs & 2D & 2019-05 & \onlinecite{scarlino_all-microwave_2019} & p4 and Fig. 5e\\
2.2 & \unit{\nano\second} & $T_2^{\text{Echo}}$ & charge & Si/SiGe & 2D & 2015-02 & \onlinecite{kim_microwave-driven_2015} & p3\\
4. & \unit{\nano\second}\textsuperscript{c} & $T_2^{\text{DynD}}$ & charge & GaAs/AlGaAs & 2D & 2013-01 & \onlinecite{cao_ultrafast_2013} & p5\\
1.5 & \unit{\nano\second} & $T_2^{\text{Rabi}}$ & charge & Si/SiGe & 2D & 2015-02 & \onlinecite{kim_microwave-driven_2015} & Fig. 1e\\
\bottomrule
\end{tabular}
\caption{Charge coherence times (part 2). Superscripts stand for the following. \textsuperscript{a}: From $\Gamma_2/2\pi \approx 3$ {GHz} \estimated from Fig. 4f. \textsuperscript{b}: From $\Gamma_2/2\pi = 9.95$ {GHz}. \textsuperscript{c}: From Landau-Zener interference.
\label{chargeCoherence2}
}
\end{table}

%% file: figuresAndTables/table-operationTime-1.txt
\rowcolors{2}{white}{gray!25}
\setcounter{rowcount}{-1}
\begin{table}
\begin{tabular}{@{}S[table-format=2.2, round-mode=figures, round-pad=false, round-precision=2, table-alignment-mode = none, table-number-alignment = center, table-column-width = 3em ]@{}lcccccccc@{ \stepcounter{rowcount} \tiny \therowcount }}
\toprule
\multicolumn{2}{c}{Time} & {Operation} & {\#Qubits} & {Qubit} & {Material} & {Host} & {Date} & {Reference} & {Source}\\
\midrule
50. & \unit{\pico\second} & gate & 1Q & charge & GaAs/AlGaAs & 2D & 2013-01 & \onlinecite{cao_ultrafast_2013} & p4\\
10.5 & \unit{\pico\second} & gate & 1Q & HY/e\textsuperscript{a} & GaAs/AlGaAs & 2D & 2013-09 & \onlinecite{medford_self-consistent_2013} & p2\\
45. & \unit{\pico\second}\textsuperscript{b} & gate & 1Q & HY/e\textsuperscript{c} & Si/SiGe & 2D & 2014-07 & \onlinecite{kim_quantum_2014} & Fig. 2b\\
4.55 & \unit{\nano\second} & gate & 1Q & HY/e & Si/SiGe & 2D & 2015-12 & \onlinecite{kim_high-fidelity_2015} & Fig. 1g\\
5.00 & \unit{\nano\second} & gate & 1Q & HY/e & GaAs/AlGaAs & 2D & 2021-06 & \onlinecite{jang_single-shot_2021} & p3 and Fig. 3a\\
5.00 & \unit{\nano\second} & gate & 1Q & HY/e\textsuperscript{d,e} & GaAs/AlGaAs & 2D & 2017-07 & \onlinecite{malinowski_symmetric_2017} & Fig. 4c\\
5.00 & \unit{\nano\second} & gate & 1Q & HY/e\textsuperscript{f,e} & GaAs/AlGaAs & 2D & 2017-07 & \onlinecite{malinowski_symmetric_2017} & Fig. 4c\\
9. & \unit{\nano\second}\textsuperscript{g} & gate & 1Q & HY/e\textsuperscript{a} & GaAs/AlGaAs & 2D & 2010-08 & \onlinecite{laird_coherent_2010} & Fig. 4\\
10. & \unit{\nano\second} & gate & 1Q & HY/e\textsuperscript{a} & $\, ^{28}\text{Si/SiGe}$ & 2D & 2019-07 & \onlinecite{andrews_quantifying_2019} & Fig. 4c caption\\
15. & \unit{\nano\second} & gate & 1Q & HY/e\textsuperscript{h} & GaAs/AlGaAs & 2D & 2013-07 & \onlinecite{medford_quantum-dot-based_2013} & Fig. 4 the blue set\\
36. & \unit{\nano\second} & gate & 1Q & HY/e\textsuperscript{i} & Si/SiGe & 2D & 2017-08 & \onlinecite{thorgrimsson_extending_2017} & Fig. 3g\\
4.07 & \unit{\nano\second} & gate & 1Q & LD/e & GaAs/AlGaAs & 2D & 2014-12 & \onlinecite{yoneda_fast_2014} & p2\\
4.81 & \unit{\nano\second} & gate & 1Q & LD/e & InSb & 1D & 2013-02 & \onlinecite{van_den_berg_fast_2013} & Fig. 2c\\
5.00 & \unit{\nano\second}\textsuperscript{j} & gate & 1Q & LD/e & CNT & 1D & 2013-07 & \onlinecite{laird_valleyspin_2013} & Fig. 3b\\
8.6 & \unit{\nano\second} & gate & 1Q & LD/e & InAs & 1D & 2010-12 & \onlinecite{nadj-perge_spinorbit_2010} & p2\\
14.86 & \unit{\nano\second} & gate & 1Q & LD/e & GaAs/AlGaAs & 2D & 2020-03 & \onlinecite{nakajima_coherence_2020} & Fig. 6\\
17. & \unit{\nano\second}\textsuperscript{k} & gate & 1Q & LD/e & InAs & 1D & 2012-10 & \onlinecite{petersson_circuit_2012} & p3 and Fig. 4f\\
20.3 & \unit{\nano\second} & gate & 1Q & LD/e & GaAs/AlGaAs & 2D & 2016-04 & \onlinecite{noiri_coherent_2016} & p4 and Fig. 3\\
40.0 & \unit{\nano\second} & gate & 1Q & LD/e\textsuperscript{l} & $\text{Si/}\text{SiO}_2$ & 1D & 2025-02 & \onlinecite{eggli_coupling_2025} & p3\\
42. & \unit{\nano\second} & gate & 1Q & LD/e & $\text{Si/}\text{SiO}_2$ & 1D & 2018-02 & \onlinecite{corna_electrically_2018} & Fig. 4b\\
50. & \unit{\nano\second} & gate & 1Q & LD/e & Si/SiGe & 2D & 2016-08 & \onlinecite{takeda_fault-tolerant_2016} & p3\\
54. & \unit{\nano\second} & gate & 1Q & LD/e & GaAs/AlGaAs & 2D & 2006-08 & \onlinecite{koppens_driven_2006} & p4\\
60. & \unit{\nano\second} & gate & 1Q & LD/e & GaAs/AlGaAs & 2D & 2008-06 & \onlinecite{koppens_spin_2008} & Fig. 2b the blue trace\\
60. & \unit{\nano\second}\textsuperscript{m} & gate & 1Q & LD/e & GaAs/AlGaAs & 2D & 2007-09 & \onlinecite{koppens_universal_2007} & p3\\
67. & \unit{\nano\second} & gate & 1Q & LD/e & GaAs/AlGaAs & 2D & 2024-01 & \onlinecite{liu_accelerated_2024} & p2 and Fig. 1d\\
70. & \unit{\nano\second} & gate & 1Q & LD/e & $\, ^{28}\text{Si/SiGe}$ & 2D & 2022-04 & \onlinecite{mills_two-qubit_2022} & p1\\
80. & \unit{\nano\second} & gate & 1Q & LD/e & Si/SiGe & 2D & 2020-01 & \onlinecite{croot_flopping-mode_2020} & p4\\
80. & \unit{\nano\second} & gate & 1Q & LD/e & Si/SiGe & 2D & 2021-06 & \onlinecite{takeda_quantum_2021} & p1\\
91.6 & \unit{\nano\second} & gate & 1Q & LD/e & GaAs/AlGaAs & 2D & 2019-04 & \onlinecite{nakajima_quantum_2019} & Fig. 2\\
94. & \unit{\nano\second} & gate & 1Q & LD/e & GaAs/AlGaAs & 2D & 2018-08 & \onlinecite{ito_four_2018} & p3 and Fig. 3b\\
100. & \unit{\nano\second} & gate & 1Q & LD/e & $\, ^{28}\text{Si/SiGe}$ & 2D & 2022-09 & \onlinecite{philips_universal_2022} & p7\\
0.1027 & \unit{\micro\second} & gate & 1Q & LD/e & $\, ^{28}\text{Si/SiGe}$ & 2D & 2022-01 & \onlinecite{noiri_fast_2022} & p4\\
0.10 & \unit{\micro\second} & gate & 1Q & LD/e & Si/SiGe & 2D & 2017-12 & \onlinecite{zajac_resonantly_2017} & p2\\
0.11 & \unit{\micro\second} & gate & 1Q & LD/e & GaAs/AlGaAs & 2D & 2007-11 & \onlinecite{nowack_coherent_2007} & p3\\
0.1 & \unit{\micro\second} & gate & 1Q & LD/e & Si/SiGe & 2D & 2022-08 & \onlinecite{takeda_quantum_2022} & {ED} Fig. 2 caption\\
0.13 & \unit{\micro\second} & gate & 1Q & LD/e & $\, ^{28}\text{Si/SiGe}$ & 2D & 2017-12 & \onlinecite{yoneda_quantum-dot_2017} & p2\\
0.145 & \unit{\micro\second} & gate & 1Q & LD/e & Si/SiGe & 2D & 2024-11 & \onlinecite{wang_pursuing_2024} & Fig. 2c\\
0.150 & \unit{\micro\second}\textsuperscript{n} & gate & 1Q & LD/e & Si/SiGe & 2D & 2014-08 & \onlinecite{kawakami_electrical_2014} & Fig. 3b\\
0.2 & \unit{\micro\second}\textsuperscript{o} & gate & 1Q & LD/e & $\, ^{28}\text{Si/SiGe}$ & 2D & 2023-04 & \onlinecite{undseth_nonlinear_2023} & Fig. 2 blue points\\
0.2 & \unit{\micro\second} & gate & 1Q & LD/e & Si/SiGe & 2D & 2018-02 & \onlinecite{watson_programmable_2018} & p1\\
0.2 & \unit{\micro\second}\textsuperscript{p} & gate & 1Q & LD/e & GaAs/AlGaAs & 2D & 2011-09 & \onlinecite{brunner_two-qubit_2011} & Fig 2b\\
0.2571 & \unit{\micro\second} & gate & 1Q & LD/e & $\, ^{28}\text{Si/}\text{SiO}_2$ & 2D & 2024-01 & \onlinecite{ma_single-spin-qubit_2024} & Fig. 2a\\
0.28 & \unit{\micro\second}\textsuperscript{q} & gate & 1Q & LD/e & GaAs/AlGaAs & 2D & 2007-12 & \onlinecite{laird_hyperfine-mediated_2007} & p3 and Fig. 3b\\
0.36 & \unit{\micro\second} & gate & 1Q & LD/e & $\, ^{28}\text{Si/}\text{SiO}_2$ & 2D & 2022-03 & \onlinecite{zwerver_qubits_2022} & {ED} Fig. 7a\\
0.3717 & \unit{\micro\second} & gate & 1Q & LD/e & Si/SiGe & 2D & 2016-10 & \onlinecite{kawakami_gate_2016} & p3\\
0.3962 & \unit{\micro\second} & gate & 1Q & LD/e & $\, ^{28}\text{Si/}\text{SiO}_2$ & 2D & 2023-03 & \onlinecite{hu_flopping-mode_2023} & Fig. 3c\\
0.5 & \unit{\micro\second} & gate & 1Q & LD/e & Si/SiGe & 2D & 2017-04 & \onlinecite{scarlino_dressed_2017} & p2\\
0.5 & \unit{\micro\second} & gate & 1Q & LD/e & $\text{Si/}\text{SiO}_2$ & 1D & 2023-10 & \onlinecite{klemt_electrical_2023} & p3 and Fig. 3c\\
0.833 & \unit{\micro\second} & gate & 1Q & LD/e & Si/SiGe & 2D & 2025-01 & \onlinecite{george_12-spin-qubit_2025} & p5 and Fig. 5e\\
1. & \unit{\micro\second}\textsuperscript{r} & gate & 1Q & LD/e & $\, ^{28}\text{Si/}\text{SiO}_2$ & 2D & 2023-01 & \onlinecite{gilbert_-demand_2023} & Fig. 3d, red point\\
\bottomrule
\end{tabular}
\caption{Operation times (part 1). Superscripts stand for the following. \textsuperscript{a}: {EO} qubit. \textsuperscript{b}: \estimated 3 oscillations in 270 ps. \textsuperscript{c}: Hybrid qubit with three electrons in two dots. \textsuperscript{d}: At a partial sweet spot. \textsuperscript{e}: Resonant exchange qubit. \textsuperscript{f}: At a full sweet spot. \textsuperscript{g}: \estimated from the uppermost trace in Fig. 4. \textsuperscript{h}: {RX} qubit. \textsuperscript{i}: Tunable between spin- and charge- qubit. \textsuperscript{j}: \estimated from Fig. 3b the rightmost point. \textsuperscript{k}: Driving through the cavity. \textsuperscript{l}: At 1 K. \textsuperscript{m}: Using $T_\mathrm{R} = 4 \pi \hbar / g \mu_B B_\mathrm{ac}$ with $B_\mathrm{ac} = 2.5$ {mT} and $g=0.44$. \textsuperscript{n}: 2 $\times$ 75 ns. \textsuperscript{o}: We take a representative value from data with strong cross-talk in Rabi frequencies when driving qubits simultaneously. \textsuperscript{p}: \estimated from Fig. 2b. \textsuperscript{q}: Nuclear-driven {EDSR}. \textsuperscript{r}: \estimated from Fig. 3d at $\Delta V_J=10$ {mV}.
\label{operationTime1}
}
\end{table}

%% file: figuresAndTables/table-operationTime-2.txt
\rowcolors{2}{white}{gray!25}
\setcounter{rowcount}{-1}
\begin{table}
\begin{tabular}{@{}S[table-format=2.2, round-mode=figures, round-pad=false, round-precision=2, table-alignment-mode = none, table-number-alignment = center, table-column-width = 3em ]@{}lcccccccc@{ \stepcounter{rowcount} \tiny \therowcount }}
\toprule
\multicolumn{2}{c}{Time} & {Operation} & {\#Qubits} & {Qubit} & {Material} & {Host} & {Date} & {Reference} & {Source}\\
\midrule
1.5 & \unit{\micro\second} & gate & 1Q & LD/e & $\, ^{28}\text{Si/}\text{SiO}_2$ & 2D & 2019-12 & \onlinecite{zhao_single-spin_2019} & Tab. 1\\
1.6 & \unit{\micro\second}\textsuperscript{a} & gate & 1Q & LD/e & $\, ^{28}\text{Si/}\text{SiO}_2$ & 2D & 2014-10 & \onlinecite{veldhorst_addressable_2014} & Fig. 4 and Fig. 2d\\
2.4 & \unit{\micro\second} & gate & 1Q & LD/e & $\, ^{28}\text{Si/}\text{SiO}_2$ & 2D & 2015-10 & \onlinecite{veldhorst_two-qubit_2015} & p5\\
3. & \unit{\micro\second}\textsuperscript{b} & gate & 1Q & LD/e & GaAs/AlGaAs & 2D & 2013-03 & \onlinecite{shafiei_resolving_2013} & p3\\
3. & \unit{\micro\second} & gate & 1Q & LD/e & $\, ^{28}\text{Si/}\text{SiO}_2$ & imp & 2015-03 & \onlinecite{muhonen_quantifying_2015} & Fig. 3c the topmost point\\
5. & \unit{\micro\second} & gate & 1Q & LD/e\textsuperscript{c} & $\text{Si/}\text{SiO}_2$ & 2D & 2020-02 & \onlinecite{leon_coherent_2020} & {SM} Fig. 4a at around 50 {nV}\\
50. & \unit{\micro\second}\textsuperscript{d} & gate & 1Q & LD/e & Si/SiGe & 2D & 2017-04 & \onlinecite{scarlino_dressed_2017} & p7\\
0.2 & \unit{\nano\second} & gate & 1Q & ST/e & GaAs/AlGaAs & 2D & 2014-01 & \onlinecite{higginbotham_coherent_2014} & Fig. 4a and Fig. 3d\\
0.350 & \unit{\nano\second} & gate & 1Q & ST/e & GaAs/AlGaAs & 2D & 2005-09 & \onlinecite{petta_coherent_2005} & p3 and Fig. 5d\\
0.833 & \unit{\nano\second}\textsuperscript{e} & gate & 1Q & ST/e & GaAs/AlGaAs & 2D & 2016-03 & \onlinecite{martins_noise_2016} & Fig. 4d\\
1.00 & \unit{\nano\second} & gate & 1Q & ST/e & GaAs/AlGaAs & 2D & 2020-07 & \onlinecite{jang_robust_2020} & p4\\
1.60 & \unit{\nano\second} & gate & 1Q & ST/e & GaAs/AlGaAs & 2D & 2020-12 & \onlinecite{jang_individual_2020} & Fig. 2a the green trace\\
2.50 & \unit{\nano\second} & gate & 1Q & ST/e & $\, ^{28}\text{Si/}\text{SiO}_2$ & 2D & 2022-02 & \onlinecite{jock_silicon_2022} & p4 and Fig. 2e the rightmost point\\
3.13 & \unit{\nano\second} & gate & 1Q & ST/e & $\, ^{28}\text{Si/SiGe}$ & 2D & 2016-03 & \onlinecite{reed_reduced_2016} & Fig. 4a\\
4.20 & \unit{\nano\second} & gate & 1Q & ST/e & GaAs/AlGaAs & 2D & 2020-08 & \onlinecite{cerfontaine_closed-loop_2020} & p4\\
4.3011 & \unit{\nano\second} & gate & 1Q & ST/e & $\, ^{28}\text{Si/SiGe}$ & 2D & 2024-08 & \onlinecite{song_coherence_2024} & p3\\
5.00 & \unit{\nano\second} & gate & 1Q & ST/e & $\, ^{28}\text{Si/SiGe}$ & 2D & 2023-03 & \onlinecite{weinstein_universal_2023} & {ED} Fig. 1c\\
15. & \unit{\nano\second} & gate & 1Q & ST/e & GaAs/AlGaAs & 2D & 2021-10 & \onlinecite{fedele_simultaneous_2021} & p3\\
36. & \unit{\nano\second} & gate & 1Q & ST/e & Si/SiGe & 2D & 2014-08 & \onlinecite{wu_two-axis_2014} & p3\\
38.2 & \unit{\nano\second} & gate & 1Q & ST/e & GaAs/AlGaAs & 2D & 2021-05 & \onlinecite{kojima_probabilistic_2021} & p3\\
44.2 & \unit{\nano\second}\textsuperscript{f} & gate & 1Q & ST/e\textsuperscript{g} & Si/SiGe & 2D & 2023-01 & \onlinecite{cai_coherent_2023} & p4\\
82.6 & \unit{\nano\second}\textsuperscript{h} & gate & 1Q & ST/e & GaAs/AlGaAs & 2D & 2022-07 & \onlinecite{kim_approaching_2022} & p4\\
87.9 & \unit{\nano\second}\textsuperscript{i} & gate & 1Q & ST/e & GaAs/AlGaAs & 2D & 2023-03 & \onlinecite{yun_probing_2023} & p4\\
91. & \unit{\nano\second} & gate & 1Q & ST/e & Si/SiGe & 2D & 2021-08 & \onlinecite{liu_magnetic-gradient-free_2021} & p3\\
0.1 & \unit{\micro\second} & gate & 1Q & ST/e & Si/SiGe & 2D & 2020-03 & \onlinecite{takeda_resonantly_2020} & p3\\
0.4082 & \unit{\nano\second} & gate & 1Q & LD/h & Ge/Si & 1D & 2023-05 & \onlinecite{liu_ultrafast_2023} & p5\\
0.923 & \unit{\nano\second} & gate & 1Q & LD/h & Ge/Si & 1D & 2022-01 & \onlinecite{wang_ultrafast_2022} & p3\\
1.15 & \unit{\nano\second} & gate & 1Q & LD/h & Ge/Si & 1D & 2021-01 & \onlinecite{froning_ultrafast_2021} & Fig. 4\\
3.40 & \unit{\nano\second}\textsuperscript{j} & gate & 1Q & LD/h\textsuperscript{j} & $\text{Si/}\text{SiO}_2$ & 1D & 2022-03 & \onlinecite{camenzind_spin_2022} & p3 and Fig. S5\\
3.85 & \unit{\nano\second} & gate & 1Q & LD/h & Ge/Si & 1D & 2025-08 & \onlinecite{carballido_compromise-free_2025} & Fig. 1 f\\
5.9 & \unit{\nano\second} & gate & 1Q & LD/h & $\text{Si/}\text{SiO}_2$ & 1D & 2016-11 & \onlinecite{maurand_cmos_2016} & p4\\
7.1 & \unit{\nano\second} & gate & 1Q & LD/h & Ge/Si & 1D & 2018-09 & \onlinecite{watzinger_germanium_2018} & Fig. 5d the point at 11 Db\\
8.8 & \unit{\nano\second} & gate & 1Q & LD/h & Ge/SiGe & 2D & 2020-07 & \onlinecite{hendrickx_single-hole_2020} & p4 and Fig. 3b\\
10. & \unit{\nano\second}\textsuperscript{k} & gate & 1Q & LD/h & Ge/SiGe & 2D & 2023-06 & \onlinecite{lawrie_simultaneous_2023} & Fig. 2g\\
11. & \unit{\nano\second}\textsuperscript{l,j} & gate & 1Q & LD/h\textsuperscript{j} & $\text{Si/}\text{SiO}_2$ & 1D & 2022-03 & \onlinecite{camenzind_spin_2022} & p3 and Fig. 3e\\
18.2 & \unit{\nano\second}\textsuperscript{m} & gate & 1Q & LD/h & $\text{Si/}\text{SiO}_2$ & 1D & 2018-03 & \onlinecite{crippa_electrical_2018} & Fig. 1d\\
20. & \unit{\nano\second} & gate & 1Q & LD/h & Ge/SiGe & 2D & 2020-01 & \onlinecite{hendrickx_fast_2020} & p2\\
33. & \unit{\nano\second} & gate & 1Q & LD/h & $\text{Si/}\text{SiO}_2$ & 1D & 2019-07 & \onlinecite{crippa_gate-reflectometry_2019} & p5 and Fig. 4d\\
50. & \unit{\nano\second} & gate & 1Q & LD/h & $\text{Si/}\text{SiO}_2$ & 1D & 2022-09 & \onlinecite{piot_single_2022} & p7\\
96. & \unit{\nano\second} & gate & 1Q & LD/h & Ge/SiGe & 2D & 2023-06 & \onlinecite{lawrie_simultaneous_2023} & p4\\
98. & \unit{\nano\second}\textsuperscript{n} & gate & 1Q & LD/h & Ge/SiGe & 2D & 2024-07 & \onlinecite{wang_operating_2024} & p2\\
0.1 & \unit{\micro\second}\textsuperscript{o} & gate & 1Q & LD/h & Ge/SiGe & 2D & 2024-02 & \onlinecite{john_bichromatic_2024} & {STab}. 2\\
0.42 & \unit{\micro\second} & gate & 1Q & LD/h & Ge/SiGe & 2D & 2026-09 & \onlinecite{kelly_identification_2026} & p6\\
3.33 & \unit{\nano\second} & gate & 1Q & ST/h & $\text{Si/}\text{SiO}_2$ & 2D & 2024-09 & \onlinecite{liles_singlet-triplet_2024} & p7\\
5.00 & \unit{\nano\second} & gate & 1Q & ST/h & Ge/SiGe & 2D & 2021-06 & \onlinecite{jirovec_singlet-triplet_2021} & p2 and Fig. 3e\\
7.1 & \unit{\nano\second} & gate & 1Q & ST/h & Ge/SiGe & 2D & 2021-06 & \onlinecite{jirovec_singlet-triplet_2021} & Fig. S15\\
0.165 & \unit{\micro\second}\textsuperscript{p} & gate & 1Q & ST/h & Ge/SiGe & 2D & 2026-01 & \onlinecite{tsoukalas_dressed_2026} & p3\\
0.15 & \unit{\micro\second} & gate & 1Q & LD/i & Si:P & imp & 2012-09 & \onlinecite{pla_single-atom_2012} & p2 and Fig. 2\\
0.31 & \unit{\micro\second} & gate & 1Q & LD/i & Si:P & imp & 2023-11 & \onlinecite{kranz_exploiting_2023} & Fig. 4c and Tab. I\\
0.829 & \unit{\micro\second} & gate & 1Q & LD/i & $\, ^{28}\text{Si:P}$ & imp & 2022-01 & \onlinecite{madzik_precision_2022} & {ED} Fig. 3 first column\\
\bottomrule
\end{tabular}
\caption{Operation times (part 2). Superscripts stand for the following. \textsuperscript{a}: \estimated from Fig. 2d: 9 oscillations in 30 \unit{\micro\second} at resonance. \textsuperscript{b}: {EDSR} without a micromagnet. \textsuperscript{c}: In a five-electron dot configuration. \textsuperscript{d}: Valley and spin flip. \textsuperscript{e}: \estimated from Fig. 4d the rightmost red point. \textsuperscript{f}: The reference calls the observed signal `Rabi oscillations' even though there is no ac field; these oscillations arise as a coherent precession upon pulsing the dot detuning. \textsuperscript{g}: $S$-$T_-$ qubit. \textsuperscript{h}: With estimation and/or feedback. \textsuperscript{i}: $Q_R$. \textsuperscript{j}: At 1.5 K. \textsuperscript{k}: \estimated Fig. 2g the leftmost point. \textsuperscript{l}: Z-gate. \textsuperscript{m}: \derived from Fig. 1d showing 11 oscillations in 400 ns. \textsuperscript{n}: $Q_A$. \textsuperscript{o}: Bichromatic driving. The value is the average of the three bichromatic Rabi frequencies given in {STab}. 2. \textsuperscript{p}: $X_{\pi/2}$.
\label{operationTime2}
}
\end{table}

%% file: figuresAndTables/table-operationTime-3.txt
\rowcolors{2}{white}{gray!25}
\setcounter{rowcount}{-1}
\begin{table}
\begin{tabular}{@{}S[table-format=2.2, round-mode=figures, round-pad=false, round-precision=2, table-alignment-mode = none, table-number-alignment = center, table-column-width = 3em ]@{}lcccccccc@{ \stepcounter{rowcount} \tiny \therowcount }}
\toprule
\multicolumn{2}{c}{Time} & {Operation} & {\#Qubits} & {Qubit} & {Material} & {Host} & {Date} & {Reference} & {Source}\\
\midrule
1.64 & \unit{\micro\second} & gate & 1Q & LD/i & $\, ^{28}\text{Si:P}$ & imp & 2024-02 & \onlinecite{reiner_high-fidelity_2024} & p5\\
2.9143 & \unit{\micro\second} & gate & 1Q & LD/i & Si:P & imp & 2025-02 & \onlinecite{thorvaldson_grovers_2025} & Fig. S2\\
3. & \unit{\micro\second} & gate & 1Q & LD/i & $\, ^{28}\text{Si:P}$ & imp & 2016-10 & \onlinecite{dehollain_optimization_2016} & p3\\
4.219 & \unit{\micro\second}\textsuperscript{a} & gate & 1Q & LD/i & $\, ^{28}\text{Si:P}$ & imp & 2023-02 & \onlinecite{savytskyy_electrically_2023} & p6\\
0.16 & \unit{\milli\second} & gate & 1Q & LD/i\textsuperscript{b} & $\, ^{28}\text{Si:P}$ & imp & 2016-10 & \onlinecite{laucht_dressed_2016} & p3 and Fig. 3d\\
74. & \unit{\pico\second} & gate & 2Q & charge & Si/SiGe & 2D & 2020-09 & \onlinecite{macquarrie_progress_2020} & p4\\
10. & \unit{\nano\second}\textsuperscript{c} & gate & 2Q & LD/e & GaAs/AlGaAs & 2D & 2011-09 & \onlinecite{brunner_two-qubit_2011} & p4\\
40. & \unit{\nano\second} & gate & 2Q & LD/e & $\, ^{28}\text{Si/SiGe}$ & 2D & 2022-04 & \onlinecite{mills_two-qubit_2022} & Fig. 2B caption\\
42.4 & \unit{\nano\second} & gate & 2Q & LD/e & $\, ^{28}\text{Si/SiGe}$ & 2D & 2024-12 & \onlinecite{dijkema_cavity-mediated_2024} & Fig. 3d\\
67. & \unit{\nano\second}\textsuperscript{d,e} & gate & 2Q & LD/e & $\, ^{28}\text{Si/SiGe}$ & 2D & 2022-11 & \onlinecite{petit_design_2022} & Tab. 1\\
89. & \unit{\nano\second}\textsuperscript{f,e} & gate & 2Q & LD/e & $\, ^{28}\text{Si/SiGe}$ & 2D & 2022-11 & \onlinecite{petit_design_2022} & Tab. 1\\
0.100 & \unit{\micro\second}\textsuperscript{g} & gate & 2Q & LD/e & Si/SiGe & 2D & 2022-08 & \onlinecite{takeda_quantum_2022} & p6\\
0.100 & \unit{\micro\second} & gate & 2Q & LD/e & $\, ^{28}\text{Si/SiGe}$ & 2D & 2022-01 & \onlinecite{xue_quantum_2022} & p6\\
0.100 & \unit{\micro\second}\textsuperscript{f} & gate & 2Q & LD/e\textsuperscript{h} & $\, ^{28}\text{Si/}\text{SiO}_2$ & 2D & 2024-09 & \onlinecite{hansen_entangling_2024} & p3\\
0.130 & \unit{\micro\second} & gate & 2Q & LD/e & Si/SiGe & 2D & 2017-12 & \onlinecite{zajac_resonantly_2017} & p2\\
0.159 & \unit{\micro\second} & gate & 2Q & LD/e & $\, ^{28}\text{Si/}\text{SiO}_2$ & 2D & 2015-10 & \onlinecite{veldhorst_two-qubit_2015} & p4\\
0.2 & \unit{\micro\second}\textsuperscript{i} & gate & 2Q & LD/e & Si/SiGe & 2D & 2018-02 & \onlinecite{watson_programmable_2018} & {ED} Fig. 9\\
0.2 & \unit{\micro\second}\textsuperscript{j,e} & gate & 2Q & LD/e & $\, ^{28}\text{Si/}\text{SiO}_2$ & 2D & 2024-03 & \onlinecite{huang_high-fidelity_2024} & {ED} Fig. 8f\\
0.270 & \unit{\micro\second} & gate & 2Q & LD/e & $\, ^{28}\text{Si/SiGe}$ & 2D & 2019-06 & \onlinecite{sigillito_site-selective_2019} & Fig. 4d\\
0.302 & \unit{\micro\second} & gate & 2Q & LD/e & $\, ^{28}\text{Si/SiGe}$ & 2D & 2019-11 & \onlinecite{sigillito_coherent_2019} & p5\\
0.53 & \unit{\micro\second} & gate & 2Q & LD/e & Si/SiGe & 2D & 2020-03 & \onlinecite{yoneda_quantum_2020} & p3\\
0.660 & \unit{\micro\second}\textsuperscript{k,e} & gate & 2Q & LD/e & $\, ^{28}\text{Si/SiGe}$ & 2D & 2022-11 & \onlinecite{petit_design_2022} & Tab. 1\\
3. & \unit{\micro\second} & gate & 2Q & LD/e & $\text{Si/}\text{SiO}_2$ & 2D & 2021-05 & \onlinecite{leon_bell-state_2021} & Fig. 2j the rightmost point\\
2. & \unit{\nano\second} & gate & 2Q & ST/e & GaAs/AlGaAs & 2D & 2019-03 & \onlinecite{malinowski_fast_2019} & Fig. 2c\\
5.5 & \unit{\nano\second}\textsuperscript{l} & gate & 2Q & ST/e & GaAs/AlGaAs & 2D & 2018-11 & \onlinecite{noiri_fast_2018} & p4 and Fig. 3\\
0.140 & \unit{\micro\second} & gate & 2Q & ST/e & GaAs/AlGaAs & 2D & 2012-04 & \onlinecite{shulman_demonstration_2012} & p3\\
9. & \unit{\nano\second} & gate & 2Q & LD/h & Ge/SiGe & 2D & 2021-03 & \onlinecite{hendrickx_four-qubit_2021} & Tab. S2\\
24. & \unit{\nano\second}\textsuperscript{m} & gate & 2Q & LD/h & $\text{Si/}\text{SiO}_2$ & 1D & 2024-05 & \onlinecite{geyer_anisotropic_2024} & p7 and Fig. 4b\\
55. & \unit{\nano\second} & gate & 2Q & LD/h & Ge/SiGe & 2D & 2020-01 & \onlinecite{hendrickx_fast_2020} & p4\\
7.6 & \unit{\nano\second}\textsuperscript{n,o} & gate & 2Q & ST/h & Ge/SiGe & 2D & 2024-10 & \onlinecite{zhang_universal_2024} & p6\\
0.800 & \unit{\nano\second} & gate & 2Q & LD/i & Si:P & imp & 2019-07 & \onlinecite{he_two-qubit_2019} & Fig. 3 and abstract\\
5.6 & \unit{\micro\second}\textsuperscript{p} & gate & 2Q & LD/i & $\, ^{28}\text{Si:P}$ & imp & 2021-01 & \onlinecite{madzik_conditional_2021} & Fig. 5c\\
0.460 & \unit{\micro\second}\textsuperscript{q} & gate & 3Q & LD/e & Si/SiGe & 2D & 2022-08 & \onlinecite{takeda_quantum_2022} & p7\\
0.400 & \unit{\micro\second}\textsuperscript{r} & measure & 1Q & charge & InAs & 1D & 2015-07 & \onlinecite{stehlik_fast_2015} & p5\\
1. & \unit{\micro\second}\textsuperscript{s} & measure & 1Q & charge & $\text{Si/}\text{SiO}_2$ & 2D & 2020-02 & \onlinecite{schaal_fast_2020} & p5 and Fig. 4c the leftmost green point\\
0.980 & \unit{\micro\second}\textsuperscript{t} & measure & 1Q & HY/e\textsuperscript{u} & $\, ^{28}\text{Si/SiGe}$ & 2D & 2022-03 & \onlinecite{blumoff_fast_2022} & p4\\
25. & \unit{\micro\second}\textsuperscript{v} & measure & 1Q & HY/e & $\, ^{28}\text{Si/SiGe}$ & 2D & 2024-10 & \onlinecite{acuna_coherent_2024} & p2\\
0.172 & \unit{\milli\second}\textsuperscript{w} & measure & 1Q & HY/e & GaAs/AlGaAs & 2D & 2021-06 & \onlinecite{jang_single-shot_2021} & p3\\
0.4 & \unit{\micro\second} & measure & 1Q & LD/e & GaAs/AlGaAs & 2D & 2003-11 & \onlinecite{hanson_zeeman_2003} & Fig. 3\\
1.500 & \unit{\micro\second}\textsuperscript{x} & measure & 1Q & LD/e & InAs & 1D & 2024-05 & \onlinecite{pita-vidal_strong_2024} & Fig. S19\\
24. & \unit{\micro\second}\textsuperscript{y} & measure & 1Q & LD/e & Si/SiGe & 2D & 2019-08 & \onlinecite{volk_fast_2019} & p7\\
0.250 & \unit{\milli\second} & measure & 1Q & LD/e & $\text{Si/}\text{SiO}_2$ & 1D & 2023-02 & \onlinecite{oakes_fast_2023} & p5 and Fig. 2j\\
0.670 & \unit{\milli\second} & measure & 1Q & LD/e & $\, ^{28}\text{Si/SiGe}$ & 2D & 2022-12 & \onlinecite{mills_high-fidelity_2022} & p4\\
0.700 & \unit{\milli\second} & measure & 1Q & LD/e & GaAs/AlGaAs & 2D & 2019-04 & \onlinecite{nakajima_quantum_2019} & Fig. 1f\\
1. & \unit{\milli\second} & measure & 1Q & LD/e & $\text{Si/}\text{SiO}_2$ & 1D & 2022-03 & \onlinecite{spence_spin-valley_2022} & p2\\
4. & \unit{\milli\second} & measure & 1Q & LD/e & Si/SiGe & 2D & 2014-08 & \onlinecite{kawakami_electrical_2014} & p1\\
49. & \unit{\milli\second}\textsuperscript{z} & measure & 1Q & LD/e & Si/SiGe & 2D & 2020-04 & \onlinecite{xue_repetitive_2020} & p3\\
0.200 & \unit{\second} & measure & 1Q & LD/e & Si/SiGe & 2D & 2011-04 & \onlinecite{simmons_tunable_2011} & Fig. 2 caption\\
50. & \unit{\nano\second}\textsuperscript{$\alpha$} & measure & 1Q & ST/e & $\text{Si/}\text{SiO}_2$ & 1D & 2021-05 & \onlinecite{ibberson_large_2021} & p5 and Fig. 3\\
0.100 & \unit{\micro\second}\textsuperscript{$\beta$} & measure & 1Q & ST/e & GaAs/AlGaAs & 2D & 2010-04 & \onlinecite{barthel_fast_2010} & p3 and Fig. 3\\
\bottomrule
\end{tabular}
\caption{Operation times (part 3). Superscripts stand for the following. \textsuperscript{a}: Hybrid electron-nuclear qubit. \textsuperscript{b}: Qubit defined in the rotating frame. \textsuperscript{c}: {SWAP} time. \textsuperscript{d}: {CPHASE}. \textsuperscript{e}: At 1 K. \textsuperscript{f}: {SWAP}. \textsuperscript{g}: {CZ} gate. \textsuperscript{h}: Qubit dressed by constant driving. \textsuperscript{i}: The reference gives also the `decay time of the {CZ} operation' of 1.6 \unit{\micro\second} in {ED} Fig. 9, which would correspond to a gate quality factor of 4.8. \textsuperscript{j}: Representative value. \textsuperscript{k}: {CROT}. \textsuperscript{l}: Gate between a single-spin and a singlet-triplet qubit. \textsuperscript{m}: Resonantly driven {CROT}. \textsuperscript{n}: {SWAP}-like gate. \textsuperscript{o}: Average of the given range 22-112 {MHz}. \textsuperscript{p}: Fidelity not measured; signal visibility was 50 \unit{\percent}. \textsuperscript{q}: Toffoli gate. \textsuperscript{r}: The value is for detecting an interdot charge transition with {SNR} 76. \textsuperscript{s}: Dot-donor transition; $\mathrm{{SNR}}\approx5$. \textsuperscript{t}: $\mathrm{{SNR}}=6.5$. \textsuperscript{u}: {EO} qubit. \textsuperscript{v}: {SPAM} fidelity. \textsuperscript{w}: 32 \unit{\micro\second}+140 \unit{\micro\second}. \textsuperscript{x}: Q1. \textsuperscript{y}: $\mathrm{{SNR}}=3.4$. \textsuperscript{z}: 15 {QND} measurements, each taking 3.263 ms. \textsuperscript{$\alpha$}: $\mathrm{{SNR}}=3.3$ for an interdot charge transition `mimicking the singlet-triplet readout scheme'. \textsuperscript{$\beta$}: $\mathrm{{SNR}}\sim$3.
\label{operationTime3}
}
\end{table}

%% file: figuresAndTables/table-operationTime-4.txt
\rowcolors{2}{white}{gray!25}
\setcounter{rowcount}{-1}
\begin{table}
\begin{tabular}{@{}S[table-format=2.2, round-mode=figures, round-pad=false, round-precision=2, table-alignment-mode = none, table-number-alignment = center, table-column-width = 3em ]@{}lcccccccc@{ \stepcounter{rowcount} \tiny \therowcount }}
\toprule
\multicolumn{2}{c}{Time} & {Operation} & {\#Qubits} & {Qubit} & {Material} & {Host} & {Date} & {Reference} & {Source}\\
\midrule
0.200 & \unit{\micro\second}\textsuperscript{a} & measure & 1Q & ST/e & GaAs/AlGaAs & 2D & 2014-01 & \onlinecite{higginbotham_coherent_2014} & {SM} Fig. S1d\\
0.800 & \unit{\micro\second} & measure & 1Q & ST/e & Si/SiGe & 2D & 2020-02 & \onlinecite{connors_rapid_2020} & Fig. 5b\\
0.8 & \unit{\micro\second} & measure & 1Q & ST/e & GaAs/AlGaAs & 2D & 2018-04 & \onlinecite{mukhopadhyay_2x2_2018} & p4\\
0.8 & \unit{\micro\second}\textsuperscript{b} & measure & 1Q & ST/e & Si/SiGe & 2D & 2020-01 & \onlinecite{noiri_radio-frequency-detected_2020} & Fig. 3c\\
1. & \unit{\micro\second} & measure & 1Q & ST/e & GaAs/AlGaAs & 2D & 2012-04 & \onlinecite{shulman_demonstration_2012} & p1\\
1. & \unit{\micro\second} & measure & 1Q & ST/e & GaAs/AlGaAs & 2D & 2014-10 & \onlinecite{shulman_suppressing_2014} & p2\\
1.65 & \unit{\micro\second} & measure & 1Q & ST/e & Si/SiGe & 2D & 2020-02 & \onlinecite{connors_rapid_2020} & p6\\
2. & \unit{\micro\second} & measure & 1Q & ST/e & GaAs/AlGaAs & 2D & 2018-09 & \onlinecite{orona_readout_2018} & Fig. 4b the rightmost point\\
2.4 & \unit{\micro\second} & measure & 1Q & ST/e & $\, ^{28}\text{Si/SiGe}$ & 2D & 2024-02 & \onlinecite{takeda_rapid_2024} & p5 and Fig. 3d\\
4. & \unit{\micro\second} & measure & 1Q & ST/e & GaAs/AlGaAs & 2D & 2017-07 & \onlinecite{nakajima_robust_2017} & p2 and {SM} p1\\
5.6 & \unit{\micro\second} & measure & 1Q & ST/e & $\text{Si/}\text{SiO}_2$ & 1D & 2023-02 & \onlinecite{oakes_fast_2023} & p6 and Fig. 4d\\
6. & \unit{\micro\second} & measure & 1Q & ST/e & Si/SiGe & 2D & 2019-08 & \onlinecite{zheng_rapid_2019} & p3 and Fig. 3f\\
6. & \unit{\micro\second} & measure & 1Q & ST/e & GaAs/AlGaAs & 2D & 2009-10 & \onlinecite{barthel_rapid_2009} & p3\\
10. & \unit{\micro\second}\textsuperscript{c} & measure & 1Q & ST/e & GaAs/AlGaAs & 2D & 2021-10 & \onlinecite{fedele_simultaneous_2021} & p3\\
10. & \unit{\micro\second} & measure & 1Q & ST/e & GaAs/AlGaAs & 2D & 2005-09 & \onlinecite{petta_coherent_2005} & p2\\
12.5 & \unit{\micro\second}\textsuperscript{d} & measure & 1Q & ST/e & Si/SiGe & 2D & 2019-07 & \onlinecite{jones_spin-blockade_2019} & p3 and p6\\
20. & \unit{\micro\second} & measure & 1Q & ST/e & GaAs/AlGaAs & 2D & 2006-11 & \onlinecite{meunier_nondestructive_2006} & p2\\
20. & \unit{\micro\second}\textsuperscript{e} & measure & 1Q & ST/e & $\text{Si/}\text{SiO}_2$ & 2D & 2022-12 & \onlinecite{niegemann_parity_2022} & p4 and Fig. 3\\
24. & \unit{\micro\second} & measure & 1Q & ST/e & $\text{Si/}\text{SiO}_2$ & 2D & 2026-07 & \onlinecite{laine_high-fidelity_2026} & p5 and Fig. 4c\\
33. & \unit{\micro\second}\textsuperscript{f} & measure & 1Q & ST/e & $\text{Si/}\text{SiO}_2$ & 1D & 2023-03 & \onlinecite{ansaloni_gate_2023} & p6 and Fig. 7e\\
0.10 & \unit{\milli\second} & measure & 1Q & ST/e & $\, ^{28}\text{Si/SiGe}$ & 2D & 2021-04 & \onlinecite{borjans_spin_2021} & p5\\
0.10 & \unit{\milli\second}\textsuperscript{g} & measure & 1Q & ST/e & $\, ^{28}\text{Si/}\text{SiO}_2$ & 2D & 2024-08 & \onlinecite{tanttu_assessment_2024} & {ED} Tab. {II}\\
0.10 & \unit{\milli\second} & measure & 1Q & ST/e & $\, ^{28}\text{Si/}\text{SiO}_2$ & 2D & 2025-04 & \onlinecite{steinacker_bell_2025} & p3\\
0.216 & \unit{\milli\second} & measure & 1Q & ST/e & GaAs/AlGaAs & 2D & 2021-02 & \onlinecite{jadot_distant_2021} & Fig. 2a caption\\
0.340 & \unit{\milli\second} & measure & 1Q & ST/e & $\text{Si/}\text{SiO}_2$ & 2D & 2026-07 & \onlinecite{laine_high-fidelity_2026} & p5 and Fig. 4c\\
0.5 & \unit{\milli\second} & measure & 1Q & ST/e & $\text{Si/}\text{SiO}_2$ & 1D & 2019-05 & \onlinecite{urdampilleta_gate-based_2019} & abstract\\
1.18 & \unit{\milli\second} & measure & 1Q & ST/e & $\text{Si/}\text{SiO}_2$ & 2D & 2024-03 & \onlinecite{ma_singlet-triplet-state_2024} & p4\\
2. & \unit{\milli\second} & measure & 1Q & ST/e & GaAs/AlGaAs & 2D & 2016-11 & \onlinecite{kiyama_single-shot_2016} & p3\\
2.6 & \unit{\milli\second} & measure & 1Q & ST/e & $\text{Si/}\text{SiO}_2$ & 2D & 2019-03 & \onlinecite{west_gate-based_2019} & p3\\
4. & \unit{\milli\second}\textsuperscript{h} & measure & 1Q & ST/e & GaAs/AlGaAs & 2D & 2010-07 & \onlinecite{petersson_charge_2010} & p4\\
20. & \unit{\milli\second} & measure & 1Q & ST/e & GaAs/AlGaAs & 2D & 2016-11 & \onlinecite{kiyama_single-shot_2016} & p3\\
10. & \unit{\micro\second} & measure & 1Q & LD/h & Ge/SiGe & 2D & 2023-06 & \onlinecite{lawrie_simultaneous_2023} & Fig. 1b caption\\
6. & \unit{\micro\second} & measure & 1Q & ST/h & Ge/SiGe & 2D & 2020-07 & \onlinecite{hendrickx_single-hole_2020} & p4\\
0.175 & \unit{\micro\second} & measure & 1Q & LD/i & Si:P & imp & 2025-04 & \onlinecite{geng_high-fidelity_2025} & Fig. 4c\\
1.5 & \unit{\micro\second} & measure & 1Q & LD/i & Si:P & imp & 2019-10 & \onlinecite{keith_single-shot_2019} & p4\\
1.5 & \unit{\micro\second}\textsuperscript{i} & measure & 1Q & LD/i & Si:P & imp & 2025-04 & \onlinecite{geng_high-fidelity_2025} & Fig. 4e\\
67. & \unit{\micro\second}\textsuperscript{j} & measure & 1Q & LD/i & Si:P & imp & 2023-02 & \onlinecite{hogg_single-shot_2023} & Tab. I\\
3. & \unit{\milli\second} & measure & 1Q & LD/i & Si:P & imp & 2015-10 & \onlinecite{watson_high-fidelity_2015} & p4\\
40. & \unit{\milli\second} & measure & 1Q & LD/i & Si:P & imp & 2013-06 & \onlinecite{buch_spin_2013} & Fig. 2b caption\\
50. & \unit{\milli\second}\textsuperscript{k} & measure & 1Q & LD/i & Si:P & imp & 2018-03 & \onlinecite{broome_two-electron_2018} & {SFig}. 2d\\
0.150 & \unit{\second}\textsuperscript{l} & measure & 1Q & LD/i & Si:P & imp & 2015-10 & \onlinecite{watson_high-fidelity_2015} & Fig. 2b-c\\
0.200 & \unit{\second} & measure & 1Q & LD/i & Si:P & imp & 2017-03 & \onlinecite{watson_atomically_2017} & p3\\
65. & \unit{\micro\second}\textsuperscript{m} & measure & 1Q & ST/i\textsuperscript{n} & Si:P & imp & 2018-05 & \onlinecite{harvey-collard_high-fidelity_2018} & Tab. {II}\\
0.150 & \unit{\milli\second} & measure & 1Q & ST/i\textsuperscript{n} & Si:P & imp & 2018-05 & \onlinecite{harvey-collard_high-fidelity_2018} & Tab. {II}\\
0.3 & \unit{\milli\second} & measure & 1Q & ST/i & Si:P & imp & 2018-11 & \onlinecite{pakkiam_single-shot_2018} & p2\\
18.4 & \unit{\milli\second}\textsuperscript{o} & measure & 1Q & ST/i & Si:P & imp & 2017-07 & \onlinecite{broome_high-fidelity_2017} & p3\\
50. & \unit{\micro\second}\textsuperscript{p} & initialize & 1Q & LD/e & Si/SiGe & 2D & 2018-02 & \onlinecite{watson_programmable_2018} & {ED} Fig. 2\\
0.150 & \unit{\milli\second}\textsuperscript{q,r} & initialize & 1Q & LD/e & $\, ^{28}\text{Si/}\text{SiO}_2$ & 2D & 2024-03 & \onlinecite{huang_high-fidelity_2024} & p3 and {ED} Fig. 3c-d\\
4. & \unit{\milli\second} & initialize & 1Q & LD/e & Si/SiGe & 2D & 2014-08 & \onlinecite{kawakami_electrical_2014} & p1\\
20. & \unit{\nano\second}\textsuperscript{s} & initialize & 1Q & ST/e & GaAs/AlGaAs & 2D & 2014-10 & \onlinecite{shulman_suppressing_2014} & p2\\
\bottomrule
\end{tabular}
\caption{Operation times (part 4). Superscripts stand for the following. \textsuperscript{a}: $\mathrm{{SNR}}=4.1$. \textsuperscript{b}: $\mathrm{{SNR}}=6$. \textsuperscript{c}: $\mathrm{{SNR}}=5$. \textsuperscript{d}: $\mathrm{{SNR}}=6.5$. \textsuperscript{e}: Measurement distinguishes $S_0$ and $T_0$ versus $T_-$ and $T_+$. \textsuperscript{f}: Detecting single electron transfer between neighboring dots. \textsuperscript{g}: Device B. \textsuperscript{h}: The reference estimates this time as required for $\mathrm{{SNR}}=1$. \textsuperscript{i}: At 3.7 K. \textsuperscript{j}: We take 1/bandwidth=1/15 {kHz}. \textsuperscript{k}: \estimated as the duration of phase 3 in {SFig}. 2d. \textsuperscript{l}: \estimated from Fig. 2b-c. \textsuperscript{m}: Latched readout. \textsuperscript{n}: Hybrid {ST} qubit: impurity -- gated dot. \textsuperscript{o}: Latched readout; $\mathrm{{SNR}}=6.4$. \textsuperscript{p}: Using spin hot-spot. \textsuperscript{q}: We take the average of the given range 100 - 200 $\mu$s. \textsuperscript{r}: At 1 K. \textsuperscript{s}: Into singlet.
\label{operationTime4}
}
\end{table}

%% file: figuresAndTables/table-operationTime-5.txt
\rowcolors{2}{white}{gray!25}
\setcounter{rowcount}{-1}
\begin{table}
\begin{tabular}{@{}S[table-format=2.2, round-mode=figures, round-pad=false, round-precision=2, table-alignment-mode = none, table-number-alignment = center, table-column-width = 3em ]@{}lcccccccc@{ \stepcounter{rowcount} \tiny \therowcount }}
\toprule
\multicolumn{2}{c}{Time} & {Operation} & {\#Qubits} & {Qubit} & {Material} & {Host} & {Date} & {Reference} & {Source}\\
\midrule
50. & \unit{\nano\second} & initialize & 1Q & ST/e & GaAs/AlGaAs & 2D & 2012-04 & \onlinecite{shulman_demonstration_2012} & p1\\
2. & \unit{\milli\second} & initialize & 1Q & ST/e & Si/SiGe & 2D & 2025-09 & \onlinecite{park_single_2025} & p3\\
20. & \unit{\milli\second}\textsuperscript{a} & initialize & 1Q & ST/e & $\text{Si/}\text{SiO}_2$ & 1D & 2019-05 & \onlinecite{urdampilleta_gate-based_2019} & Fig. 3\\
10. & \unit{\milli\second} & initialize & 1Q & LD/i & $\, ^{28}\text{Si:P}$ & imp & 2022-10 & \onlinecite{johnson_beating_2022} & p6\\
\bottomrule
\end{tabular}
\caption{Operation times (part 5). Superscripts stand for the following. \textsuperscript{a}: Into triplet.
\label{operationTime5}
}
\end{table}

%% file: figuresAndTables/table-operationFidelity-1.txt
\rowcolors{2}{white}{gray!25}
\setcounter{rowcount}{-1}
\begin{table}
\begin{tabular}{@{}S[table-format=2.4\unit{\percent}\textsuperscript{x}, table-align-text-after=true, round-mode=none, round-pad=false, round-precision=0]cccccccc@{ \stepcounter{rowcount} \tiny \therowcount }}
\toprule
{Fidelity} & {Operation} & {\#Qubits} & {Qubit} & {Material} & {Host} & {Date} & {Reference} & {Source}\\
\midrule
99.89 \unit{\percent}\textsuperscript{a} & gate & 1Q & HY/e & $\, ^{28}\text{Si/SiGe}$ & 2D & 2023-03 & \onlinecite{weinstein_universal_2023} & p8 and {ED} Fig. 2a\\
99.88 \unit{\percent}\textsuperscript{b} & gate & 1Q & HY/e & Si/SiGe & 2D & 2021-11 & \onlinecite{ha_flexible_2021} & p4\\
99.84 \unit{\percent} & gate & 1Q & HY/e & $\, ^{28}\text{Si/SiGe}$ & 2D & 2024-10 & \onlinecite{acuna_coherent_2024} & abstract and p4\\
99.65 \unit{\percent} & gate & 1Q & HY/e\textsuperscript{c} & $\, ^{28}\text{Si/SiGe}$ & 2D & 2019-07 & \onlinecite{andrews_quantifying_2019} & abstract\\
94.5 \unit{\percent}\textsuperscript{d} & gate & 1Q & HY/e & Si/SiGe & 2D & 2015-12 & \onlinecite{kim_high-fidelity_2015} & p5 and Fig. 4e and g\\
91. \unit{\percent}\textsuperscript{e} & gate & 1Q & HY/e\textsuperscript{f} & Si/SiGe & 2D & 2014-07 & \onlinecite{kim_quantum_2014} & p3\\
85. \unit{\percent} & gate & 1Q & HY/e\textsuperscript{g} & Si/SiGe & 2D & 2019-11 & \onlinecite{penthorn_two-axis_2019} & p4\\
99.96 \unit{\percent}\textsuperscript{h} & gate & 1Q & LD/e & $\, ^{28}\text{Si/SiGe}$ & 2D & 2022-09 & \onlinecite{philips_universal_2022} & Fig. 2d\\
99.957 \unit{\percent} & gate & 1Q & LD/e & $\, ^{28}\text{Si/}\text{SiO}_2$ & 2D & 2019-04 & \onlinecite{yang_silicon_2019} & abstract\\
99.956 \unit{\percent} & gate & 1Q & LD/e & $\, ^{28}\text{Si/SiGe}$ & 2D & 2022-12 & \onlinecite{mills_high-fidelity_2022} & p5 and Fig. 4c\\
99.95 \unit{\percent} & gate & 1Q & LD/e & $\, ^{28}\text{Si/}\text{SiO}_2$ & imp & 2015-03 & \onlinecite{muhonen_quantifying_2015} & p6\\
99.93 \unit{\percent} & gate & 1Q & LD/e & $\, ^{28}\text{Si/}\text{SiO}_2$ & 2D & 2023-01 & \onlinecite{gilbert_-demand_2023} & p5 and Fig. 4\\
99.926 \unit{\percent} & gate & 1Q & LD/e & $\, ^{28}\text{Si/SiGe}$ & 2D & 2017-12 & \onlinecite{yoneda_quantum-dot_2017} & p4 and Fig. 5\\
99.91 \unit{\percent} & gate & 1Q & LD/e & $\, ^{28}\text{Si/}\text{SiO}_2$ & 2D & 2018-10 & \onlinecite{chan_assessment_2018} & p3\\
99.91 \unit{\percent} & gate & 1Q & LD/e & Si/SiGe & 2D & 2021-06 & \onlinecite{takeda_quantum_2021} & p1 for Q3\\
99.906 \unit{\percent}\textsuperscript{i} & gate & 1Q & LD/e & $\, ^{28}\text{Si/SiGe}$ & 2D & 2022-09 & \onlinecite{noiri_shuttling-based_2022} & p3 and Fig. 1f\\
99.85 \unit{\percent}\textsuperscript{j,k} & gate & 1Q & LD/e & $\, ^{28}\text{Si/SiGe}$ & 2D & 2023-10 & \onlinecite{undseth_hotter_2023} & p9\\
99.85 \unit{\percent}\textsuperscript{l,m} & gate & 1Q & LD/e & $\, ^{28}\text{Si/}\text{SiO}_2$ & 2D & 2024-03 & \onlinecite{huang_high-fidelity_2024} & p5 and Fig. 3d-e\\
99.84 \unit{\percent} & gate & 1Q & LD/e & $\, ^{28}\text{Si/SiGe}$ & 2D & 2022-01 & \onlinecite{noiri_fast_2022} & p4 and Fig. 2c\\
99.80 \unit{\percent} & gate & 1Q & LD/e & $\, ^{28}\text{Si/SiGe}$ & 2D & 2025-03 & \onlinecite{katiraee-far_unified_2025} & p6\\
99.8 \unit{\percent} & gate & 1Q & LD/e & $\, ^{28}\text{Si/SiGe}$ & 2D & 2018-10 & \onlinecite{takeda_optimized_2018} & p5\\
99.73 \unit{\percent}\textsuperscript{n} & gate & 1Q & LD/e & $\, ^{28}\text{Si/SiGe}$ & 2D & 2022-04 & \onlinecite{mills_two-qubit_2022} & p1 and Fig. 1C\\
99.72 \unit{\percent}\textsuperscript{o} & gate & 1Q & LD/e & Si/SiGe & 2D & 2022-08 & \onlinecite{takeda_quantum_2022} & {ED} Fig. 2f-h\\
99.72 \unit{\percent} & gate & 1Q & LD/e & $\, ^{28}\text{Si/SiGe}$ & 2D & 2022-01 & \onlinecite{xue_quantum_2022} & p2\\
99.7 \unit{\percent} & gate & 1Q & LD/e & Si/SiGe & 2D & 2017-12 & \onlinecite{zajac_resonantly_2017} & p2\\
99.7 \unit{\percent} & gate & 1Q & LD/e\textsuperscript{p} & $\text{Si/}\text{SiO}_2$ & 2D & 2020-02 & \onlinecite{leon_coherent_2020} & {SM} Fig. 3\\
99.69 \unit{\percent} & gate & 1Q & LD/e\textsuperscript{q} & Si/SiGe & 2D & 2021-05 & \onlinecite{xue_cmos-based_2021} & p4\\
99.65 \unit{\percent}\textsuperscript{r} & gate & 1Q & LD/e & $\, ^{28}\text{Si/}\text{SiO}_2$ & 2D & 2024-08 & \onlinecite{tanttu_assessment_2024} & {ED} Tab. {II}\\
99.6 \unit{\percent} & gate & 1Q & LD/e & $\, ^{28}\text{Si/}\text{SiO}_2$ & 2D & 2014-10 & \onlinecite{veldhorst_addressable_2014} & p2 and Fig. 4\\
99.6 \unit{\percent} & gate & 1Q & LD/e & Si/SiGe & 2D & 2016-08 & \onlinecite{takeda_fault-tolerant_2016} & p4\\
99.54 \unit{\percent}\textsuperscript{s} & gate & 1Q & LD/e & $\, ^{28}\text{Si/SiGe}$ & 2D & 2025-07 & \onlinecite{unseld_baseband_2025} & p3 and Fig.2\\
99.51 \unit{\percent}\textsuperscript{t} & gate & 1Q & LD/e & $\, ^{28}\text{Si/SiGe}$ & 2D & 2026-05 & \onlinecite{matsumoto_two-qubit_2026} & Fig. S2d\\
99.5 \unit{\percent} & gate & 1Q & LD/e & $\, ^{28}\text{Si/}\text{SiO}_2$ & 2D & 2019-05 & \onlinecite{huang_fidelity_2019} & {ED} Tab. 2\\
99.50 \unit{\percent}\textsuperscript{u} & gate & 1Q & LD/e & $\, ^{28}\text{Si/SiGe}$ & 2D & 2025-07 & \onlinecite{unseld_baseband_2025} & abstract and Fig.4f\\
99.47 \unit{\percent}\textsuperscript{v} & gate & 1Q & LD/e & $\, ^{28}\text{Si/}\text{SiO}_2$ & 2D & 2025-04 & \onlinecite{steinacker_bell_2025} & Fig. 2d-e\\
99.34 \unit{\percent} & gate & 1Q & LD/e & $\, ^{28}\text{Si/SiGe}$ & 2D & 2024-06 & \onlinecite{de_smet_high-fidelity_2024} & p5\\
99.3 \unit{\percent} & gate & 1Q & LD/e\textsuperscript{m} & $\, ^{28}\text{Si/}\text{SiO}_2$ & 2D & 2020-04 & \onlinecite{petit_universal_2020} & p3 and Fig. 2g\\
99.3 \unit{\percent} & gate & 1Q & LD/e\textsuperscript{w} & $\, ^{28}\text{Si/}\text{SiO}_2$ & 2D & 2015-11 & \onlinecite{veldhorst_spin-orbit_2015} & p4 and Fig. 4c\\
99.19 \unit{\percent}\textsuperscript{x} & gate & 1Q & LD/e & Si/SiGe & 2D & 2024-11 & \onlinecite{wang_pursuing_2024} & p4 and Fig. 2d\\
99.12 \unit{\percent} & gate & 1Q & LD/e & $\, ^{28}\text{Si/}\text{SiO}_2$ & 2D & 2024-01 & \onlinecite{ma_single-spin-qubit_2024} & p3 and Fig. 3b\\
99.1 \unit{\percent} & gate & 1Q & LD/e & $\, ^{28}\text{Si/}\text{SiO}_2$ & 2D & 2022-03 & \onlinecite{zwerver_qubits_2022} & p5 and S Fig. 15f\\
99.1 \unit{\percent}\textsuperscript{y,z} & gate & 1Q & LD/e & $\, ^{28}\text{Si/}\text{SiO}_2$ & 2D & 2022-09 & \onlinecite{hansen_implementation_2022} & p7 and Fig. 6d\\
99.1 \unit{\percent}\textsuperscript{$\alpha$} & gate & 1Q & LD/e & $\text{Si/}\text{SiO}_2$ & 2D & 2026-01 & \onlinecite{kuno_robust_2026} & p5 and Fig. 4d\\
99. \unit{\percent} & gate & 1Q & LD/e & Si/SiGe & 2D & 2016-10 & \onlinecite{kawakami_gate_2016} & p5\\
98.8 \unit{\percent} & gate & 1Q & LD/e & Si/SiGe & 2D & 2018-02 & \onlinecite{watson_programmable_2018} & p1 and {ED} Fig. 4\\
98.8 \unit{\percent} & gate & 1Q & LD/e & Si/SiGe & 2D & 2019-04 & \onlinecite{xue_benchmarking_2019} & p7\\
98.6 \unit{\percent}\textsuperscript{$\beta$,z} & gate & 1Q & LD/e & $\, ^{28}\text{Si/}\text{SiO}_2$ & 2D & 2022-09 & \onlinecite{hansen_implementation_2022} & p6 and Fig. 6b\\
97.5 \unit{\percent} & gate & 1Q & LD/e & GaAs/AlGaAs & 2D & 2020-03 & \onlinecite{nakajima_coherence_2020} & p3 and Fig. 4e\\
96.6 \unit{\percent} & gate & 1Q & LD/e & GaAs/AlGaAs & 2D & 2014-12 & \onlinecite{yoneda_fast_2014} & p2\\
95. \unit{\percent}\textsuperscript{$\gamma$} & gate & 1Q & LD/e & $\text{Si/}\text{SiO}_2$ & 2D & 2026-01 & \onlinecite{kuno_robust_2026} & p5 and Fig. 4d\\
\bottomrule
\end{tabular}
\caption{Operation fidelities (part 1). Superscripts stand for the following. \textsuperscript{a}: Average single-qubit Clifford fidelity. \textsuperscript{b}: Per Clifford gate. \textsuperscript{c}: {EO} qubit. \textsuperscript{d}: Average of $\mathcal{F}[X_\pi]=93$ \unit{\percent} and $\mathcal{F}[Z_\pi]=96$ \unit{\percent}. \textsuperscript{e}: The average of the six fidelity values given. \textsuperscript{f}: Hybrid qubit with three electrons in two dots. \textsuperscript{g}: Electron-valley qubit. \textsuperscript{h}: Q3. \textsuperscript{i}: Left qubit. \textsuperscript{j}: At 200 {mK}. \textsuperscript{k}: We take the average of the range given by the article. \textsuperscript{l}: Clifford gate set fidelity. \textsuperscript{m}: At 1 K. \textsuperscript{n}: The average of the six values given in Fig. 1C. \textsuperscript{o}: The average over the three qubits. \textsuperscript{p}: In a five-electron dot configuration. \textsuperscript{q}: With the control chip inside the fridge. \textsuperscript{r}: Device B. \textsuperscript{s}: {EDSR} gate. \textsuperscript{t}: Q5. \textsuperscript{u}: Hopping gate. \textsuperscript{v}: Average X-gate fidelity for the two qubits. \textsuperscript{w}: A three-electron qubit. \textsuperscript{x}: Q1. \textsuperscript{y}: Qubit dressed with an oscillatory drive-amplitude. \textsuperscript{z}: Fidelity of Clifford gates. \textsuperscript{$\alpha$}: Driven qubit. \textsuperscript{$\beta$}: Qubit dressed with a constant drive-amplitude. \textsuperscript{$\gamma$}: Bare qubit.
\label{operationFidelity1}
}
\end{table}

%% file: figuresAndTables/table-operationFidelity-2.txt
\rowcolors{2}{white}{gray!25}
\setcounter{rowcount}{-1}
\begin{table}
\begin{tabular}{@{}S[table-format=2.4\unit{\percent}\textsuperscript{x}, table-align-text-after=true, round-mode=none, round-pad=false, round-precision=0]cccccccc@{ \stepcounter{rowcount} \tiny \therowcount }}
\toprule
{Fidelity} & {Operation} & {\#Qubits} & {Qubit} & {Material} & {Host} & {Date} & {Reference} & {Source}\\
\midrule
82.3 \unit{\percent} & gate & 1Q & LD/e & $\, ^{28}\text{Si/}\text{SiO}_2$ & 2D & 2025-03 & \onlinecite{ni_swap_2025} & p4\\
81. \unit{\percent} & gate & 1Q & LD/e & InSb & 1D & 2013-02 & \onlinecite{van_den_berg_fast_2013} & p3\\
73. \unit{\percent} & gate & 1Q & LD/e & GaAs/AlGaAs & 2D & 2006-08 & \onlinecite{koppens_driven_2006} & p5\\
48. \unit{\percent} & gate & 1Q & LD/e & InAs & 1D & 2010-12 & \onlinecite{nadj-perge_spinorbit_2010} & p2\\
99.76 \unit{\percent}\textsuperscript{a} & gate & 1Q & ST/e & $\, ^{28}\text{Si/}\text{SiO}_2$ & 2D & 2020-04 & \onlinecite{yang_operation_2020} & Fig. 3d\\
99.6 \unit{\percent} & gate & 1Q & ST/e & Si/SiGe & 2D & 2020-03 & \onlinecite{takeda_resonantly_2020} & p4\\
99.5 \unit{\percent} & gate & 1Q & ST/e & GaAs/AlGaAs & 2D & 2020-08 & \onlinecite{cerfontaine_closed-loop_2020} & Fig. 3\\
99.18 \unit{\percent} & gate & 1Q & ST/e & Si/SiGe & 2D & 2024-12 & \onlinecite{walelign_dynamically_2024} & p5\\
99.05 \unit{\percent}\textsuperscript{b,c,d} & gate & 1Q & ST/e & GaAs/AlGaAs & 2D & 2022-07 & \onlinecite{kim_approaching_2022} & p5\\
98.6 \unit{\percent}\textsuperscript{e} & gate & 1Q & ST/e & $\, ^{28}\text{Si/}\text{SiO}_2$ & 2D & 2020-04 & \onlinecite{yang_operation_2020} & Fig. 3h\\
98.6 \unit{\percent} & gate & 1Q & ST/e & GaAs/AlGaAs & 2D & 2017-01 & \onlinecite{nichol_high-fidelity_2017} & p3 and Fig. 2c\\
98.5 \unit{\percent} & gate & 1Q & ST/e & GaAs/AlGaAs & 2D & 2016-06 & \onlinecite{cerfontaine_feedback-tuned_2016} & p4\\
99.992 \unit{\percent} & gate & 1Q & LD/h & Ge/SiGe & 2D & 2023-06 & \onlinecite{lawrie_simultaneous_2023} & p3 and Fig. 2k\\
99.97 \unit{\percent}\textsuperscript{f} & gate & 1Q & LD/h & Ge/SiGe & 2D & 2024-07 & \onlinecite{wang_operating_2024} & p3 and Fig. 1I\\
99.94 \unit{\percent} & gate & 1Q & LD/h & Ge/SiGe & 2D & 2024-05 & \onlinecite{hendrickx_sweet-spot_2024} & p5 and Fig. 6c\\
99.9 \unit{\percent} & gate & 1Q & LD/h & Ge/SiGe & 2D & 2021-03 & \onlinecite{hendrickx_four-qubit_2021} & Fig. S4 dot 3\\
99.7 \unit{\percent}\textsuperscript{g} & gate & 1Q & LD/h & Ge/SiGe & 2D & 2024-05 & \onlinecite{hendrickx_sweet-spot_2024} & p5 and Fig. 6d\\
99.3 \unit{\percent} & gate & 1Q & LD/h & Ge/SiGe & 2D & 2020-01 & \onlinecite{hendrickx_fast_2020} & p2 and Fig. 2d\\
98.9 \unit{\percent}\textsuperscript{e} & gate & 1Q & LD/h\textsuperscript{e} & $\text{Si/}\text{SiO}_2$ & 1D & 2022-03 & \onlinecite{camenzind_spin_2022} & p3 and Fig. 3b\\
98.5 \unit{\percent}\textsuperscript{h} & gate & 1Q & LD/h & Ge/SiGe & 2D & 2023-06 & \onlinecite{lawrie_simultaneous_2023} & Fig. 2g\\
99.68 \unit{\percent} & gate & 1Q & ST/h & Ge/SiGe & 2D & 2026-01 & \onlinecite{tsoukalas_dressed_2026} & p3 and Fig. 2d\\
99.63 \unit{\percent} & gate & 1Q & ST/h\textsuperscript{i} & Ge/SiGe & 2D & 2026-01 & \onlinecite{tsoukalas_dressed_2026} & Fig. 3d\\
99.942 \unit{\percent} & gate & 1Q & LD/i & $\, ^{28}\text{Si:P}$ & imp & 2016-10 & \onlinecite{dehollain_optimization_2016} & p6 and Fig. 3\\
99.78 \unit{\percent} & gate & 1Q & LD/i & $\, ^{28}\text{Si:P}$ & imp & 2024-02 & \onlinecite{reiner_high-fidelity_2024} & Fig. 5d\\
99.6 \unit{\percent} & gate & 1Q & LD/i & $\, ^{28}\text{Si:P}$ & imp & 2014-10 & \onlinecite{muhonen_storing_2014} & p2\\
99.4 \unit{\percent} & gate & 1Q & LD/i & Si:P & imp & 2015-04 & \onlinecite{laucht_electrically_2015} & p4 and Fig. 4a\\
99. \unit{\percent}\textsuperscript{j} & gate & 1Q & LD/i & $\, ^{28}\text{Si:P}$ & imp & 2022-01 & \onlinecite{madzik_precision_2022} & p3\\
99. \unit{\percent} & gate & 1Q & LD/i & $\, ^{28}\text{Si:P}$ & imp & 2024-09 & \onlinecite{stemp_tomography_2024} & p3\\
98.4 \unit{\percent}\textsuperscript{k} & gate & 1Q & LD/i & $\, ^{28}\text{Si:P}$ & imp & 2023-02 & \onlinecite{savytskyy_electrically_2023} & Fig. 3D\\
68. \unit{\percent} & gate & 2Q & charge & GaAs/AlGaAs & 2D & 2015-07 & \onlinecite{li_conditional_2015} & p4\\
63. \unit{\percent} & gate & 2Q & charge & Si/SiGe & 2D & 2020-09 & \onlinecite{macquarrie_progress_2020} & p4\\
96.3 \unit{\percent}\textsuperscript{l} & gate & 2Q & HY/e & $\, ^{28}\text{Si/SiGe}$ & 2D & 2023-03 & \onlinecite{weinstein_universal_2023} & Fig. 4a\\
99.81 \unit{\percent}\textsuperscript{m} & gate & 2Q & LD/e & $\, ^{28}\text{Si/SiGe}$ & 2D & 2022-04 & \onlinecite{mills_two-qubit_2022} & p3\\
99.65 \unit{\percent}\textsuperscript{n} & gate & 2Q & LD/e & $\, ^{28}\text{Si/SiGe}$ & 2D & 2022-01 & \onlinecite{xue_quantum_2022} & p3 and {ED} Fig. 5\\
99.51 \unit{\percent}\textsuperscript{o} & gate & 2Q & LD/e & $\, ^{28}\text{Si/SiGe}$ & 2D & 2022-01 & \onlinecite{noiri_fast_2022} & p4 and Fig. 2g\\
99.48 \unit{\percent} & gate & 2Q & LD/e & $\, ^{28}\text{Si/SiGe}$ & 2D & 2024-01 & \onlinecite{wu_hamiltonian_2024} & Fig. 3d\\
99.17 \unit{\percent}\textsuperscript{p} & gate & 2Q & LD/e & $\, ^{28}\text{Si/}\text{SiO}_2$ & 2D & 2024-08 & \onlinecite{tanttu_assessment_2024} & p6\\
99.09 \unit{\percent}\textsuperscript{q} & gate & 2Q & LD/e & $\, ^{28}\text{Si/}\text{SiO}_2$ & 2D & 2025-04 & \onlinecite{steinacker_bell_2025} & Fig. 2f\\
98.86 \unit{\percent}\textsuperscript{r} & gate & 2Q & LD/e & $\, ^{28}\text{Si/SiGe}$ & 2D & 2026-05 & \onlinecite{matsumoto_two-qubit_2026} & p5 and Fig. S3a\\
98.62 \unit{\percent}\textsuperscript{s} & gate & 2Q & LD/e & $\, ^{28}\text{Si/SiGe}$ & 2D & 2022-04 & \onlinecite{mills_two-qubit_2022} & p3\\
98.2 \unit{\percent} & gate & 2Q & LD/e & $\, ^{28}\text{Si/}\text{SiO}_2$ & 2D & 2022-02 & \onlinecite{evans_fast_2022} & p4 and Fig. 3a\\
98. \unit{\percent} & gate & 2Q & LD/e & $\, ^{28}\text{Si/}\text{SiO}_2$ & 2D & 2019-05 & \onlinecite{huang_fidelity_2019} & p4 and Fig. 4\\
92.72 \unit{\percent}\textsuperscript{m} & gate & 2Q & LD/e & $\, ^{28}\text{Si/SiGe}$ & 2D & 2022-09 & \onlinecite{noiri_shuttling-based_2022} & p5 and Fig. 4f\\
92. \unit{\percent} & gate & 2Q & LD/e & Si/SiGe & 2D & 2019-04 & \onlinecite{xue_benchmarking_2019} & p7\\
91. \unit{\percent}\textsuperscript{m} & gate & 2Q & LD/e & Si/SiGe & 2D & 2024-11 & \onlinecite{wang_pursuing_2024} & p6\\
86.1 \unit{\percent} & gate & 2Q & LD/e\textsuperscript{t} & $\, ^{28}\text{Si/}\text{SiO}_2$ & 2D & 2020-04 & \onlinecite{petit_universal_2020} & p3 and Fig. 3d\\
84. \unit{\percent}\textsuperscript{u} & gate & 2Q & LD/e & $\, ^{28}\text{Si/SiGe}$ & 2D & 2019-11 & \onlinecite{sigillito_coherent_2019} & p6\\
83.1 \unit{\percent}\textsuperscript{v} & gate & 2Q & LD/e & $\, ^{28}\text{Si/SiGe}$ & 2D & 2024-12 & \onlinecite{dijkema_cavity-mediated_2024} & p5 and Fig. 9\\
56. \unit{\percent}\textsuperscript{w} & gate & 2Q & LD/e & Si/SiGe & 2D & 2017-12 & \onlinecite{zajac_resonantly_2017} & p3\\
90. \unit{\percent} & gate & 2Q & ST/e & GaAs/AlGaAs & 2D & 2017-01 & \onlinecite{nichol_high-fidelity_2017} & p4 and Fig. 4e\\
\bottomrule
\end{tabular}
\caption{Operation fidelities (part 2). Superscripts stand for the following. \textsuperscript{a}: At 0.04 K. \textsuperscript{b}: We take the values obtained by {GST}. \textsuperscript{c}: X/2 gate. \textsuperscript{d}: With estimation and/or feedback. \textsuperscript{e}: At 1.5 K. \textsuperscript{f}: $Q_A$. \textsuperscript{g}: At 1.1 K. \textsuperscript{h}: \estimated Fig. 2g the leftmost point. \textsuperscript{i}: Qubit defined in the rotating frame. \textsuperscript{j}: Lower limit. \textsuperscript{k}: Hybrid electron-nuclear qubit. \textsuperscript{l}: Average two-qubit Clifford fidelity 97.1\unit{\percent}. \textsuperscript{m}: {CZ} gate. \textsuperscript{n}: {CZ}. \textsuperscript{o}: {CNOT}. \textsuperscript{p}: Average over three devices. \textsuperscript{q}: {DCZ} gate. \textsuperscript{r}: Shuttling-based {CZ} gate. \textsuperscript{s}: {CNOT} gate synthetized using the {CZ} gate. \textsuperscript{t}: At 1 K. \textsuperscript{u}: {SWAP}. \textsuperscript{v}: {iSWAP}. \textsuperscript{w}: Resonantly induced {CNOT}.
\label{operationFidelity2}
}
\end{table}

%% file: figuresAndTables/table-operationFidelity-3.txt
\rowcolors{2}{white}{gray!25}
\setcounter{rowcount}{-1}
\begin{table}
\begin{tabular}{@{}S[table-format=2.4\unit{\percent}\textsuperscript{x}, table-align-text-after=true, round-mode=none, round-pad=false, round-precision=0]cccccccc@{ \stepcounter{rowcount} \tiny \therowcount }}
\toprule
{Fidelity} & {Operation} & {\#Qubits} & {Qubit} & {Material} & {Host} & {Date} & {Reference} & {Source}\\
\midrule
72. \unit{\percent}\textsuperscript{a} & gate & 2Q & ST/e & GaAs/AlGaAs & 2D & 2012-04 & \onlinecite{shulman_demonstration_2012} & p3\\
99.3 \unit{\percent} & gate & 2Q & LD/h & Ge/SiGe & 2D & 2024-07 & \onlinecite{wang_operating_2024} & p4 and Fig. 2e\\
90. \unit{\percent} & gate & 2Q & LD/i & Si:P & imp & 2019-07 & \onlinecite{he_two-qubit_2019} & p3\\
82. \unit{\percent}\textsuperscript{b} & gate & 2Q & LD/i & $\, ^{28}\text{Si:P}$ & imp & 2024-09 & \onlinecite{stemp_tomography_2024} & {SM} Tab. 10\\
99.975 \unit{\percent}\textsuperscript{c,d} & measure & 1Q & HY/e\textsuperscript{e} & $\, ^{28}\text{Si/SiGe}$ & 2D & 2022-03 & \onlinecite{blumoff_fast_2022} & p8 and Fig. 7\\
99.97 \unit{\percent}\textsuperscript{f} & measure & 1Q & HY/e & BLG & 2D & 2024-01 & \onlinecite{garreis_long-lived_2024} & p4 and Fig. 3d\\
96.4 \unit{\percent} & measure & 1Q & HY/e & GaAs/AlGaAs & 2D & 2021-06 & \onlinecite{jang_single-shot_2021} & p4 and Fig. 2d\\
96. \unit{\percent} & measure & 1Q & HY/e & $\, ^{28}\text{Si/SiGe}$ & 2D & 2023-03 & \onlinecite{weinstein_universal_2023} & p3 and {ED} Fig. 2\\
95.3 \unit{\percent}\textsuperscript{g} & measure & 1Q & HY/e & $\, ^{28}\text{Si/SiGe}$ & 2D & 2024-10 & \onlinecite{acuna_coherent_2024} & p4\\
75. \unit{\percent} & measure & 1Q & HY/e\textsuperscript{e} & GaAs/AlGaAs & 2D & 2013-09 & \onlinecite{medford_self-consistent_2013} & p4\\
99.56 \unit{\percent} & measure & 1Q & LD/e & $\, ^{28}\text{Si/SiGe}$ & 2D & 2022-12 & \onlinecite{mills_high-fidelity_2022} & p4 and Fig. 3\\
99.54 \unit{\percent} & measure & 1Q & LD/e & $\text{Si/}\text{SiO}_2$ & 1D & 2023-02 & \onlinecite{oakes_fast_2023} & p5 and Fig. 2j\\
99. \unit{\percent}\textsuperscript{h,i} & measure & 1Q & LD/e & $\, ^{28}\text{Si/SiGe}$ & 2D & 2022-09 & \onlinecite{philips_universal_2022} & Fig. 2e\\
99. \unit{\percent}\textsuperscript{j} & measure & 1Q & LD/e & $\, ^{28}\text{Si/SiGe}$ & 2D & 2022-04 & \onlinecite{mills_two-qubit_2022} & p1\\
97.8 \unit{\percent}\textsuperscript{k,l} & measure & 1Q & LD/e & $\, ^{28}\text{Si/}\text{SiO}_2$ & 2D & 2024-03 & \onlinecite{huang_high-fidelity_2024} & p4 and Fig. S3\\
97. \unit{\percent} & measure & 1Q & LD/e & GaAs/AlGaAs & 2D & 2016-01 & \onlinecite{baart_single-spin_2016} & abstract\\
95. \unit{\percent} & measure & 1Q & LD/e & Si/SiGe & 2D & 2014-08 & \onlinecite{kawakami_electrical_2014} & p1\\
95. \unit{\percent} & measure & 1Q & LD/e & $\, ^{28}\text{Si/}\text{SiO}_2$ & 2D & 2014-10 & \onlinecite{veldhorst_addressable_2014} & p2\\
95. \unit{\percent} & measure & 1Q & LD/e & Si/SiGe & 2D & 2020-03 & \onlinecite{yoneda_quantum_2020} & p5\\
94.5 \unit{\percent} & measure & 1Q & LD/e & Si/SiGe & 2D & 2020-04 & \onlinecite{xue_repetitive_2020} & abstract and p3\\
92. \unit{\percent} & measure & 1Q & LD/e & $\text{Si/}\text{SiO}_2$ & 1D & 2022-03 & \onlinecite{spence_spin-valley_2022} & p2\\
89.9 \unit{\percent} & measure & 1Q & LD/e & Si/SiGe & 2D & 2021-06 & \onlinecite{takeda_quantum_2021} & p1 for Q3\\
89. \unit{\percent} & measure & 1Q & LD/e & GaAs/AlGaAs & 2D & 2019-04 & \onlinecite{nakajima_quantum_2019} & p3\\
86. \unit{\percent} & measure & 1Q & LD/e & GaAs/AlGaAs & 2D & 2011-08 & \onlinecite{nowack_single-shot_2011} & p4\\
82.5 \unit{\percent}\textsuperscript{m} & measure & 1Q & LD/e & GaAs/AlGaAs & 2D & 2004-07 & \onlinecite{elzerman_single-shot_2004} & p4\\
81. \unit{\percent} & measure & 1Q & LD/e & Si/SiGe & 2D & 2018-02 & \onlinecite{watson_programmable_2018} & p1\\
80. \unit{\percent} & measure & 1Q & LD/e & InAs & 1D & 2010-12 & \onlinecite{nadj-perge_spinorbit_2010} & p1\\
75. \unit{\percent}\textsuperscript{n} & measure & 1Q & LD/e & InAs & 1D & 2024-05 & \onlinecite{pita-vidal_strong_2024} & p29 and Fig. S19\\
99.96 \unit{\percent}\textsuperscript{o} & measure & 1Q & ST/e & $\, ^{28}\text{Si/SiGe}$ & 2D & 2024-02 & \onlinecite{takeda_rapid_2024} & p4 and Fig. 3d\\
99.96 \unit{\percent} & measure & 1Q & ST/e & $\, ^{28}\text{Si/}\text{SiO}_2$ & 2D & 2025-04 & \onlinecite{steinacker_bell_2025} & Fig. 2c\\
99.92 \unit{\percent} & measure & 1Q & ST/e & $\text{Si/}\text{SiO}_2$ & 2D & 2026-07 & \onlinecite{laine_high-fidelity_2026} & p5 and Fig. 4c\\
99.9 \unit{\percent}\textsuperscript{p} & measure & 1Q & ST/e & $\text{Si/}\text{SiO}_2$ & 2D & 2022-12 & \onlinecite{niegemann_parity_2022} & p4 and Fig. 3\\
99.85 \unit{\percent}\textsuperscript{q} & measure & 1Q & ST/e & BLG & 2D & 2024-01 & \onlinecite{garreis_long-lived_2024} & p4 and Fig. 3d\\
99.5 \unit{\percent}\textsuperscript{r} & measure & 1Q & ST/e & GaAs/AlGaAs & 2D & 2017-07 & \onlinecite{nakajima_robust_2017} & p4\\
99.3 \unit{\percent} & measure & 1Q & ST/e & $\, ^{28}\text{Si/}\text{SiO}_2$ & 2D & 2019-12 & \onlinecite{zhao_single-spin_2019} & p2\\
99.21 \unit{\percent} & measure & 1Q & ST/e & $\text{Si/}\text{SiO}_2$ & 1D & 2023-02 & \onlinecite{oakes_fast_2023} & p6 and Fig. 4d\\
99.2 \unit{\percent} & measure & 1Q & ST/e & $\, ^{28}\text{Si/SiGe}$ & 2D & 2021-04 & \onlinecite{borjans_spin_2021} & p5\\
99. \unit{\percent} & measure & 1Q & ST/e & Si/SiGe & 2D & 2020-02 & \onlinecite{connors_rapid_2020} & p6\\
99. \unit{\percent} & measure & 1Q & ST/e & $\, ^{28}\text{Si/}\text{SiO}_2$ & 2D & 2018-10 & \onlinecite{fogarty_integrated_2018} & p2\\
99. \unit{\percent}\textsuperscript{s} & measure & 1Q & ST/e & $\, ^{28}\text{Si/}\text{SiO}_2$ & 2D & 2024-08 & \onlinecite{tanttu_assessment_2024} & {ED} Tab. {II}\\
99. \unit{\percent} & measure & 1Q & ST/e & $\text{Si/}\text{SiO}_2$ & 2D & 2026-07 & \onlinecite{laine_high-fidelity_2026} & p5 and Fig. 4c\\
98.45 \unit{\percent}\textsuperscript{t} & measure & 1Q & ST/e & $\text{Si/}\text{SiO}_2$ & 1D & 2019-05 & \onlinecite{urdampilleta_gate-based_2019} & abstract and p2\\
98.3 \unit{\percent}\textsuperscript{u,v} & measure & 1Q & ST/e & GaAs/AlGaAs & 2D & 2022-07 & \onlinecite{kim_approaching_2022} & p5\\
98.2 \unit{\percent} & measure & 1Q & ST/e & Si/SiGe & 2D & 2020-02 & \onlinecite{connors_rapid_2020} & Fig. 5b\\
98. \unit{\percent} & measure & 1Q & ST/e & Si/SiGe & 2D & 2019-08 & \onlinecite{zheng_rapid_2019} & p3 and Fig. 3f\\
98. \unit{\percent} & measure & 1Q & ST/e & GaAs/AlGaAs & 2D & 2014-10 & \onlinecite{shulman_suppressing_2014} & p2\\
97.59 \unit{\percent} & measure & 1Q & ST/e & $\text{Si/}\text{SiO}_2$ & 2D & 2024-03 & \onlinecite{ma_singlet-triplet-state_2024} & p4\\
97.5 \unit{\percent} & measure & 1Q & ST/e & GaAs/AlGaAs & 2D & 2006-11 & \onlinecite{meunier_high_2006} & p3\\
97. \unit{\percent}\textsuperscript{w} & measure & 1Q & ST/e & GaAs/AlGaAs & 2D & 2016-11 & \onlinecite{kiyama_single-shot_2016} & p4\\
96. \unit{\percent} & measure & 1Q & ST/e & GaAs/AlGaAs & 2D & 2018-04 & \onlinecite{mukhopadhyay_2x2_2018} & p4\\
\bottomrule
\end{tabular}
\caption{Operation fidelities (part 3). Superscripts stand for the following. \textsuperscript{a}: Fidelity of a Bell state produced using {CPHASE}. \textsuperscript{b}: The average fidelity of {CROT}[$X_{\pi/2}$] gate. \textsuperscript{c}: The corresponding infidelity includes both preparation and measurement errors. \textsuperscript{d}: Reference reports $F_{{BC}}$, standing for `benchmark contrast fidelity' and advocates using it instead of `assignment fidelity' $F_A$ which corresponds to our Eq.~\eqref{eq:fidelityMeasurement}. \textsuperscript{e}: {EO} qubit. \textsuperscript{f}: Valley degree of freedom. \textsuperscript{g}: {SPAM} fidelity. \textsuperscript{h}: Q3. \textsuperscript{i}: Lower limit \derived from visibility V=98\unit{\percent}. \textsuperscript{j}: The average figure of merit for the two qubits measured. \textsuperscript{k}: We take the average of the two values given as 99.34 \unit{\percent} and 96.15 \unit{\percent}. \textsuperscript{l}: At 1 K. \textsuperscript{m}: The average of 93 \unit{\percent} and 72 \unit{\percent}. \textsuperscript{n}: Q1. \textsuperscript{o}: \estimated infidelity $4\times 10^{-4}$ as the red curve minimum in Fig. 3d. \textsuperscript{p}: Measurement distinguishes $S_0$ and $T_0$ versus $T_-$ and $T_+$. \textsuperscript{q}: The average of the given range. \textsuperscript{r}: Latched readout. \textsuperscript{s}: Device B. \textsuperscript{t}: The average of 99.6 \unit{\percent} and 97.3 \unit{\percent}; latched readout. \textsuperscript{u}: We take the values obtained by {GST}. \textsuperscript{v}: With estimation and/or feedback. \textsuperscript{w}: S/T binary result.
\label{operationFidelity3}
}
\end{table}

%% file: figuresAndTables/table-operationFidelity-4.txt
\rowcolors{2}{white}{gray!25}
\setcounter{rowcount}{-1}
\begin{table}
\begin{tabular}{@{}S[table-format=2.4\unit{\percent}\textsuperscript{x}, table-align-text-after=true, round-mode=none, round-pad=false, round-precision=0]cccccccc@{ \stepcounter{rowcount} \tiny \therowcount }}
\toprule
{Fidelity} & {Operation} & {\#Qubits} & {Qubit} & {Material} & {Host} & {Date} & {Reference} & {Source}\\
\midrule
95. \unit{\percent} & measure & 1Q & ST/e & GaAs/AlGaAs & 2D & 2021-02 & \onlinecite{jadot_distant_2021} & p1\\
95. \unit{\percent} & measure & 1Q & ST/e & GaAs/AlGaAs & 2D & 2017-06 & \onlinecite{fujita_coherent_2017} & p2\\
93. \unit{\percent} & measure & 1Q & ST/e & GaAs/AlGaAs & 2D & 2020-06 & \onlinecite{qiao_conditional_2020} & p5 and {SM} Fig. 11\\
91. \unit{\percent} & measure & 1Q & ST/e & $\text{Si/}\text{SiO}_2$ & 2D & 2026-06 & \onlinecite{hamonic_single-shot_2026} & p5\\
90. \unit{\percent} & measure & 1Q & ST/e & GaAs/AlGaAs & 2D & 2009-10 & \onlinecite{barthel_rapid_2009} & p3\\
90. \unit{\percent} & measure & 1Q & ST/e & GaAs/AlGaAs & 2D & 2020-07 & \onlinecite{jang_robust_2020} & p4\\
87. \unit{\percent} & measure & 1Q & ST/e & GaAs/AlGaAs & 2D & 2006-11 & \onlinecite{meunier_nondestructive_2006} & p2\\
80. \unit{\percent} & measure & 1Q & ST/e & GaAs/AlGaAs & 2D & 2021-05 & \onlinecite{kojima_probabilistic_2021} & p2 and p3\\
80. \unit{\percent} & measure & 1Q & ST/e & GaAs/AlGaAs & 2D & 2018-09 & \onlinecite{orona_readout_2018} & p4\\
80. \unit{\percent} & measure & 1Q & ST/e & GaAs/AlGaAs & 2D & 2017-09 & \onlinecite{flentje_coherent_2017} & p2\\
80. \unit{\percent} & measure & 1Q & ST/e & GaAs/AlGaAs & 2D & 2015-08 & \onlinecite{bertrand_quantum_2015} & p3\\
76. \unit{\percent}\textsuperscript{a} & measure & 1Q & ST/e & GaAs/AlGaAs & 2D & 2016-11 & \onlinecite{kiyama_single-shot_2016} & p4\\
75. \unit{\percent} & measure & 1Q & ST/e & $\text{Si/}\text{SiO}_2$ & 2D & 2019-03 & \onlinecite{west_gate-based_2019} & p3\\
97. \unit{\percent}\textsuperscript{b} & measure & 1Q & LD/h & Ge/SiGe & 2D & 2026-09 & \onlinecite{kelly_identification_2026} & p6\\
87. \unit{\percent}\textsuperscript{c} & measure & 1Q & LD/h & Ge/Si & 1D & 2018-10 & \onlinecite{vukusic_single-shot_2018} & p3\\
78. \unit{\percent}\textsuperscript{d} & measure & 1Q & ST/h & Ge/SiGe & 2D & 2020-07 & \onlinecite{hendrickx_single-hole_2020} & p4\\
99.8 \unit{\percent} & measure & 1Q & LD/i & Si:P & imp & 2017-03 & \onlinecite{watson_atomically_2017} & p3 and p4\\
99.8 \unit{\percent} & measure & 1Q & LD/i & Si:P & imp & 2015-10 & \onlinecite{watson_high-fidelity_2015} & p3 and Fig. 4a\\
99.44 \unit{\percent} & measure & 1Q & LD/i & Si:P & imp & 2025-04 & \onlinecite{geng_high-fidelity_2025} & Fig. 4d\\
99. \unit{\percent}\textsuperscript{e} & measure & 1Q & LD/i & Si:P & imp & 2022-09 & \onlinecite{keith_ramped_2022} & p2\\
98.7 \unit{\percent} & measure & 1Q & LD/i & Si:P & imp & 2015-10 & \onlinecite{watson_high-fidelity_2015} & p3 and Fig. 4b\\
97.87 \unit{\percent}\textsuperscript{f} & measure & 1Q & LD/i & Si:P & imp & 2025-04 & \onlinecite{geng_high-fidelity_2025} & Fig. 4f\\
97.7 \unit{\percent} & measure & 1Q & LD/i & Si:P & imp & 2019-01 & \onlinecite{koch_spin_2019} & p3\\
97.6 \unit{\percent} & measure & 1Q & LD/i & Si:P & imp & 2018-03 & \onlinecite{broome_two-electron_2018} & p2 and {STable} 1 line $F_\mathrm{M}$ qubit R\\
97. \unit{\percent} & measure & 1Q & LD/i & $\, ^{28}\text{Si:P}$ & imp & 2014-10 & \onlinecite{muhonen_storing_2014} & p2\\
97. \unit{\percent} & measure & 1Q & LD/i & Si:P & imp & 2019-10 & \onlinecite{keith_single-shot_2019} & p4\\
96. \unit{\percent} & measure & 1Q & LD/i & Si:P & imp & 2010-09 & \onlinecite{morello_single-shot_2010} & p4 and Fig. 4c\\
95. \unit{\percent}\textsuperscript{g} & measure & 1Q & LD/i & Si:P & imp & 2013-06 & \onlinecite{buch_spin_2013} & p3 and Fig. S2\\
95. \unit{\percent}\textsuperscript{h} & measure & 1Q & LD/i & Si:P & imp & 2023-02 & \onlinecite{hogg_single-shot_2023} & Fig. 2d\\
92.3 \unit{\percent} & measure & 1Q & LD/i & Si:P & imp & 2023-11 & \onlinecite{kranz_exploiting_2023} & Tab. I\\
90. \unit{\percent} & measure & 1Q & LD/i & Si:P & imp & 2012-09 & \onlinecite{pla_single-atom_2012} & p3\\
80. \unit{\percent} & measure & 1Q & LD/i & $\, ^{28}\text{Si:P}$ & imp & 2022-01 & \onlinecite{madzik_precision_2022} & p9\\
99.864 \unit{\percent}\textsuperscript{i} & measure & 1Q & ST/i\textsuperscript{j} & Si:P & imp & 2018-05 & \onlinecite{harvey-collard_high-fidelity_2018} & Tab. {II}\\
99.3 \unit{\percent} & measure & 1Q & ST/i\textsuperscript{j} & Si:P & imp & 2018-05 & \onlinecite{harvey-collard_high-fidelity_2018} & Tab. {II}\\
98.4 \unit{\percent}\textsuperscript{i} & measure & 1Q & ST/i & Si:P & imp & 2017-07 & \onlinecite{broome_high-fidelity_2017} & p3 and abstract\\
95. \unit{\percent} & measure & 1Q & ST/i & Si:P & imp & 2014-06 & \onlinecite{dehollain_single-shot_2014} & p3\\
82.9 \unit{\percent} & measure & 1Q & ST/i & Si:P & imp & 2018-11 & \onlinecite{pakkiam_single-shot_2018} & p4\\
84. \unit{\percent}\textsuperscript{k} & measure & 2Q & ST/i & GaAs/AlGaAs & 2D & 2023-03 & \onlinecite{nurizzo_complete_2023} & p6\\
99.975 \unit{\percent}\textsuperscript{l,m} & initialize & 1Q & HY/e\textsuperscript{n} & $\, ^{28}\text{Si/SiGe}$ & 2D & 2022-03 & \onlinecite{blumoff_fast_2022} & p8 and Fig. 7\\
99.76 \unit{\percent} & initialize & 1Q & LD/e & $\, ^{28}\text{Si/SiGe}$ & 2D & 2022-12 & \onlinecite{mills_high-fidelity_2022} & p5\\
99.6 \unit{\percent}\textsuperscript{o} & initialize & 1Q & LD/e & $\, ^{28}\text{Si/}\text{SiO}_2$ & 2D & 2024-03 & \onlinecite{huang_high-fidelity_2024} & p3\\
99.27 \unit{\percent} & initialize & 1Q & LD/e & $\, ^{28}\text{Si/}\text{SiO}_2$ & 2D & 2025-04 & \onlinecite{steinacker_bell_2025} & Fig. 2b\\
99. \unit{\percent}\textsuperscript{p,q} & initialize & 1Q & LD/e & $\, ^{28}\text{Si/SiGe}$ & 2D & 2022-09 & \onlinecite{philips_universal_2022} & Fig. 2e\\
99. \unit{\percent}\textsuperscript{r} & initialize & 1Q & LD/e & Si/SiGe & 2D & 2018-02 & \onlinecite{watson_programmable_2018} & p1\\
99. \unit{\percent} & initialize & 1Q & LD/e & Si/SiGe & 2D & 2020-03 & \onlinecite{yoneda_quantum_2020} & p5\\
99. \unit{\percent} & initialize & 1Q & LD/e & Si/SiGe & 2D & 2021-06 & \onlinecite{takeda_quantum_2021} & p1\\
98.4 \unit{\percent}\textsuperscript{s} & initialize & 1Q & LD/e & $\, ^{28}\text{Si/SiGe}$ & 2D & 2022-04 & \onlinecite{mills_two-qubit_2022} & p1\\
98.33 \unit{\percent} & initialize & 1Q & LD/e & Si/SiGe & 2D & 2023-06 & \onlinecite{kobayashi_feedback-based_2023} & p5\\
95. \unit{\percent} & initialize & 1Q & LD/e & Si/SiGe & 2D & 2014-08 & \onlinecite{kawakami_electrical_2014} & p1\\
95. \unit{\percent} & initialize & 1Q & LD/e & $\, ^{28}\text{Si/SiGe}$ & 2D & 2019-11 & \onlinecite{sigillito_coherent_2019} & p1\\
\bottomrule
\end{tabular}
\caption{Operation fidelities (part 4). Superscripts stand for the following. \textsuperscript{a}: $S$/$T_0$/$T_+$ ternary result. \textsuperscript{b}: {SPAM} fidelity. \textsuperscript{c}: The average of 90.7 \unit{\percent} and 83.3 \unit{\percent}. \textsuperscript{d}: \derived from visibility, $1-\mathcal{F}=[1-v]/2$ with $v=0.56$. \textsuperscript{e}: The reference states `above 99\unit{\percent}  spin readout fidelity'. \textsuperscript{f}: At 3.7 K. \textsuperscript{g}: \derived from visibility $V=93.4$ \unit{\percent}. \textsuperscript{h}: Dot D2. \textsuperscript{i}: Latched readout. \textsuperscript{j}: Hybrid {ST} qubit: impurity -- gated dot. \textsuperscript{k}: The four states $S$, $T_0$, $T_+$, and $T_-$ of a two-spin qubit are discriminated in three consecutive measurements. The value is the average of 99.57\unit{\percent}, 84.1\unit{\percent}, 68.2\unit{\percent} for the fidelities of $S$, $T_0$, and $T_+$, which could be quantified. \textsuperscript{l}: The corresponding infidelity includes both preparation and measurement errors. \textsuperscript{m}: Reference reports $F_{{BC}}$, standing for `benchmark contrast fidelity' and advocates using it instead of `assignment fidelity' $F_A$ which corresponds to our Eq.~\eqref{eq:fidelityMeasurement}. \textsuperscript{n}: {EO} qubit. \textsuperscript{o}: At 1 K. \textsuperscript{p}: Q3. \textsuperscript{q}: Lower limit \derived from visibility V=98\unit{\percent}. \textsuperscript{r}: Using spin hot-spot. \textsuperscript{s}: The average figure of merit for the two qubits measured.
\label{operationFidelity4}
}
\end{table}

%% file: figuresAndTables/table-operationFidelity-5.txt
\rowcolors{2}{white}{gray!25}
\setcounter{rowcount}{-1}
\begin{table}
\begin{tabular}{@{}S[table-format=2.4\unit{\percent}\textsuperscript{x}, table-align-text-after=true, round-mode=none, round-pad=false, round-precision=0]cccccccc@{ \stepcounter{rowcount} \tiny \therowcount }}
\toprule
{Fidelity} & {Operation} & {\#Qubits} & {Qubit} & {Material} & {Host} & {Date} & {Reference} & {Source}\\
\midrule
92. \unit{\percent} & initialize & 1Q & LD/e & $\, ^{28}\text{Si/}\text{SiO}_2$ & 2D & 2014-10 & \onlinecite{veldhorst_addressable_2014} & p2\\
99.7 \unit{\percent}\textsuperscript{a,b} & initialize & 1Q & ST/e & GaAs/AlGaAs & 2D & 2022-07 & \onlinecite{kim_approaching_2022} & p5\\
99.6 \unit{\percent}\textsuperscript{c} & initialize & 1Q & ST/e & $\text{Si/}\text{SiO}_2$ & 2D & 2022-12 & \onlinecite{niegemann_parity_2022} & p4\\
98.5 \unit{\percent}\textsuperscript{d} & initialize & 1Q & ST/e & $\text{Si/}\text{SiO}_2$ & 1D & 2019-05 & \onlinecite{urdampilleta_gate-based_2019} & p3\\
98. \unit{\percent} & initialize & 1Q & ST/e & Si/SiGe & 2D & 2025-09 & \onlinecite{park_single_2025} & p3\\
95. \unit{\percent} & initialize & 1Q & ST/e & GaAs/AlGaAs & 2D & 2021-02 & \onlinecite{jadot_distant_2021} & p1\\
95. \unit{\percent} & initialize & 1Q & ST/e & $\text{Si/}\text{SiO}_2$ & 2D & 2026-06 & \onlinecite{hamonic_single-shot_2026} & p4\\
89. \unit{\percent} & initialize & 1Q & ST/e & GaAs/AlGaAs & 2D & 2020-06 & \onlinecite{qiao_conditional_2020} & p5\\
99.81 \unit{\percent}\textsuperscript{e} & initialize & 1Q & LD/i & Si:P & imp & 2022-09 & \onlinecite{keith_ramped_2022} & p5\\
98.9 \unit{\percent} & initialize & 1Q & LD/i & $\, ^{28}\text{Si:P}$ & imp & 2022-10 & \onlinecite{johnson_beating_2022} & p6 and Fig. 4\\
96.6 \unit{\percent}\textsuperscript{f,g} & initialize & 2Q & LD/e & $\, ^{28}\text{Si/SiGe}$ & 2D & 2022-04 & \onlinecite{mills_two-qubit_2022} & p2\\
96. \unit{\percent}\textsuperscript{h} & initialize & 2Q & LD/e & $\, ^{28}\text{Si/}\text{SiO}_2$ & 2D & 2022-02 & \onlinecite{evans_fast_2022} & p6 and Fig. 3b\\
94.1 \unit{\percent}\textsuperscript{i} & initialize & 2Q & LD/e & Si/SiGe & 2D & 2021-06 & \onlinecite{takeda_quantum_2021} & p2\\
91.5 \unit{\percent}\textsuperscript{j,k} & initialize & 2Q & LD/e & $\, ^{28}\text{Si/SiGe}$ & 2D & 2022-09 & \onlinecite{philips_universal_2022} & Fig. 4h\\
90. \unit{\percent}\textsuperscript{i} & initialize & 2Q & LD/e & $\text{Si/}\text{SiO}_2$ & 2D & 2021-05 & \onlinecite{leon_bell-state_2021} & Fig. 3 caption\\
89. \unit{\percent}\textsuperscript{i} & initialize & 2Q & LD/e & Si/SiGe & 2D & 2018-02 & \onlinecite{watson_programmable_2018} & p3\\
89. \unit{\percent}\textsuperscript{i} & initialize & 2Q & LD/e & $\, ^{28}\text{Si/}\text{SiO}_2$ & 2D & 2019-05 & \onlinecite{huang_fidelity_2019} & p3\\
86.5 \unit{\percent}\textsuperscript{j,l} & initialize & 2Q & LD/e & $\, ^{28}\text{Si/SiGe}$ & 2D & 2022-09 & \onlinecite{philips_universal_2022} & Fig. 4h\\
93. \unit{\percent}\textsuperscript{i} & initialize & 2Q & ST/e & GaAs/AlGaAs & 2D & 2017-01 & \onlinecite{nichol_high-fidelity_2017} & p4\\
88. \unit{\percent}\textsuperscript{m} & initialize & 3Q & LD/e & Si/SiGe & 2D & 2021-06 & \onlinecite{takeda_quantum_2021} & p4\\
86.6 \unit{\percent}\textsuperscript{n} & initialize & 3Q & LD/e & Si/SiGe & 2D & 2022-08 & \onlinecite{takeda_quantum_2022} & Fig. 2\\
76. \unit{\percent}\textsuperscript{o,k} & initialize & 3Q & LD/e & $\, ^{28}\text{Si/SiGe}$ & 2D & 2022-09 & \onlinecite{philips_universal_2022} & Fig. 5g\\
59.25 \unit{\percent}\textsuperscript{o,l} & initialize & 3Q & LD/e & $\, ^{28}\text{Si/SiGe}$ & 2D & 2022-09 & \onlinecite{philips_universal_2022} & Fig. 5g\\
\bottomrule
\end{tabular}
\caption{Operation fidelities (part 5). Superscripts stand for the following. \textsuperscript{a}: We take the values obtained by {GST}. \textsuperscript{b}: With estimation and/or feedback. \textsuperscript{c}: Into singlet. \textsuperscript{d}: Into triplet. \textsuperscript{e}: Into spin down state. \textsuperscript{f}: The average figure of merit for the two qubits measured. \textsuperscript{g}: The average fidelity of the four Bell-states; there was no correction for {SPAM}. \textsuperscript{h}: Fidelity of a Bell state; the article gives a range 94.6--98.3 \unit{\percent}. The value 96 \unit{\percent} was \estimated from Fig. 3b. \textsuperscript{i}: Into a Bell state using a two-qubit algorithm. \textsuperscript{j}: Average of five nearest-neighbors qubit pairs. \textsuperscript{k}: Corrected for {SPAM} errors. \textsuperscript{l}: Not corrected for {SPAM} errors. \textsuperscript{m}: Initialization into the {GHZ} state through a quantum algorithm. \textsuperscript{n}: {GHZ} state; fidelity compensated for readout errors. \textsuperscript{o}: Average of four nearest-neighbors qubit triples.
\label{operationFidelity5}
}
\end{table}

%% file: figuresAndTables/table-qualityFactorMerged-1.txt
\rowcolors{2}{white}{gray!25}
\setcounter{rowcount}{-1}
\begin{table}
\begin{tabular}{@{}S[round-mode=none, round-pad=false, round-precision=3, table-alignment-mode = marker, table-number-alignment = center, table-align-text-after=false ]cccccccc@{ \stepcounter{rowcount} \tiny \therowcount }}
\toprule
{QFactor} & {QFactorType} & {\#Qubits} & {Qubit} & {Material} & {Host} & {Date} & {Reference} & {Source}\\
\midrule
558. \textsuperscript{a} & gate & 1Q & ST/e & $\, ^{28}\text{Si/SiGe}$ & 2D & 2024-08 & \onlinecite{song_coherence_2024} & p3 and Fig. 2b\\
444. \textsuperscript{b} & gate & 1Q & LD/e & $\, ^{28}\text{Si/SiGe}$ & 2D & 2017-12 & \onlinecite{yoneda_quantum-dot_2017} & p2\\
200. \textsuperscript{c} & qubit & 2Q & LD/e & $\text{Si/}\text{SiO}_2$ & 2D & 2021-05 & \onlinecite{leon_bell-state_2021} & Fig. 2j the rightmost point\\
95.22 \textsuperscript{d} & qubit & 1Q & LD/e & Si/SiGe & 2D & 2024-11 & \onlinecite{wang_pursuing_2024} & p3 and Fig. 2c\\
70. \textsuperscript{e} & gate & 1Q & LD/e & Si/SiGe & 2D & 2016-08 & \onlinecite{takeda_fault-tolerant_2016} & p3\\
52.  & qubit & 1Q & ST/h & Ge/SiGe & 2D & 2021-06 & \onlinecite{jirovec_singlet-triplet_2021} & p4 and {SM} Fig. 15c\\
50. \textsuperscript{f} & gate & 1Q & ST/e & $\, ^{28}\text{Si/SiGe}$ & 2D & 2023-03 & \onlinecite{weinstein_universal_2023} & p8 and {ED} Fig. 1c\\
50. \textsuperscript{g,h} & gate & 2Q & LD/e & $\, ^{28}\text{Si/}\text{SiO}_2$ & 2D & 2024-03 & \onlinecite{huang_high-fidelity_2024} & p6 and {ED} Fig. 8g-h\\
50. \textsuperscript{i} & gate & 1Q & LD/e & $\, ^{28}\text{Si/}\text{SiO}_2$ & 2D & 2025-01 & \onlinecite{chu_single-qubit_2025} & p3\\
44.  & gate & 1Q & ST/e & $\, ^{28}\text{Si/SiGe}$ & 2D & 2016-03 & \onlinecite{reed_reduced_2016} & Fig. 4b the topmost red point\\
43.5 \textsuperscript{j,k} & gate & 1Q & LD/h\textsuperscript{k} & $\text{Si/}\text{SiO}_2$ & 1D & 2022-03 & \onlinecite{camenzind_spin_2022} & p3 and Fig. 1e\\
42.5 \textsuperscript{l} & gate & 1Q & LD/e & GaAs/AlGaAs & 2D & 2020-03 & \onlinecite{nakajima_coherence_2020} & Fig. 6\\
40. \textsuperscript{m} & qubit & 1Q & ST/e & GaAs/AlGaAs & 2D & 2022-07 & \onlinecite{kim_approaching_2022} & p4\\
40.  & qubit & 1Q & ST/h & $\text{Si/}\text{SiO}_2$ & 2D & 2024-09 & \onlinecite{liles_singlet-triplet_2024} & p7\\
35. \textsuperscript{n} & gate & 1Q & ST/e & GaAs/AlGaAs & 2D & 2016-03 & \onlinecite{martins_noise_2016} & p2 and Fig. 4d\\
32. \textsuperscript{o,p} & qubit & 1Q & LD/e & $\, ^{28}\text{Si/SiGe}$ & 2D & 2025-07 & \onlinecite{unseld_baseband_2025} & p3\\
28.56  & qubit & 1Q & ST/e & $\text{Si/}\text{SiO}_2$ & 2D & 2026-06 & \onlinecite{hamonic_single-shot_2026} & Fig. 3c and p5\\
26. \textsuperscript{q} & gate & 2Q & LD/e & $\, ^{28}\text{Si/}\text{SiO}_2$ & 2D & 2015-10 & \onlinecite{veldhorst_two-qubit_2015} & p4\\
22.8 \textsuperscript{r} & qubit & 1Q & LD/h & Ge/Si & 1D & 2022-01 & \onlinecite{wang_ultrafast_2022} & p4\\
20.  & qubit & 1Q & ST/e & $\, ^{28}\text{Si/}\text{SiO}_2$ & 2D & 2022-02 & \onlinecite{jock_silicon_2022} & p4 and Fig. 2g the rightmost point\\
19. \textsuperscript{s} & qubit & 2Q & ST/e & GaAs/AlGaAs & 2D & 2018-11 & \onlinecite{noiri_fast_2018} & p4\\
18.8 \textsuperscript{t,k} & qubit & 1Q & LD/h\textsuperscript{k} & $\text{Si/}\text{SiO}_2$ & 1D & 2022-03 & \onlinecite{camenzind_spin_2022} & {SM} Fig. S11\\
17. \textsuperscript{u} & qubit & 1Q & LD/e\textsuperscript{v} & $\text{Si/}\text{SiO}_2$ & 2D & 2020-02 & \onlinecite{leon_coherent_2020} & {SM} Fig. 4c\\
15. \textsuperscript{w} & qubit & 1Q & ST/e & GaAs/AlGaAs & 2D & 2014-01 & \onlinecite{higginbotham_coherent_2014} & abstract and Fig. 4a\\
14. \textsuperscript{u} & qubit & 1Q & ST/e & GaAs/AlGaAs & 2D & 2020-07 & \onlinecite{jang_robust_2020} & p4\\
\bottomrule
\end{tabular}
\caption{Quality factors (part 1). Superscripts stand for the following. \textsuperscript{a}: Operation induced by a micromagnet and a baseband pulse. \textsuperscript{b}: The reference defines $Q=2 f_\mathrm{R} \TRabi$. To convert to our definition, we divide 2, and use $f_\mathrm{R}=3.9$ {MHz} and $\TRabi=113$ \unit{\micro\second} to get $Q=444$. \textsuperscript{c}: The reference defines $Q=J T_2^\mathrm{{CZ}}$, with $T_2^\mathrm{{CZ}}$ the decay time of the controlled-Z two-qubit gate. \textsuperscript{d}: The reference defines $Q^*=2 f_R T_2^R$. To convert to our definition, we divide by 2 and use the values $T_2^R=27.6\, \mu s$ and $f_R=3.45$ {MHz} given in Fig. 2c. \textsuperscript{e}: The reference defines $Q=2f_\mathrm{R} \TRabi$. To convert to our definition, we divide by 2. \textsuperscript{f}: At $J=100$ Mhz. The figure shows oscillation counts up to 1600 at different values of  $J$. \textsuperscript{g}: {CZ} gate. \textsuperscript{h}: At 1 K. \textsuperscript{i}: The reference defines $Q=2f_R T_2^*$. To convert to our definition, we divide by 2. \textsuperscript{j}: The reference defines $Q=2 f_R \TRabi$ and finds $Q \gg 87$. To convert to our definition, we divide by 2 and take the stated lower limit. \textsuperscript{k}: At 1.5 K. \textsuperscript{l}: The reference defines $Q=2f_R \TRabi$. To convert to our definition, we divide by 2. \textsuperscript{m}: With estimation and/or feedback. \textsuperscript{n}: The reference defines the quality factor as "the number of [exchange-induced singlet-triplet] oscillations before the [oscillation-signal] amplitude decays to $1/e$ of its initial value", corresponding qualitatively to $f_\mathrm{R} \TRabi$. \textsuperscript{o}: {EDSR}. \textsuperscript{p}: The reference uses $Q=T_2^\mathrm{R}/t_{180}$. To convert to our definition, we divide $Q$ by 2. \textsuperscript{q}: The reference defines $N_\mathrm{{CZ}}=f_\mathrm{{CZ}} T_2^\mathrm{{CZ}}$ with $T_2^\mathrm{{CZ}} = 8.3$ \unit{\micro\second}. \textsuperscript{r}: \derived the reference defines $Q=2f_\mathrm{R} T_2^*$. To convert to our definition, we divide by 2 and use $f_\mathrm{R}=542$ {MHz} and $T_2^*=42$ ns. \textsuperscript{s}: The reference defines $Q=2J T_2^*/h$. To convert to our definition, we divide by 2. \textsuperscript{t}: The reference defines $Q^*=2 f_R T_2^*$ and reports up to $Q^*=37.6$ \estimated as the rightmost blue point in Fig.~S11. To convert to our definition, we divide by 2. \textsuperscript{u}: The reference defines $Q=T_2^*/T_\mathrm{op}$. To convert to our definition, we divide by 2. \textsuperscript{v}: In a five-electron dot configuration. \textsuperscript{w}: The reference defines $Q=[J/2\pi \hbar]/\Gamma_\Sigma$ with $J/2\pi \hbar \approx 1.5$ {GHz} and $\Gamma_\Sigma \approx$ 100 {MHz} from Fig. 3d.
\label{qualityFactorMerged1}
}
\end{table}

%% file: figuresAndTables/table-qualityFactorMerged-2.txt
\rowcolors{2}{white}{gray!25}
\setcounter{rowcount}{-1}
\begin{table}
\begin{tabular}{@{}S[round-mode=none, round-pad=false, round-precision=3, table-alignment-mode = marker, table-number-alignment = center, table-align-text-after=false ]cccccccc@{ \stepcounter{rowcount} \tiny \therowcount }}
\toprule
{QFactor} & {QFactorType} & {\#Qubits} & {Qubit} & {Material} & {Host} & {Date} & {Reference} & {Source}\\
\midrule
12.5 \textsuperscript{a,b} & gate & 1Q & LD/e & $\text{Si/}\text{SiO}_2$ & 2D & 2026-01 & \onlinecite{kuno_robust_2026} & p4\\
12. \textsuperscript{c} & gate & 1Q & ST/e & GaAs/AlGaAs & 2D & 2022-07 & \onlinecite{kim_approaching_2022} & p4\\
11. \textsuperscript{d} & gate & 1Q & ST/e & GaAs/AlGaAs & 2D & 2023-03 & \onlinecite{yun_probing_2023} & p4\\
9.1 \textsuperscript{e} & qubit & 1Q & LD/h & Ge/Si & 1D & 2018-09 & \onlinecite{watzinger_germanium_2018} & p4\\
9. \textsuperscript{f} & gate & 1Q & LD/e & Si/SiGe & 2D & 2020-01 & \onlinecite{croot_flopping-mode_2020} & p4\\
8.5 \textsuperscript{g,h} & qubit & 2Q & LD/e & $\, ^{28}\text{Si/SiGe}$ & 2D & 2025-07 & \onlinecite{unseld_baseband_2025} & p3\\
8. \textsuperscript{i} & gate & 1Q & HY/e\textsuperscript{j,k} & GaAs/AlGaAs & 2D & 2017-07 & \onlinecite{malinowski_symmetric_2017} & Fig. 4c\\
7. \textsuperscript{l} & qubit & 1Q & LD/e & $\, ^{28}\text{Si/}\text{SiO}_2$ & 2D & 2023-01 & \onlinecite{gilbert_-demand_2023} & Fig. 3d, orange point\\
7. \textsuperscript{m} & qubit & 1Q & ST/e & GaAs/AlGaAs & 2D & 2024-02 & \onlinecite{berritta_real-time_2024} & p3 and Fig. 2d\\
7. \textsuperscript{n} & qubit & 2Q & ST/h & Ge/SiGe & 2D & 2024-10 & \onlinecite{zhang_universal_2024} & p6 and p7\\
6. \textsuperscript{o} & qubit & 1Q & ST/e & GaAs/AlGaAs & 2D & 2013-04 & \onlinecite{dial_charge_2013} & p3\\
5.1  & qubit & 1Q & ST/e & GaAs/AlGaAs & 2D & 2021-10 & \onlinecite{fedele_simultaneous_2021} & Fig. 2c Qubit 2\\
5. \textsuperscript{p} & gate & 1Q & LD/e & InAs & 1D & 2010-12 & \onlinecite{nadj-perge_spinorbit_2010} & p2 and Fig. 2d\\
4. \textsuperscript{i} & gate & 1Q & HY/e\textsuperscript{q,k} & GaAs/AlGaAs & 2D & 2017-07 & \onlinecite{malinowski_symmetric_2017} & Fig. 4c\\
1.35 \textsuperscript{r,b} & gate & 1Q & LD/e & $\text{Si/}\text{SiO}_2$ & 2D & 2026-01 & \onlinecite{kuno_robust_2026} & Fig. S1c\\
\bottomrule
\end{tabular}
\caption{Quality factors (part 2). Superscripts stand for the following. \textsuperscript{a}: Driven qubit. \textsuperscript{b}: The reference defines $Q=2f_R T_2^*$. To convert to our definition, we divide by 2. \textsuperscript{c}: With estimation and/or feedback. \textsuperscript{d}: $Q_R$. \textsuperscript{e}: The reference defines "the ratio $T_2^* / T_\mathrm{op}$", without using the term "quality factor" explicitly. To convert to our definition, we divide by 2 and use $f_\mathrm{R}=70$ {MHz} and $T_2^*=130$ ns. \textsuperscript{f}: The reference defines $Q=2 \TRabi f_R$. To convert to our definition, we divide by 2. \textsuperscript{g}: $\mathrm{{CZ}}_{12}$. \textsuperscript{h}: The reference uses $Q=T_2^\mathrm{{DCZ}}/t_\mathrm{{CZ}}$. To convert to our definition, we divide $Q$ by 2. \textsuperscript{i}: Approximate value at $f_\mathrm{R}=100$ {MHz}. \textsuperscript{j}: At a full sweet spot. \textsuperscript{k}: Resonant exchange qubit. \textsuperscript{l}: \estimated from Fig. 3d at $\Delta V_J=10$ {mV}. \textsuperscript{m}: With feedback. \textsuperscript{n}: We use Eq.~\eqref{eq:qualityFactorGate} with representative values $f=66$ {MHz} and $T_2^\mathrm{gate}=108$ ns. \textsuperscript{o}: The reference defines $Q=\TEcho J/2\pi \hbar$, giving the maximal value $Q=600$ for $\TEcho = 9$ \unit{\micro\second}. Converting to our definition results in the quality factor at least hundred times smaller, since the maximal $T_2^*$ was 90 ns, see Fig. 4b. \textsuperscript{p}: Rough approximation, given as `number of resolved Rabi oscillation periods'. \textsuperscript{q}: At a partial sweet spot. \textsuperscript{r}: Bare qubit.
\label{qualityFactorMerged2}
}
\end{table}

%% file: figuresAndTables/table-qubitArrays-1.txt
\rowcolors{2}{white}{gray!25}
\setcounter{rowcount}{-1}
\begin{table}
\begin{tabular}{@{}ccccccc@{ \stepcounter{rowcount} \tiny \therowcount }}
\toprule
{Dimensions} & {Functionality} & {Qubit} & {Material} & {Host} & {Date} & {Reference}\\
\midrule
$4 \times 4$ & device & LD/h & Ge/SiGe & 2D & 2023-08 & \onlinecite{borsoi_shared_2023}\\
$12$ & device & LD/e & Si/SiGe & 2D & 2024-05 & \onlinecite{neyens_probing_2024}\\
$12$ & device & LD/e & Si/SiGe & 2D & 2025-01 & \onlinecite{george_12-spin-qubit_2025}\\
$10$ & device & charge & Si:P & imp & 2022-06 & \onlinecite{kiczynski_engineering_2022}\\
$10$ & device & LD/h & Ge/SiGe & 2D & 2024-07 & \onlinecite{wang_operating_2024}\\
$9$ & device & LD/e & Si/SiGe & 2D & 2016-11 & \onlinecite{zajac_scalable_2016}\\
$9$ & device & LD/e & $\, ^{28}\text{Si/SiGe}$ & 2D & 2019-03 & \onlinecite{mills_shuttling_2019}\\
$3 \times 3$ & device & LD/e & GaAs/AlGaAs & 2D & 2020-12 & \onlinecite{mortemousque_coherent_2020}\\
$3 \times 3$ & device & LD/e & $\text{Si/}\text{SiO}_2$ & 2D & 2021-01 & \onlinecite{ruffino_integrated_2021}\\
$3 \times 3$ & device & LD/e & GaAs/AlGaAs & 2D & 2021-08 & \onlinecite{mortemousque_enhanced_2021}\\
$8$ & device & LD/e & GaAs/AlGaAs & 2D & 2019-04 & \onlinecite{volk_loading_2019}\\
$4 \times 2$ & device & LD/e & $\text{Si/}\text{SiO}_2$ & 1D & 2020-08 & \onlinecite{chanrion_charge_2020}\\
$4 \times 2$ & device & LD/h & Ge/SiGe & 2D & 2024-03 & \onlinecite{hsiao_exciton_2024}\\
$6$ & device & LD/e & GaAs/AlGaAs & 2D & 2022-07 & \onlinecite{knorzer_long-range_2022}\\
$6$ & device & LD/e & $\, ^{28}\text{Si/SiGe}$ & 2D & 2023-10 & \onlinecite{undseth_hotter_2023}\\
$6$ & device & LD/e & $\, ^{28}\text{Si/SiGe}$ & 2D & 2024-06 & \onlinecite{de_smet_high-fidelity_2024}\\
$6$ & device & LD/h & Ge/SiGe & 2D & 2026-09 & \onlinecite{kelly_identification_2026}\\
$6$ & device & ST/h & Ge/SiGe & 2D & 2026-01 & \onlinecite{tsoukalas_dressed_2026}\\
$5$ & device & LD/e & GaAs/AlGaAs & 2D & 2016-12 & \onlinecite{ito_detection_2016}\\
$5$ & device & LD/e & Si/SiGe & 2D & 2020-02 & \onlinecite{lawrie_quantum_2020}\\
$4$ & device & LD/e & GaAs/AlGaAs & 2D & 2014-03 & \onlinecite{takakura_single_2014}\\
$4$ & device & LD/e & GaAs/AlGaAs & 2D & 2014-05 & \onlinecite{delbecq_full_2014}\\
$4$ & device & LD/e & GaAs/AlGaAs & 2D & 2016-07 & \onlinecite{baart_nanosecond-timescale_2016}\\
$4$ & device & LD/e & GaAs/AlGaAs & 2D & 2016-08 & \onlinecite{otsuka_single-electron_2016}\\
$4$ & device & LD/e & GaAs/AlGaAs & 2D & 2017-06 & \onlinecite{fujita_coherent_2017}\\
$4$ & device & LD/e & GaAs/AlGaAs & 2D & 2018-08 & \onlinecite{ito_four_2018}\\
$4$ & device & LD/e & GaAs/AlGaAs & 2D & 2021-01 & \onlinecite{van_diepen_electron_2021}\\
$4$ & device & LD/e & $\, ^{28}\text{Si/SiGe}$ & 2D & 2023-07 & \onlinecite{zwerver_shuttling_2023}\\
$2 \times 2$ & device & LD/e & GaAs/AlGaAs & 2D & 2012-09 & \onlinecite{thalineau_few-electron_2012}\\
$2 \times 2$ & device & LD/e & GaAs/AlGaAs & 2D & 2018-04 & \onlinecite{mukhopadhyay_2x2_2018}\\
$2 \times 2$ & device & LD/e & $\text{Si/}\text{SiO}_2$ & 1D & 2020-09 & \onlinecite{duan_remote_2020}\\
$2 \times 2$ & device & LD/e & $\text{Si/}\text{SiO}_2$ & 1D & 2020-10 & \onlinecite{gilbert_single-electron_2020}\\
$2 \times 2$ & device & LD/e & $\text{Si/}\text{SiO}_2$ & 1D & 2020-12 & \onlinecite{ansaloni_single-electron_2020}\\
$2 \times 2$ & device & LD/e & $\text{Si/}\text{SiO}_2$ & 2D & 2022-12 & \onlinecite{niegemann_parity_2022}\\
$2 \times 2$ & device & LD/e & $\text{Si/}\text{SiO}_2$ & 1D & 2023-03 & \onlinecite{ansaloni_gate_2023}\\
$2 \times 2$ & device & LD/e & $\, ^{28}\text{Si/SiGe}$ & 2D & 2023-08 & \onlinecite{unseld_2d_2023}\\
$2 \times 2$ & device & LD/e & $\, ^{28}\text{Si/SiGe}$ & 2D & 2023-12 & \onlinecite{meyer_single-electron_2023}\\
$2 \times 2$ & device & LD/e & Si/SiGe & 2D & 2024-10 & \onlinecite{wang_highly_2024}\\
$2 \times 2$ & device & LD/e & $\text{Si/}\text{SiO}_2$ & 2D & 2025-07 & \onlinecite{hamonic_combining_2025}\\
$2 \times 2$ & device & ST/e & GaAs/AlGaAs & 2D & 2021-10 & \onlinecite{fedele_simultaneous_2021}\\
$2 \times 2$ & device & LD/h & Ge/SiGe & 2D & 2020-02 & \onlinecite{lawrie_quantum_2020}\\
$2 \times 2$ & device & LD/h & Ge/SiGe & 2D & 2020-07 & \onlinecite{hendrickx_single-hole_2020}\\
$2 \times 2$ & device & LD/h & Ge/SiGe & 2D & 2021-01 & \onlinecite{van_riggelen_two-dimensional_2021}\\
$2 \times 2$ & device & LD/h & Ge/SiGe & 2D & 2023-06 & \onlinecite{wang_probing_2023}\\
$2 \times 2$ & device & LD/h & Ge/SiGe & 2D & 2023-06 & \onlinecite{lawrie_simultaneous_2023}\\
$2 \times 2$ & device & LD/h & Ge/SiGe & 2D & 2024-07 & \onlinecite{ivlev_coupled_2024}\\
$3$ & device & charge & Si/SiGe & 2D & 2020-08 & \onlinecite{holman_microwave_2020}\\
$3$ & device & LD/e & GaAs/AlGaAs & 2D & 2009-11 & \onlinecite{gaudreau_tunable_2009}\\
$3$ & device & LD/e & GaAs/AlGaAs & 2D & 2013-02 & \onlinecite{busl_bipolar_2013}\\
$3$ & device & LD/e & GaAs/AlGaAs & 2D & 2016-01 & \onlinecite{baart_single-spin_2016}\\
\bottomrule
\end{tabular}
\caption{Qubit arrays (part 1).
\label{qubitArray1}
}
\end{table}

%% file: figuresAndTables/table-qubitArrays-2.txt
\rowcolors{2}{white}{gray!25}
\setcounter{rowcount}{-1}
\begin{table}
\begin{tabular}{@{}ccccccc@{ \stepcounter{rowcount} \tiny \therowcount }}
\toprule
{Dimensions} & {Functionality} & {Qubit} & {Material} & {Host} & {Date} & {Reference}\\
\midrule
$3$ & device & LD/e & GaAs/AlGaAs & 2D & 2017-06 & \onlinecite{flentje_linear_2017}\\
$3$ & device & LD/e & GaAs/AlGaAs & 2D & 2017-09 & \onlinecite{flentje_coherent_2017}\\
$3$ & device & LD/e & GaAs/AlGaAs & 2D & 2017-10 & \onlinecite{noiri_cotunneling_2017}\\
$3$ & device & LD/e & Si/SiGe & 2D & 2019-07 & \onlinecite{jones_spin-blockade_2019}\\
$3$ & device & LD/e & $\text{Si/}\text{SiO}_2$ & 2D & 2020-02 & \onlinecite{lawrie_quantum_2020}\\
$3$ & device & LD/e & $\, ^{28}\text{Si/}\text{SiO}_2$ & 2D & 2021-01 & \onlinecite{chan_exchange_2021}\\
$3$ & device & LD/e & GaAs/AlGaAs & 2D & 2021-05 & \onlinecite{kojima_probabilistic_2021}\\
$3$ & device & LD/e & GaAs/AlGaAs & 2D & 2022-08 & \onlinecite{nurizzo_controlled_2022}\\
$3$ & device & LD/e & $\, ^{28}\text{Si/SiGe}$ & 2D & 2022-09 & \onlinecite{noiri_shuttling-based_2022}\\
$3$ & device & ST/e & GaAs/AlGaAs & 2D & 2020-12 & \onlinecite{jang_individual_2020}\\
$3$ & device & LD/h & Ge/Si & 1D & 2018-08 & \onlinecite{froning_single_2018}\\
$2$ & device & charge & Si/SiGe & 2D & 2019-12 & \onlinecite{neyens_measurements_2019}\\
$2$ & device & HY/e & $\, ^{28}\text{Si/SiGe}$ & 2D & 2021-04 & \onlinecite{chen_detuning_2021}\\
$2$ & device & HY/e & Si/SiGe & 2D & 2021-11 & \onlinecite{ha_flexible_2021}\\
$2$ & device & ST/e & GaAs/AlGaAs & 2D & 2011-07 & \onlinecite{van_weperen_charge-state_2011}\\
$2$ & device & ST/e & GaAs/AlGaAs & 2D & 2018-10 & \onlinecite{croot_device_2018}\\
$2$ & device & ST/e & GaAs/AlGaAs & 2D & 2023-03 & \onlinecite{yun_probing_2023}\\
$2$ & device & LD/h & Ge/SiGe & 2D & 2019-03 & \onlinecite{hardy_single_2019}\\
$4$ & simulator & LD/e & $\, ^{28}\text{Si/SiGe}$ & 2D & 2019-06 & \onlinecite{sigillito_site-selective_2019}\\
$4$ & simulator & LD/e & GaAs/AlGaAs & 2D & 2019-09 & \onlinecite{kandel_coherent_2019}\\
$4$ & simulator & LD/e & GaAs/AlGaAs & 2D & 2020-06 & \onlinecite{qiao_conditional_2020}\\
$4$ & simulator & LD/e & GaAs/AlGaAs & 2D & 2020-07 & \onlinecite{qiao_coherent_2020}\\
$4$ & simulator & LD/e & GaAs/AlGaAs & 2D & 2021-01 & \onlinecite{qiao_long-distance_2021}\\
$4$ & simulator & LD/e & GaAs/AlGaAs & 2D & 2021-04 & \onlinecite{kandel_adiabatic_2021}\\
$4$ & simulator & LD/e & GaAs/AlGaAs & 2D & 2021-11 & \onlinecite{van_diepen_quantum_2021}\\
$4$ & simulator & ST/h & Ge/SiGe & 2D & 2024-10 & \onlinecite{zhang_universal_2024}\\
$2 \times 2$ & simulator & LD/e & GaAs/AlGaAs & 2D & 2020-03 & \onlinecite{dehollain_nagaoka_2020}\\
$3$ & simulator & LD/e & $\, ^{28}\text{Si/SiGe}$ & 2D & 2015-05 & \onlinecite{eng_isotopically_2015}\\
$3$ & simulator & LD/e & GaAs/AlGaAs & 2D & 2016-10 & \onlinecite{baart_coherent_2016}\\
$3$ & simulator & LD/e & GaAs/AlGaAs & 2D & 2018-11 & \onlinecite{noiri_fast_2018}\\
$3$ & simulator & LD/e & GaAs/AlGaAs & 2D & 2019-04 & \onlinecite{nakajima_quantum_2019}\\
$3$ & simulator & LD/e & GaAs/AlGaAs & 2D & 2020-03 & \onlinecite{nakajima_coherence_2020}\\
$2$ & simulator & LD/e & GaAs/AlGaAs & 2D & 2011-08 & \onlinecite{nowack_single-shot_2011}\\
$2$ & simulator & LD/e & GaAs/AlGaAs & 2D & 2011-09 & \onlinecite{brunner_two-qubit_2011}\\
$2$ & simulator & ST/e & GaAs/AlGaAs & 2D & 2012-04 & \onlinecite{shulman_demonstration_2012}\\
$2$ & simulator & ST/e & GaAs/AlGaAs & 2D & 2019-03 & \onlinecite{malinowski_fast_2019}\\
$6$ & processor & LD/e & $\, ^{28}\text{Si/SiGe}$ & 2D & 2022-09 & \onlinecite{philips_universal_2022}\\
$2 \times 2$ & processor & LD/e & $\, ^{28}\text{Si/SiGe}$ & 2D & 2025-07 & \onlinecite{unseld_baseband_2025}\\
$2 \times 2$ & processor & LD/h & Ge/SiGe & 2D & 2021-03 & \onlinecite{hendrickx_four-qubit_2021}\\
$3$ & processor & LD/e & Si/SiGe & 2D & 2021-06 & \onlinecite{takeda_quantum_2021}\\
$3$ & processor & LD/e & Si/SiGe & 2D & 2022-08 & \onlinecite{takeda_quantum_2022}\\
$2$ & processor & HY/e & $\, ^{28}\text{Si/SiGe}$ & 2D & 2023-03 & \onlinecite{weinstein_universal_2023}\\
$2$ & processor & LD/e & $\, ^{28}\text{Si/}\text{SiO}_2$ & 2D & 2015-10 & \onlinecite{veldhorst_two-qubit_2015}\\
$2$ & processor & LD/e & Si/SiGe & 2D & 2017-12 & \onlinecite{zajac_resonantly_2017}\\
$2$ & processor & LD/e & Si/SiGe & 2D & 2018-02 & \onlinecite{watson_programmable_2018}\\
$2$ & processor & LD/e & $\text{Si/}\text{SiO}_2$ & 2D & 2021-05 & \onlinecite{leon_bell-state_2021}\\
$2$ & processor & LD/e & $\, ^{28}\text{Si/SiGe}$ & 2D & 2022-04 & \onlinecite{mills_two-qubit_2022}\\
$2$ & processor & LD/h & Ge/SiGe & 2D & 2020-01 & \onlinecite{hendrickx_fast_2020}\\
\bottomrule
\end{tabular}
\caption{Qubit arrays (part 2).
\label{qubitArray2}
}
\end{table}

%% file: figuresAndTables/vocabulary.txt
\item[\val{1D}] The quantum dot host is quasi-one-dimensional, such as a nanowire. This is one of the values of \key{Host}.
\item[\val{2D}] The quantum dot host is quasi-two-dimensional, such as a two-dimensional electron (or hole) gas of a quantum well, an epilayer, or a similar heterostructure. This is one of the values of \key{Host}.
\item[\val{charge}] The qubit basis is represented by two different orbitals of a confined particle. Most often, the orbitals differ in their centers (positions), for example a pair of states localized each in one minimum of a double-well potential of a double dot. Unlike for spin-related qubits, we do not discriminate the carrier, be it electron, hole, or impurity. However, most charge-qubit experiments use electrons. This is one of the values of \key{Qubit}.
\item[\key{Coherence}] It describes in what type of experiment the corresponding coherence time has been measured. This attribute can have the following values: \val{$T_1$}, \val{$T_2^*$}, \val{$T_2^{\text{Echo}}$}, \val{$T_2^{\text{DynD}}$}, and \val{$T_2^{\text{Rabi}}$}.
\item[\key{Date}] Publication date. In tables, the year and month are given. In plots, the full date down to a day (if available) is used in sorting and horizontal shifts. If the reference is unpublished (only an arxiv version exists), we use the submission date of the first arvix version.
\item[\val{device}] The array functionality is at the lowest level out of those spanning the range from fabricating a sample to implementing a fully functional quantum processor. See Sec.~\ref{sec:arrayFunctionality} for details. This is one of the values of \key{Functionality}.
\item[\key{Dimensions}] The single dimension of a one-dimensional array, or the two dimensions of a two-dimensional array starting with the larger one.
\item[\key{Fidelity}] Operation fidelity. See Sec.~\ref{sec:fidelityDefinition} for details.
\item[\key{Functionality}] What functionality is available for the qubit-array initialization, manipulation, and measurement. See Sec.~\ref{sec:arrayFunctionality}. This attribute can have the following values: \val{device}, \val{simulator}, and \val{processor}.
\item[\val{gate}] The number of gate-signal oscillations before the signal amplitude drops to 1/e. See Sec.~\ref{sec:qualityFactor} for details. This is one of the values of \key{QFactorType}.
\item[\val{gate}] The operation is a gate. We use the following acronyms: the "C" in the two-qubit gates CZ, CPHASE, and CNOT stands for "controlled". SWAP is a two-qubit gate exchanging the two states. See \cite{mermin_quantum_2007} for more on how such gates are use in basic algorithms. Single qubit gates are often induced by ESR (electron spin resonance) or EDSR (electric dipole spin resonance). This is one of the values of \key{Operation}.
\item[\key{Host}] Where are the dots defined. This attribute can have the following values: \val{2D}, \val{1D}, and \val{imp}.
\item[\val{HY/e}] The qubit basis is represented by electron states having hybrid character, most often differing in both the spin and charge degrees of freedom. This is one of the values of \key{Qubit}.
\item[\val{imp}] The quantum dot host is an impurity. In this review we include only experiments on gated impurities accessed electrically. This is one of the values of \key{Host}.
\item[\val{initialize}] The operation is an initialization. Concerning three-qubit initializations, "GHZ" stands for Greenberger-Horne-Zeilinger state. This is one of the values of \key{Operation}.
\item[\val{LD/e}] The qubit basis is represented by the spin-up and spin-down state of a confined conduction-band electron (or many-electron) state. The acronym stands for the names of Daniel Loss and David DiVincenzo, who established the field of spin qubits by their Ref.~\cite{loss_quantum_1998}. This is one of the values of \key{Qubit}.
\item[\val{LD/h}] The same as \val{LD/e} but using valence-band holes instead of conduction-band electrons. This is one of the values of \key{Qubit}.
\item[\val{LD/i}] The same as \val{LD/e} but using impurity-bound electrons instead of conduction-band electrons confined by gates \cite{kane_silicon-based_1998}. This is one of the values of \key{Qubit}.
\item[\key{Material}] Material of the device. Most of the structures used in experiments are composites, for example a heterostructure with GaAs and AlGaAs layers is given as GaAs/AlGaAs. In some plots we merge groups of materials by retaining only the primary material where the qubit host particle is located. For example, under such a contraction both "Si/SiO\textsubscript{2}" and "Si/Ge" become "Si". When merging in this way, we further merge binary and ternary alloys of Al, Ga, In, As, and Sb under the key "III-V". For carbon-based materials, we use "SLG" for single-layer graphene (usually etched), "BLG" for bi-layer graphene, and "CNT" for a carbon nanotube. When merging, carbon-based materials become "C". We use "Si:P" and "Si:B" for a phosphorus and a boron impurity in silicon \recheck{(at the moment, we have no data on other impurities)}. When merging, they become "Si:X".
\item[\val{measure}] The operation is a measurement. This is one of the values of \key{Operation}.
\item[\key{Note}] Additional information concerning the \val{value} exists, indicated by a superscript. The correspodning note is given in the table caption.
\item[\key{Operation}] The operation type. This attribute can have the following values: \val{gate}, \val{measure}, and \val{initialize}.
\item[\val{processor}] The array functionality is at the highest level out of those spanning the range from fabricating a sample to implementing a fully functional quantum processor. See Sec.~\ref{sec:arrayFunctionality} for details. This is one of the values of \key{Functionality}.
\item[\key{QFactor}] The number of gate-signal oscillations within a certain characteristic time.
\item[\key{QFactorType}] It describes what type of quality factor has been evaluated. This attribute can have the following values: \val{qubit}, and \val{gate}.
\item[\val{qubit}] The number of gate-signal oscillations within the inhomogeneous dephasing time. See Sec.~\ref{sec:qualityFactor} for details. This is one of the values of \key{QFactorType}.
\item[\key{Qubit}] Which degree of freedom defines the qubit. It includes the carrier (such as the electron, hole, or impurity) and the subspace representing the qubit states' pair, such as the spin-1/2 (the LD qubit), the singlet and triplet (the ST qubit), orbitals (the charge qubit), and their hybrids, among others the exchange-only (EO) and resonant-exchange (RX) qubits. This attribute can have the following values: \val{charge}, \val{HY/e}, \val{LD/e}, \val{ST/e}, \val{LD/h}, \val{ST/h}, \val{LD/i}, and \val{ST/i}.
\item[\key{\#Qubits}] Number of qubits that the operation involves.
\item[\key{Reference}] The number referring to the bibliography at the end of the document. The bibliography is sorted alphabetically in first authors' names.
\item[\val{simulator}] The array functionality is at the second level out of those spanning the range from fabricating a sample to implementing a fully functional quantum processor. See Sec.~\ref{sec:arrayFunctionality} for details. This is one of the values of \key{Functionality}.
\item[\key{Source}] Where in the reference the value can be found. Typical sources are page X, denoted as pX, or figure Y, denoted as Fig.~Y. Multiple places might be given, if relevant. An example of the value source is "p3 and Fig. 2a".
\item[\val{ST/e}] The qubit basis is represented by the spin-singlet and spin-triplet state of a confined conduction-band electron pair \cite{levy_universal_2002}. This is one of the values of \key{Qubit}.
\item[\val{ST/h}] The same as \val{ST/e} but using valence-band holes instead of conduction-band electrons. This is one of the values of \key{Qubit}.
\item[\val{ST/i}] The same as \val{ST/e} but using impurity-bound electrons instead of conduction-band electrons confined by gates. This is one of the values of \key{Qubit}.
\item[\val{$T_1$}] Energy relaxation. The decay of population of the energy basis state(s) was measured. Even though relaxation is physically a different process than decoherence, to simplify the database design we include the relaxation time among the coherence times. This is one of the values of \key{Coherence}.
\item[\val{$T_2^*$}] Inhomogeneous dephasing. The experiment measured the decay of phase of an idling qubit without any echos. Most often, it corresponds to a Ramsey sequence: initialization -- $\pi/2$ pulse -- free precession for time $\tau$ -- $\pi/2$ pulse -- measurement. This is one of the values of \key{Coherence}.
\item[\val{$T_2^{\text{DynD}}$}] Dephasing under a dynamical decoupling protocol. The experiment measured the decay of phase of a qubit applying more than one echo-pulse. While there are several families of pulse sequences, we do not discriminate them. This is one of the values of \key{Coherence}.
\item[\val{$T_2^{\text{Echo}}$}] Dephasing under Hahn echo. The experiment measured the decay of phase of a qubit applying a single echo-pulse. A typical sequence is: initialization -- $\pi/2$ pulse -- free precession for time $\tau/2$ -- $\pi$ pulse -- free precession for time $\tau/2$ -- $\pi/2$ pulse -- measurement. This is one of the values of \key{Coherence}.
\item[\val{$T_2^{\text{Rabi}}$}] Dephasing of a driven qubit. Strictly speaking, this time should describe the decay of the relative phase of the two quasi-energy states in the rotating frame of reference. Usually, it is extracted as the observed decay time of Rabi oscillations. This is one of the values of \key{Coherence}.
\item[\key{Time}] (In tables on coherence). The timescale of the coherence decay observed in a certain type of experiment. The type is described by \key{Coherence}.
\item[\key{Time}] (In tables on operation times) The duration of the operation. If the reference gives a frequency $f$ for a gate signal, we convert it to time $t$ using $t = 1/2f$, see Eq.~\ref{eq:gateTime}.
\item[\key{Value}] This keyword wraps the value of Terminal oNode.